\documentclass[prb,twocolumn,amsmath,amssymb,showpacs,nofootinbib,floatfix]{revtex4}

\usepackage{graphicx,bm}

\makeatletter
\def\graphicscale{\twocolumn@sw{0.3}{0.4}}
\def\graphicthreescale{\twocolumn@sw{0.3}{0.4}}

\begin{document}

\title{Finite-size scaling at quantum transitions}

\author{Massimo Campostrini,$^1$ Andrea Pelissetto,$^2$ and Ettore Vicari$^1$} 

\address{$^1$ Dipartimento di Fisica dell'Universit\`a di Pisa
        and INFN, Largo Pontecorvo 3, I-56127 Pisa, Italy}
\address{$^2$ Dipartimento di Fisica dell'Universit\`a di Roma ``La Sapienza"
        and INFN, Sezione di Roma I, I-00185 Roma, Italy}

\date{\today}

\begin{abstract}

We develop the finite-size scaling (FSS) theory at quantum
transitions, considering generic boundary conditions, such as open and
periodic boundary conditions, and also the corrections to the leading
FSS behaviors.  Using renormalization-group (RG) theory, we generalize
Wegner's scaling Ansatz to the quantum case, classifying the different
sources of scaling corrections.  We identify nonanalytic corrections
due to irrelevant (bulk and boundary) RG perturbations and analytic
contributions due to regular backgrounds and analytic expansions of
the nonlinear scaling fields.

To check the general predictions, we consider the quantum XY chain in
a transverse field. For this model exact or numerically accurate
results can be obtained by exploiting its fermionic quadratic
representation.  We study the FSS of several observables, such as the
free energy, the energy differences between low-energy levels,
correlation functions of the order parameter, etc., confirming the
general predictions in all cases.  Moreover, we consider bipartite
entanglement entropies, which are characterized by the presence of
additional scaling corrections, as predicted by conformal field
theory.

\end{abstract}

\pacs{05.30.Rt,64.60.an,05.10.Cc,05.70.Jk}

\maketitle



\section{Introduction}

Finite-size effects in critical phenomena have been the object of
theoretical studies for a long
time.\cite{FB-72,Barber-83,Privman-90,PHA-91,PV-02} Finite-size
scaling (FSS) describes the critical behavior around a critical point,
when the correlation length $\xi$ of the critical modes becomes
comparable to the size $L$ of the system. For large sizes, this regime
presents universal features, shared by all systems whose transition
belongs to the same universality class.  Although formulated in the
classical framework, FSS also holds at zero-temperature quantum
transitions,\cite{SGCS-97} in which the critical behavior is driven by
quantum fluctuations.

The FSS approach is one of the most effective techniques for the
numerical determination of the critical quantities.  While
infinite-volume methods require that the condition $\xi \ll L$ is
satisfied, FSS applies to the less demanding regime $\xi\sim L$. More
precisely, FSS theory provides the asymptotic scaling behavior when
both $L,\xi\to\infty$ keeping their ratio $\xi/L$ fixed.  However
knowledge of the asymptotic behavior may not be enough to estimate the
critical parameters, because data are generally available for limited
ranges of parameter values and system sizes, which are often
relatively small.  Under these conditions, the asymptotic FSS
predictions are affected by sizable scaling corrections. Thus,
reliably accurate estimates of the critical parameters need a robust
control of the corrections to the asymptotic behavior.  This is also
important for a conclusive identification of the universality class of
the quantum critical behavior when it is {\em a priori} uncertain.

An understanding of the finite-size effects is also relevant for
experiments, when relatively small systems are considered, see, e.g.,
Ref.~\onlinecite{GKMD-08}, or in particle systems trapped by external
(usually harmonic) forces, as in recent cold-atom experiments, see,
e.g., Refs.~\onlinecite{CW-02,Ketterle-02,DRBOKS-07,BDZ-08,CV-09}.

In this paper we study FSS at quantum transitions,~\cite{Sachdev-book}
within the framework of the renormalization-group (RG) theory.  We
generalize Wegner's scaling Ansatz~\cite{Wegner-76} to quantum
systems. This allows us to characterize the corrections to the
asymptotic FSS behavior.  We predict nonanalytic scaling corrections
due to the irrelevant RG perturbations and analytic contributions
which are due to regular backgrounds and to the expansions of the
nonlinear scaling fields in terms of the Hamiltonian parameters.

To verify the RG predictions, we consider the quantum XY chain in a
transverse field, which represents a standard theoretical laboratory
for the understanding of quantum transitions. Its Hamiltonian can be
rewritten as a quadratic Hamiltonian of spinless
fermions.\cite{LSM-61,Katsura-62} Using this representation, several
quantities can be computed either exactly or very precisely by
numerical methods. This allows us to check the FSS predictions for the
scaling corrections of several observables. We consider the free
energy, the energy differences between the lowest-energy levels, and
the correlation functions of the order parameter, confirming the RG
results in all cases.

Finally, we discuss the FSS behavior of the bipartite entanglement
entropies of one-dimensional systems with an isolated critical point
with $z=1$.  We perform a detailed study in the XY model, verifying
the presence of further peculiar corrections,~\cite{CC-10,CCP-10}
beside those associated with the usual bulk and boundary RG irrelevant
perturbations.

The paper is organized as follows.  In Sec.~\ref{generalFSS} we
discuss the general RG approach to the study of FSS at quantum
transitions, considering, in particular, the case of isolated quantum
critical points between different quantum phases.  In
Sec.~\ref{XYchain} we present a thorough FSS analysis of the quantum
XY chain in a transverse field at the Ising transition, checking the
general asymptotic FSS predictions for several physically interesting
quantities.  Sec.~\ref{entang} is devoted to a FSS analysis of the
bipartite entanglement entropy at the quantum transition of the
quantum XY chain, focusing again on the nature of the scaling
corrections.  Finally, in Sec.~\ref{conclusions} we summarize our main
results and draw some conclusions.  A few appendices report some
formulas which are used in the paper.

\section{Finite-size scaling at a quantum transition}
\label{generalFSS}

In this section we summarize the main RG ideas behind FSS, providing
the framework to analyze critical finite-size effects in continuous
quantum transitions.  As their classical counterparts, they are
characterized by a diverging length scale $\xi$ and show universal
scaling properties, which can be analyzed in the framework of RG
theory.  Guided by the quantum-to-classical mapping, we generalize
Wegner's scaling Ansatz~\cite{Wegner-76} to quantum systems, and then
use it to predict the type of subleading corrections that are expected
in finite systems and/or at finite $T$, close to a continuous
transition.

We consider the standard case in which the quantum zero-temperature
transition of a $d$-dimensional system is characterized by two
relevant parameters $\mu$ and $h$, defined so that they vanish at the
critical point. Therefore, the quantum critical point is at
\begin{equation}
T=0,\quad \mu=0,\quad h=0.
\label{qcp}
\end{equation}
We assume the presence of a parity-like ${\mathbb Z}_2$ symmetry, as
it occurs, for instance, in Ising or O$(N)$ transitions which separate
a paramagnetic phase with $\mu > 0$ from a ferromagnetic phase with
$\mu<0$.  The parameter $\mu$ is coupled to a RG perturbation that is
invariant under the symmetry, while $h$ is associated with the leading
odd perturbation, generally related to the order parameter of the
transition.  As usual, we express the RG dimensions of the
perturbations associated with $\mu$ and $h$ in terms of the critical
exponents $\nu$ and $\eta$, as
\begin{equation}
y_\mu\equiv {1\over \nu},\quad
y_h \equiv {1\over 2} (d+z+2-\eta), 
\label{ymuh}
\end{equation}
where $z$ is the dynamic critical exponent associated with time and
temperature. At the critical point the low-energy scales vanish: the
gap $\Delta$ behaves as $\Delta\sim |\mu|^{z\nu}$ at $T=0$ and
$h=0$. The length scale $\xi$ associated with the critical modes
diverges as $\xi\sim |\mu|^{-\nu}$ at $T=0$ and $h=0$, and as $\xi\sim
T^{-1/z}$ at $\mu=0$ and $h=0$.

\subsection{Scaling law of the free energy}
\label{freeen}

Under the quantum-to-classical mapping, the inverse temperature
corresponds to the system size in an imaginary time direction.  Thus,
the temperature scaling at a quantum critical point in $d$ dimensions
is analogous to FSS in a corresponding $d+1$ classical system. If
$z=1$, which holds for transitions described by 2D conformal field
theories (CFTs)~\cite{Affleck-86} and for paramagnetic-ferromagnetic
transitions in $d$-dimensional O($N$) symmetric spin
systems~\cite{CSY-94} (Ising systems correspond to $N=1$), the quantum
transition corresponds to a classical $(d+1)$-dimensional equilibrium
transition, in which $1/T$ plays the role of an additional spatial
dimension.  There are also interesting cases in which $z\not=1$.  For
instance, superfluid-to-vacuum and Mott transitions of lattice
particle systems described by the Hubbard and Bose-Hubbard models have
$z=2$ when driven by the chemical potential.\cite{Sachdev-book} In
this case, the classical system is strongly anisotropic since $L\to
\lambda L$, $1/T\to \lambda^z/T$ under a RG rescaling.  Also for this
class of transitions, which include many dynamic off-equilibrium
transitions,\cite{Zia-review} FSS is quite well established. We can,
therefore, extend those results to the quantum case.

According to RG, close to a continuous transition the free energy
satisfies a general scaling law.  Extending the classical FSS Ansatz,
\cite{Wegner-76,Diehl-86,Privman-90,PHA-91,SGCS-97,SS-00} we
generally write the free-energy density as
\begin{eqnarray}
F(L,T,\mu,h) &=& F_{\rm reg}(L,T,\mu,h) 
\label{Gsing-RG-1}\\
&+& F_{\rm sing}(u_l,u_t,u_\mu,u_h,\{v_i\},\{\widetilde{v}_{i}\}),\nonumber
\end{eqnarray}
where $F_{\rm reg}$ is a nonuniversal function which is analytic at
the critical point, and $F_{\rm sing}$ bears the nonanalyticity of the
critical behavior and its universal features.  The arguments of
$F_{\rm sing}$ are the so-called nonlinear scaling fields.
\cite{Wegner-76} They are analytic nonlinear functions of the model
parameters, which are associated with the eigenoperators that
diagonalize the RG flow close to the RG fixed point.

The scaling fields $u_\mu$ and $u_h$ are the relevant scaling fields
related to the model parameters $\mu$ and $h$.  The scaling fields
$u_l$ and $u_t$ are also relevant, with RG dimensions 
\begin{equation}
y_l = 1,\qquad  y_t = z, 
\label{ylz}
\end{equation}
respectively, and are associated with the finite spatial size $L$
($u_l\sim 1/L$) and with the temperature ($u_t\sim T$).

Beside the relevant scaling fields, we should also consider an
infinite number of irrelevant scaling fields with negative RG
dimensions.  We distinguish them into two families, the bulk scaling
fields $\{v_i\}$ and the surface scaling fields
$\{\widetilde{v}_{i}\}$, with RG dimensions $y_i$ and
$\widetilde{y}_i$, respectively. The first set is the only one that
occurs in the infinite-volume limit and whenever there are no
boundaries in the system, for instance, for periodic boundary
conditions (PBC). They are responsible for the scaling corrections to
the leading critical behavior in the infinite-volume limit. Using
standard notation, assuming that they are ordered so that $|y_1| \le
|y_2| \le \ldots$, we set
\begin{equation}
\omega = -y_1.
\label{defomega}
\end{equation}
In the presence of a surface, an additional set of boundary RG
perturbations should be included. Their RG dimensions depend on the
type of boundary conditions and, in particular, on the type of surface
critical behavior one is considering. As before, we set
\begin{equation}
\omega_s=-\widetilde{y}_1, 
\label{defomegas}
\end{equation}
where $\widetilde{y}_1$ is the dimension of the leading boundary
operator.

The singular part of the free energy is expected to satisfy the 
scaling equation
\begin{eqnarray}
&& F_{\rm sing}(u_l,u_t,u_\mu,u_h,\{v_i\},\{\widetilde{v}_i\})  =  
\label{Fsing-scaling}\\
&& \lambda^{-(d+z)} 
   F_{\rm sing}(\lambda u_l,\lambda^z u_t,\lambda^{y_\mu} u_\mu,
    \lambda^{y_h} u_h,\{\lambda^{y_i} v_i\},
    \{\lambda^{\widetilde{y}_i}\widetilde{v}_i\}),
\nonumber 
\end{eqnarray}
where $\lambda$ is arbitrary. In the FSS case it is useful to take
$\lambda = 1/u_l$, obtaining
\begin{eqnarray}
F_{\rm sing} =  
u_l^{d+z} {\cal F}\left[ {u_t\over u_l^z}, {u_\mu\over u_l^{y_\mu}}, 
{u_h\over u_l^{y_h}}, 
\left\{v_i\over u_l^{y_i}\right\},
\left\{\widetilde{v}_{i}\over u_l^{\widetilde{y}_{i}}\right\} \right].
\label{scalsing}
\end{eqnarray}
An important question concerns the universality of the function $\cal
F$. Since scaling fields are arbitrarily normalized, universality
holds apart from a normalization of each argument and an overall
constant. Therefore, given two different models, if ${\cal F}_1$ and
${\cal F}_2$ are the corresponding scaling functions, we have
\begin{eqnarray}
&& {\cal F}_1(x_1,x_2,x_3,\{y_i\} ,\{\widetilde{y}_i\}) 
\nonumber \\ 
&& \quad = A 
{\cal F}_2(c_1 x_1,c_2 x_2,c_3 x_3,\{d_i y_i\} ,
\{\widetilde{d}_i \widetilde{y}_i\}),
\end{eqnarray}
where all constants $A$, $c_i$, $d_i$, $\widetilde{d}_i$,
are nonuniversal.

To go further we must discuss how the nonlinear scaling fields depend
on the control parameters $\mu$, $h$, $L$, and $T$.  First, it is
natural to assume that the bulk scaling fields $u_\mu$, $u_h$, and
$v_i$ do not depend on the temperature and the size of the system,
i.e. they do not mix with $1/L$ and $T$.  This hypothesis is quite
natural for systems with short-range interactions. Under a RG
transformation, the transformed bulk couplings only depend on the
local Hamiltonian, hence they are independent of the boundary. Taking
into account the assumed ${\mathbb Z}_2$ symmetry and the even/odd
properties of $\mu$/$h$, close to the critical point the relevant
scaling fields $u_\mu$ and $u_h$ can be generally expanded as
\begin{eqnarray}
&&u_\mu = \mu + b_\mu \mu^2 + O(\mu^3,h^2\mu),   \label{umu}\\
&&u_h = h + b_h \mu h + O(h^3,\mu^2 h),   \label{uh}
\end{eqnarray}
where $b_\mu$ and $b_h$ are nonuniversal constants.  As for the
irrelevant scaling fields, they are usually nonvanishing at the
critical point.

Let us now discuss the scaling fields $u_l$ and $u_t$, which are
associated with the size of the $(d+1)$ dimensional system.  For
classical systems in a box of size $L^d$ with PBC and, more generally,
for translation-invariant boundary conditions, it is usually assumed
that $u_l=1/L$, exactly.  This assumption, which has been verified in
many instances---for instance, in the two-dimensional Ising
model---and extensively discussed in Ref.~\onlinecite{SS-00}, can be
justified as follows. Consider a lattice system and a decimation
transformation which reduces the number of lattice sites by a factor
$2^d$.  In the absence of boundaries and for short-range interactions,
the new (translation-invariant) couplings are only functions of the
couplings of the original lattice and are independent of $L$, while
$L\to L/2$. Since the flow of $L$ is independent of the flow of the
couplings, we expect
\begin{equation}
u_l = L^{-1}\qquad {\rm for}\;\; {\rm PBC}. 
\label{ulpbc}
\end{equation}
This condition does not generally hold for non-translation invariant
systems.  We thus assume that $u_l$ is an arbitrary function of
$1/L$.  For $L\to\infty$ it can be expanded as
\begin{equation}
u_l = L^{-1} + b L^{-2} + \ldots
\label{ull}
\end{equation}
Note that, if we define an effective size 
\begin{equation}
L_{e}= L - b,
\label{defleff}
\end{equation}
the scaling field becomes 
\begin{equation}
u_l = 1/L_{e} + O(L_{e}^{-3}).
\label{defulobc}
\end{equation}
Hence, by using $L_{e}$, all subleading corrections due to $b/L^2$ are
eliminated in {\em any} observable. Of course, this does not imply
that $1/L$ corrections are absent in {\em any} observable, as such
type of corrections may have other sources (we will come back to this
point in Sec.~\ref{III.E}).  Such an observable-independent shift is
often considered in FSS studies of systems with boundary conditions
that differ from the periodic ones, see, e.g.,
Refs.~\onlinecite{Hasenbusch-09,Hasenbusch-12,DGHHRS-12}, which also
provide some evidence of the presence of $L^{-2}$, $L^{-3}$
corrections in the scaling field $u_l$.

Let us now consider the thermal scaling field $u_t$. To clarify the
issue, let us first assume that $z=1$, so that the quantum system is
equivalent to a classical $(d+1)$-dimensional system. The classical
system is, however, weakly anisotropic: couplings in the thermal
direction differ from those in the spatial one. Moreover, the
anisotropy depends on the model parameters. In classical weakly
anisotropic systems universality is obtained only after transforming
to an isotropic system by means of a scale transformation, see
Refs.~\onlinecite{DC-09,Kastening-12} and references therein.
Therefore, we define
\begin{equation}
u_t={T\over c(\mu,h)} \approx {T\over c_0} [1 + b_t \mu + O(\mu^2,h^2)],
\label{utt}
\end{equation}
where $c(\mu,h)$ is an appropriate nonuniversal function, $c_0\equiv
c(0,0)$ and $b_t$ is a constant.  The function $c$ may be identified
with the speed of sound. More precisely, if $E({\bf k})$ is the
dispersion relation of the model, which is assumed to be spatially
isotropic, we define $c = |\nabla_{\bf k} E|_{k_{\rm min}}$, where
$k_{\rm min}$ is the value of $k$ where the energy has an absolute
minimum.  Relation (\ref{utt}) is expected to hold also when
$z\not=1$, although in this case the rescaling factor is not related
to the sound velocity.

The scaling variables $v_i u_l^{-y_i}$ and $\widetilde{v}_i
u_l^{-\widetilde{y}_i}$, corresponding to the irrelevant scaling
fields, vanish for $L\to\infty$ since $y_i$ and $\widetilde{y}_i$ are
negative.  Thus, provided that $F_{\rm sing}$ is finite and
nonvanishing in this limit, we can expand the singular part of the
free energy as
\begin{eqnarray} 
F_{\rm sing} &\approx&
u_l^{d+z} {\cal F}( u_t/u_l^z, u_\mu/u_l^{y_\mu}, u_h/u_l^{y_h})
+\label{Fsing-RG}\\
&+&v_1 u_l^{d+z+\omega} {\cal F}_\omega( u_t/u_l^z, u_\mu/u_l^{y_\mu}, u_h/u_l^{y_h})
+ ...\nonumber\\
&+&\widetilde{v}_1 
   u_l^{d+z+\omega_s} {\cal F}_s( u_t/u_l^z, u_\mu/u_l^{y_\mu}, u_h/u_l^{y_h})
+ ...,
\nonumber
\end{eqnarray}
where we retain only the contributions of the dominant (least)
irrelevant bulk and surface scaling fields, of RG dimensions $-\omega$
and $-\omega_s$, respectively.  Note that the expansion
(\ref{Fsing-RG}) is only possible below the upper critical
dimension.\cite{Fisher-74-Temple} Above it, $F_{\rm sing}$ is singular
and cannot be expanded as in Eq.~(\ref{Fsing-RG}).  The breakdown of
this expansion causes a breakdown of the hyperscaling relations and
allows us to obtain the mean-field exponents.

The scaling functions ${\cal F}$, ${\cal F}_{\omega}$, and ${\cal
  F}_s$ are expected to be universal, apart from multiplicative
normalizations and normalizations of the scaling fields. This implies
that, within the given universality class, they are independent of the
microscopic features of the model.  However, they depend on the nature
of the boundary conditions. Note also the presence of the variable
$u_t/u_l^z$, which corresponds to the so-called shape factor in
classical transitions: the universal scaling functions depend on the
shape of the finite system that is considered.

Finally, we should also take into account the regular part $F_{\rm
  reg}$ of the free energy, see Eq.~(\ref{Gsing-RG-1}).  For classical
systems, in the absence of boundaries, e.g., for PBC, $F_{\rm reg}$ is
assumed to be independent of $L$, or, more plausibly, to depend on $L$
only through exponentially small
corrections.\cite{Privman-90,PHA-91,SS-00} Extending this result to
the quantum case, we assume that $F_{\rm reg}$ does not depend on $T$.
Instead, we see no reason why $F_{\rm reg}$ should not depend on $L$
for generic spatial boundary conditions.  Therefore, we assume a
regular expansion in powers of $1/L$ such as
\begin{eqnarray}
F_{\rm reg}(\mu,h,L) = F_{{\rm reg},0}(\mu,h) 
+ {1\over L} F_{{\rm reg},1}(\mu,h) + \ldots
\label{favb}
\end{eqnarray}
where $F_{{\rm reg},0}(\mu,h)$ is the bulk contribution, the only one
present when PBC are considered.

Expansions (\ref{Fsing-RG}) and (\ref{favb}) allow us to compute all
scaling corrections. As usual, we introduce the scaling variables
\begin{equation}
w \equiv \mu L^{1/\nu},\qquad \kappa \equiv h L^{y_h},
\qquad \tau \equiv {1\over c_0} T L^z,
\label{scalvar}
\end{equation}
and 
\begin{equation}
w_{e} \equiv \mu L^{1/\nu}_{e}, 
\qquad \kappa_{e} \equiv h L^{y_h}_{e},
\qquad \tau_{e} \equiv {1\over c_0} T L_{e}^z,
\end{equation}
where $L_{e}$ is defined above in Eq.~(\ref{defleff}).  Then, we have
\begin{eqnarray}
{u_\mu\over u_l^{y_\mu}} &\approx& 
   w \left(1 - {b_1\over \nu} {1\over L}\right) + 
   {b_\mu\over L^{1/\nu}} w^2 \approx w_{e} + {b_\mu\over L^{1/\nu}} 
       w_{e}^2, 
\label{espansione-rapporti}\\
{u_h\over u_l^{y_h}} &\approx& 
   \kappa \left(1 - {y_h b_1} {1\over L}\right) + 
   {b_h\over L^{1/\nu}} w\kappa \approx \kappa_{e} + 
    {b_h\over L^{1/\nu}} \kappa_{e}  w_{e}, 
\nonumber \\
{u_t\over u_l^{z}} &\approx& 
   \tau \left(1 - {z b_1} {1\over L}\right) + 
   {b_t\over L^{1/\nu}} \tau w \approx \tau_{e} + 
   {b_t\over L^{1/\nu}} \tau_{e}  w_{e}, 
\nonumber 
\end{eqnarray}
where we have included the leading scaling correction. If $\nu < 1$
and $w$, $\kappa$, and $\tau$ are used as FSS variables, the leading
correction is of order $1/L$.  If instead, one uses $w_{e}$,
$\kappa_{e}$, and $\tau_{e}$, the leading correction decreases faster,
as $L^{-1/\nu}$.

Collecting all terms and using $L_{e}$ as basic length scale, we can
write
\begin{eqnarray}
F(L,T,\mu,h)&=&
F_{{\rm reg},0}(\mu,h) + 
L_{e}^{-(d+z)} {\cal F}( \tau_{e}, w_{e}, \kappa_{e})
\nonumber \\
&+& v_1 L_{e}^{-(d+z+\omega)} 
   {\cal F}_\omega(\tau_{e}, w_{e}, \kappa_{e})
\label{F-expansion-final}\\
&+& \widetilde{v}_1 
   L_{e}^{-(d+z+\omega_s)} 
  {\cal F}_s(\tau_{e}, w_{e}, \kappa_{e})
\nonumber \\
&+&  {1\over L} F_{{\rm reg},1}(\mu,h) + 
    {1\over L^2} F_{{\rm reg},2}(\mu,h) +
    \ldots
\nonumber 
\end{eqnarray}
where $v_1$ and $\widetilde{v}_1$ are computed at the critical point.
The missing corrections are of order (relative to the leading singular
term $L_{e}^{-(d+z)}$) $L_{e}^{-1/\nu}$, $L_{e}^{-|y_2|}$,
$L_{e}^{-|\widetilde{y}_2|}$ (they are due to the singular part of the
free energy), and of order $L^{d+z-3}$ (they are due to the regular
part of the free energy). The last three terms appearing in
Eq.~(\ref{F-expansion-final}) represent boundary contributions, hence
they should not be present for PBC. Morever, in this case we also have
$L_{e} = L$. Finally, note that, since the corrections of order
$L_{e}^{-1/\nu}$ are due to the expansion of the scaling fields, they
are always proportional to $w$, see Eq.~(\ref{espansione-rapporti}),
thus they vanish for $\mu = 0$.

To summarize, the RG expansion (\ref{F-expansion-final}) provides
information on the corrections to the asymptotic behavior.  There are
several different sources of scaling corrections:
\begin{itemize}
\item[(i)] The irrelevant RG perturbations which give generally rise to
$O(L^{-\omega})$ corrections, where $\omega$ is a universal exponent
associated with the leading irrelevant RG perturbation.
\item[(ii)] Corrections arising from the expansion of the scaling fields
$u_\mu$, $u_h$, and $u_t$ in terms of the Hamiltonian parameters. 
They give rise to corrections of order $L^{-1/\nu}$ and 
are absent for $\mu = 0$. 
\item[(iii)] Corrections arising from the analytic background term of the
free energy.
\item[(iv)] The irrelevant RG perturbations asssociated with the boundary
conditions, which are of order $L^{-\omega_s}$.  They are
absent in the absence of boundaries, such as PBC.
\item[(v)] The $O(1/L)$ boundary corrections arising from the nontrivial
analytic $L$-dependence of the scaling field $u_l$, Eq.~(\ref{ull}).
They are absent in the absence of boundaries.  The leading correction can be
taken into account by simply redefining the length scale $L$, i.e., by using 
$L_{e}$ instead of $L$, cf. Eq.~(\ref{defleff}).
\end{itemize}
Eqs.~(\ref{scalsing}) and (\ref{Fsing-RG}) give the generic scaling
form of the free-energy density.  However, in certain cases the
behavior is more complex due to the appearance of logarithmic
terms.\cite{Wegner-76} They may be due to the presence of marginal RG
perturbations, as it happens in Berezinskii-Kosterlitz-Thouless
transitions in U(1)-symmetric
systems,\cite{BKT-73,AGG-80,Hasenbusch-05,PV-13} or to resonances
between the RG eigenvalues, as it occurs in transitions belonging to
the 2D Ising universality class~\cite{Wegner-76,CHPV-02} or to the 3D
O$(N)$-vector universality class in the large-$N$
limit.\cite{DGHHRS-12,PV-99} We should also mention that peculiar FSS
behaviors, for instance, a modulation of the leading amplitudes, are
observed in quantum particle systems at fixed chemical potential when
an infinite number of level crossings occurs as the system size
varies, and in the so-called XX chain in a transverse external
field.~\cite{CV-10-XX,CMV-11-jstat,CTV-12}

Several interesting quantities can be obtained by taking derivatives
of the free energy. For example, in particle systems whose relevant
parameter $\mu$ is a linear function of the chemical potential, the
FSS of the particle density is obtained by differentiating
Eq.~(\ref{F-expansion-final}) with respect to $\mu$, i.e.  $\rho\sim
{\partial F/\partial\mu}$. Therefore, for $h=0$, we obtain
\begin{eqnarray}
\rho &=&
 \rho_{\rm reg}(\mu) +  {1\over L} \rho_{\rm reg,1}(\mu) + ...
\nonumber \\
&& + L_{e}^{-(d+z-y_\mu)} {\cal D}( w_{e}, \tau_{e})+ ...
\label{pade}
\end{eqnarray}
We note that the regular term represents the leading term when
$d+z-y_\mu>0$, which is the case for most physically interesting
systems. The compressibility can be obtained by taking an additional
derivative with respect to $\mu$.

\subsection{Scaling law in the infinite-volume limit}
\label{freeeninfv}

We can also write down general scaling laws for the quantum critical
behavior in the infinite-volume limit. We start again from
Eq.~(\ref{Fsing-scaling}), setting in this case $u_l = 0$ and $\lambda
= u_t^{-1/z}$.  The free-energy density scales as
\begin{eqnarray}
F &=& F_{\rm reg}(\mu,h) +  \label{scalsingiv}\\
&+& u_t^{d/z+1} {\cal F}\left( u_\mu u_t^{-y_\mu/z}, u_h u_t^{-y_h/z},
\left\{ {v_i u_t^{-y_i/z}}\right\} \right), \nonumber
\end{eqnarray}
where, as explained above, we assume that $F_{\rm reg}(\mu,h)$ is 
$T$ independent.
For $h=0$ and $u_t\to 0$, we can expand the free energy as 
\begin{eqnarray} 
F &\approx&
u_t^{d/z+1} {\cal A}\left(u_\mu u_t^{-1/(z\nu)} \right)  
+\label{Fsingiv-RG}\\
&+&u_t^{d/z+1+\omega/z} v_1 {\cal A}_\omega
      \left( u_\mu u_t^{-1/(z\nu)} \right) + 
F_{\rm reg}(\mu,0) + \ldots
\nonumber
\end{eqnarray}
The specific heat is obtained by differentiating the previous expression:
\begin{eqnarray}
C_V &\equiv& T {\partial^2 F\over \partial T^2} =
{u_t^{d/z} \over c} \left[ {\cal C}\left( u_\mu u_t^{-1/(z\nu)} \right)  +
\right. 
\label{cviv}\\
&+& \left. 
u_t^{\omega/z} {\cal C}_\omega\left( u_\mu u_t^{-1/(z\nu)}  \right) 
+ \ldots \right]
\nonumber
\end{eqnarray}
Notice that there are no contributions from the regular part of the
free energy. At the critical point $\mu = 0$, Eq.~(\ref{cviv}) predicts
\begin{equation}
C_V \sim T^{d/z}\left[ 1 + O(T^{\omega/z})\right].
\label{cvnbeh}
\end{equation}

\subsection{FSS of the low-energy scales}
\label{lowespco}

At $T=0$ and $h=0$, any low-energy scale, and, in particular, the
energy difference of the lowest-energy levels, is expected to show the
asymptotic FSS behavior
\begin{eqnarray}
c(\mu) \Delta(L,\mu)&=&
L_{e}^{-z} \left[ {\cal D}( w_{e})
+v_1 L_{e}^{-\omega} 
   {\cal D}_\omega(w_{e}) \right.
\nonumber \\
&+&\left. \widetilde{v}_1 
   L_{e}^{-\omega_s} 
  {\cal D}_s(w_{e}) + \ldots\right],
\label{Deltasca}
\end{eqnarray}
where $c$ is the function providing the relation of $u_t$ with $T$,
cf. Eq.~(\ref{utt}).  Such a factor is needed to take into account
that energies are expressed in terms of the temperature $T$, while the
right-hand side contains the spatial dimension $L_{e}$.  The neglected
corrections are of order $L^{-2}$, $L^{-1/\nu}$, $L^{-|y_2|}$,
$L^{-|\widetilde{y}_2|}$.  The scaling functions ${\cal D}_\#$ are
universal, apart from multiplicative factors and a normalization of
their argument.  For $w_{e}\to \infty$, ${\cal D}(w)\sim w^{z\nu}$ to
ensure $\Delta\sim \mu^{z\nu}$ for $\mu>0$ (paramagnetic phase) in the
infinite-voume limit.

\subsection{FSS of the two-point correlation function}
\label{twopcf}

We now consider the correlation functions of the order-parameter field
$\phi({\bf x},t)$, for example, the equal-time two-point function,
\begin{equation}
G({\bf x},{\bf y})=\langle \phi({\bf x},t) \phi({\bf y},t) \rangle.
\label{gxy}
\end{equation}
For vanishing magnetic field, the leading scaling behavior is given by
\begin{eqnarray}
&&G({\bf x},{\bf y};T,\mu,L) \approx u_l^{d+z-2+\eta} \times 
\label{gxysca}\\
&& \times \; {\cal G}(u_l {\bf x}, u_l {\bf y}; u_t/u_l^z, 
u_\mu/u_l^{y_\mu}).
\nonumber
\end{eqnarray}
Eq.~(\ref{gxysca}) is only valid for $L\to \infty$, $|{\bf x}-{\bf
  y}|\to \infty$ with $|{\bf x} - {\bf y}|/L$ fixed.  Instead, if one
takes the limit at fixed $|{\bf x}-{\bf y}|$, no singular behavior is
observed in the FSS limit.  Corrections to Eq.~(\ref{gxysca}) arise
from two different sources.  First of all, there are the corrections
due to the scaling fields with negative RG dimensions.  Moreover,
there are corrections which we will call {\em field-mixing}
terms. Indeed, the order-parameter field $\phi$ is in general a linear
combination,
\begin{equation}
\phi = \sum_{i=1} a_i {\cal O}_{h,i},
\label{phiexp}
\end{equation}
of the odd fixed-point operators ${\cal O}_{h,i}$, which satisfy
\begin{equation}
\langle {\cal O}_{h,i}({\bf r}) \; {\cal O}_{h,j}({\bf s}) \rangle \sim |{\bf r}
- {\bf s}|^{-d_i-d_j}
\label{ohico}
\end{equation}
at the critical point.  The associated dimensions $d_i$ (we assume
here $d_1 < d_2 < d_3 \ldots$) are related to the RG dimensions
defined before by $d_i = d + z - y_{h,i}$.  The leading odd operator
${\cal O}_h\equiv {\cal O}_{h1}$ is associated with the leading
nonlinear scaling field of RG dimension $y_h$ given in
Eq.~(\ref{ymuh}).  Eq.~(\ref{gxysca}) represents the contribution of
the leading operator ${\cal O}_h$ since
\begin{equation}
d + z - y_h = {1\over 2} (d +z - 2 + \eta). 
\label{dzeta}
\end{equation}
Beside, we should also consider the contributions of all subleading
operators that have the same symmetry properties of the order
parameter. Hence, we end up with the expansion
\begin{eqnarray}
&&G({\bf x},{\bf y};T,\mu,L) \approx \sum_{jk} u_l^{2 (d+z)-y_{hj}-y_{hk}} \times 
\label{gxysca-2}\\
&& \times \; {\cal G}_{jk}(u_l {\bf x}, u_l {\bf y}; u_t/u_l^z, 
u_\mu/u_l^{y_\mu}, \{v_i u_l^{-y_i}\}, 
                   \{\widetilde{v}_i u_l^{-\widetilde{y}_i}\}).
\nonumber 
\end{eqnarray}
In the case of OBC, also boundary operators should be considered.

Let us consider the space integral of the correlation function
(\ref{gxy}), defined as
\begin{equation}
\chi_{\bf y} \equiv  \sum_{{\bf x}} G({\bf y},{\bf x}).\label{chidef}
\end{equation}
In the case of PBC, since translation invariance holds, $\chi_{\bf y}$
is independent of ${\bf y}$.  In the presence of a boundary, this is
no longer the case.  As long as ${\bf y}$ is fixed in the FSS
limit, the leading scaling behavior is always the same, while scaling
corrections are expected to depend on ${\bf y}$.  The asymptotic FSS
expansion of $\chi_{\bf y}$ for $h=0$ and $T=0$ is expected to be
\begin{eqnarray}
&&\chi_{\bf y}(\mu,L) = L^{2-z-\eta} \Bigl[ {\cal X}(w) + 
L^{-\omega} {\cal X}_\omega(w)  + L^{-1} {\cal X}_{s1}(w) 
\nonumber\\
&&\;+  L^{-\omega_s} {\cal X}_{s2}(w) +
L^{-1/\nu} {\cal X}_u(w) + L^{y_{h2} - y_h} {\cal X}_h(w) + \ldots
\Bigr] 
\nonumber \\ 
&&\; + B_\chi(\mu,L) ,
\label{chisclaw}
\end{eqnarray}
where $w$ is defined in Eq.~(\ref{scalvar}), the scaling functions
${\cal X}_\#$ are universal apart from multiplicative factors and a
normalization of the argument, and $y_{h2}$ is the RG dimension of the
next-to-leading operator which is odd under $h\to -h$ (this term is
due to the field mixing).  The corrections of order $L^{-1/\nu}$ arise
from the expansion (\ref{umu}) of the scaling field $u_\mu$.  Finally,
$B_\chi$ is an analytic background term which represents the
contribution to the integral of points $\bf x$ such that $|{\bf
  x}-{\bf y}|\ll L$. It is the analogue of the analytic part of the
free energy, see Eq.~(\ref{Gsing-RG-1}).  Therefore, the leading
scaling corrections for $\chi$ scale as $L^{-\zeta}$ with
\begin{equation}
\zeta = {\rm min}\,[\omega,\,1,\, \omega_s,\,1/\nu,\,2-z-\eta,
        y_h - y_{h2}].
\label{expzeta}
\end{equation}
It is important to note that $\chi_{\bf y}$ should not be confused
with the magnetic susceptibility, which is a macroscopic quantity
obtained by differentiating the free energy with respect to the
magnetic field.

One can also consider a correlation length $\xi$ associated with the
critical modes. Since $\xi$ has RG dimension 1, in the FSS limit we
obtain an expansion analogous to Eq.~(\ref{chisclaw}), i.e.
\begin{eqnarray}
&&\xi(\mu,L) = L \Bigl[ {\cal Y}(w) + 
L^{-\omega} {\cal Y}_\omega(w)  + L^{-1} {\cal Y}_{s1}(w) 
\nonumber\\
&&\;+  L^{-\omega_s} {\cal Y}_{s2}(w) +
L^{-1/\nu} {\cal Y}_u(w) + L^{y_{h2} - y_h} {\cal Y}_h(w) + \ldots
\Bigr] \nonumber \\
&&\; + B_\xi(\mu,L).
\label{xisca}
\end{eqnarray}
Here $B_\xi(\mu,L)$ is a background term depending on the explicit
definition of the correlation length.  For example, we may define a
second-moment correlation length by using the two-point function
(\ref{gxy}), as
\begin{eqnarray}
\xi^2 = {1\over 2d\chi_{\bf 0}} \sum_{\bf x} {\bf x}^2 G({\bf 0},{\bf x})
\label{xidef} 
\end{eqnarray}
where the point ${\bf y}=0$ is at the center of the system.
In the case of PBC, one may consider the more convenient definition
\begin{equation}
\xi^2 \equiv  {1\over 4 \sin^2 (p_{\rm min}/2)} 
{\widetilde{G}({\bf 0}) - \widetilde{G}({\bf p})\over 
\widetilde{G}({\bf p})},
\label{xidefpb}
\end{equation}
where ${\bf p} = (p_{\rm min},0,...)$, $p_{\rm min} \equiv 2 \pi/L$,
and $\widetilde{G}({\bf p})$ is the Fourier transform of $G({\bf x})$.
For these definitions there are two background contributions.  One
contribution is due to $\chi_{\bf 0}$ and scales as $L^{\eta+z-2}$.  A
second one is due the sum appearing in the numerator of expression
(\ref{xidef}) and scales as $L^{\eta+z-4}$.  This second contribution
is subleading with respect to the first one, hence
\begin{equation}
B_{\xi}(\mu,L) =L^{\eta+z-1} B_{\chi}(\mu,L).
\label{bxil}
\end{equation}
We thus conclude that scaling
corrections are analogous to those for $\chi$, i.e. scale as
$L^{-\zeta}$, where $\zeta$ is given in Eq.~(\ref{expzeta}).

\subsection{Dimensionless RG invariant quantities}
\label{drgi}

Dimensionless RG invariant quantities are particularly useful to
investigate the critical region. Examples of such quantities are the
ratio
\begin{equation}
R_\xi\equiv \xi/L,
\label{rxidef}
\end{equation}
where $\xi$ is any length scale related to the critical modes, for
example the one defined in Eq.~(\ref{xidef}), and ratios of the
correlation function $G$ at different distances, e.g.,
\begin{equation}
R_g({\bf X},{\bf Y}) = \ln[G({\bf 0},{\bf X}L)/G({\bf 0},{\bf Y}L)]
\label{rgxy}
\end{equation}
where the point ${\bf x}=0$ is at the center of the system.
We denote them generically by $R$.

According to FSS, at $T=0$ and $h=0$, they must behave as
\begin{eqnarray}
R(\mu,L) &=& {\cal R}(w) + L^{-1/\nu}\, {\cal R}_u(w) + 
L^{-\omega} \,{\cal R}_\omega(w) 
\nonumber \\
&+&
 L^{-1}\, {\cal R}_{s1}(w) +  L^{-2}\, {\cal R}_{s2}(w) + 
 L^{-\omega_s}\, {\cal R}_{\omega_s}(w) 
\nonumber \\
&+& L^{-(y_h-y_{h2})} {\cal R}_{h}(w) 
 + \ldots,
\label{rscaf}
\end{eqnarray}
where $w=\mu L^{1/\nu}$.  Note the presence of the corrections of
order $L^{-1}$, which are related to the fact that $L$ is used as a
normalizing length scale in Eqs.~(\ref{rxidef}) and (\ref{rgxy}).  One
could have equally used $L_{e}$ or $u_l^{-1}$, obtaining RG invariant
quantities that have the same universal scaling behavior, but that
differ by corrections of order $1/L$.

The scaling function ${\cal R}(w)$ is universal apart from a trivial
normalization of the argument.  In particular, the limit
\begin{equation}
\lim_{L\to\infty} R(0,L) =  {\cal R}(0)
\label{rzero}
\end{equation}
is universal within the given universality class, i.e., it is
independent of the microscopic details of the model, although it
depends on the shape of the finite volume and on the boundary
conditions.  Since ${\cal R}_u$ arises from the next-to-leading
$O(\mu^2)$ term of the expansion (\ref{umu}) of the scaling fields, we
have ${\cal R}_u \sim w^2 {\cal R}'(w)$ (with an unknown coefficient
because the expansion of the scaling field is usually unknown). Thus,
this term does not contribute at $\mu=0$.  Note also that the boundary
term is absent for PBC.  Moreover, in the case of $R_\xi$ with $\xi$
defined as in Eq.~(\ref{xidef}), there is also a $L^{-2+z+\eta}$
correction due to the background $B_\xi$ [this term is absent in the
  case of $R_g$ as defined in Eq.~(\ref{rgxy})].

A popular method to determine the critical point uses the finite-size
behavior of $R$ as a function of $L$ and $\mu$. Indeed, if
\begin{equation}
\lim_{\mu\to0^-}\lim_{L\to\infty}R(\mu,L)>
\lim_{L\to\infty}R(0,L)>
\lim_{\mu\to0^+}\lim_{L\to\infty}R(\mu,L)
\label{monbeh}
\end{equation}
or viceversa, one can define $\mu_{\rm cross}$ by requiring
\begin{equation}
R(\mu_{\rm cross},L) = R(\mu_{\rm cross},2L).
\label{rcross}
\end{equation}
The crossing point $\mu_{\rm cross}$ converges to $\mu=0$ with
corrections of order $L^{-1/\nu-\zeta}$.  Here $\zeta={\rm
  min}[\omega,1,\omega_s,(y_h-y_{h2})]$ for generic boundary
conditions breaking translation invariance and $\zeta={\rm
  min}[\omega,(y_h-y_{h2})]$ for PBC. In the presence of backgrounds,
we should also include the background corrections.  For instance, in
the case of $R_\xi$, cf. Eq.~(\ref{rxidef}), we have
\begin{equation}
\zeta={\rm  min}[\omega,1,\omega_s,2-z-\eta,y_h-y_{h2}].
\label{zetarxi}
\end{equation}

\section{FSS in the quantum XY chain}
\label{XYchain}

\subsection{The 1D XY model} \label{III.A}

The quantum XY chain in a transverse field is a standard theoretical
laboratory for quantum transitions.  Its Hamiltonian is
\begin{eqnarray}
&&H(J,g) = - \sum_{x=-L/2+1}^{L/2} {\cal H}_x, \label{Isc}\\
&&{\cal H}_x =  {J\over 2}[ (1+\gamma) \sigma^{(1)}_x \sigma^{(1)}_{x+1}
+ (1-\gamma) \sigma^{(2)}_x \sigma^{(2)}_{x+1}] + g \sigma^{(3)}_x,
\nonumber
\end{eqnarray}
where $\sigma^{(i)}$ are the Pauli matrices.  We set $J=1$ and
consider chains with open and periodic boundary conditions (OBC and
PBC, respectively). We always take $L$ even, setting the origin at the
center of the domain, more precisely at one of the two central sites,
so that $-L/2+1\le x \le L/2$.  For $\gamma=0$ we recover the
so-called XX chain in a transverse external field.

For any $\gamma\not=0$ the model undergoes a
quantum transition at
\begin{equation}
\mu\equiv g - 1 = 0,
\label{tauc}
\end{equation}
separating a quantum paramagnetic phase for $\mu>0$ from a quantum
ferromagnetic phase for $\mu<0$.  The transition belongs to the 2D
Ising universality class, hence its critical behavior is associated
with a 2D conformal field theory (CFT) with central charge
$c=1/2$. The critical exponents take the values $z=1$, $\nu=1$, and
$\eta=1/4$. The structure of the subleading corrections for Ising
systems was discussed in
Refs.~\onlinecite{Henkel-87,Reinicke-87a,Reinicke-87b,CHPV-02}.  In
particular, Reinicke\cite{Reinicke-87b} analyzed the subleading
corrections for the XY chain at the critical point.  If the finite
system is translation invariant---this is the case of PBC---the most
relevant subleading operators have RG dimension $-2$.  They belong to
the identity family and can be expressed by using the Virasoro
generators as $Q_2^I \bar{Q}_2^I$ and $Q_4^{I} + \bar{Q}_4^I$, where
$Q_2 = L_{-2}|I\rangle$, $Q_4 = (L_{-2}^2 - {3\over 5} L_{-4})
|I\rangle$.  The analysis of Ref.~\onlinecite{Reinicke-87b} shows that
the spin-zero operator $Q_2^I \bar{Q}_2^I$ (which can be related to
the energy-momentun tensor) is absent, as it also occurs in the
classical 2D Ising model.\cite{CHPV-02} The second (spin-four)
operator gives instead rise to scaling corrections that are
proportional to $3/4 - \gamma^2$.  The primary field associated with
the energy family controls the off-critical behavior.  The
corresponding scaling field is $u_\mu \sim
\mu/\gamma$.\cite{Reinicke-87a,footnote-Reinicke}

According to the analysis of Ref.~\onlinecite{CHPV-02}, the next
subleading operator ($Q_4^\epsilon + \overline{Q}_4^\epsilon$ in their
notations) gives correction of order $L^{-3}$. Such an operator is odd
under duality transformations, which also hold for the XY model to
some extent, as we discuss below. Hence, we expect it to contribute
only at quadratic order (hence it gives corrections of order
$L^{-6}$), as in occurs in the 2D Ising model. \cite{CHPV-02} If this
term is absent, the next-to-leading correction is related to the
leading spin-6 operator in the identity family, which has RG dimension
$-4$.

The quantum XY Hamiltonian (\ref{Isc}) can be mapped onto a quadratic
Hamiltonian of spinless fermions by a Jordan-Wigner
transformation,\cite{LSM-61,Katsura-62} which can be straightforwardly
diagonalized. One obtains\cite{Katsura-62}
\begin{equation}
H = \sum_k {E}(k) \left(a^+_k a_k - {1\over2}\right),
\label{H-harmonic}
\end{equation}
where $a^+_k$ and $a_k$ are fermionic creation-annihilation operators and 
\begin{equation}
E(k) = 
   2 \left[ g^2 + \gamma^2 - 2 g \cos k + (1 - \gamma^2) \cos^2 k \right]^{1/2}.
\label{Ek-XY}
\end{equation}
The set of values of $k$ which must be summed over and the allowed states
depend on the boundary conditions.\cite{LSM-61,Katsura-62,Pfeuty-70,BG-85}

In the limit $T\to 0$, the relevant modes are those with the lowest
energy, i.e., those with $k\approx 0$. For $k\to 0$, the energy $E(k)$
can be expanded as
\begin{equation}
E(k)^2 = c(\mu,\gamma)^2 
    \left[u_\mu(\mu,\gamma)^2 + k^2 + v_1(\mu,\gamma) k^4 + O(k^6)\right],
\label{dispersion-E}
\end{equation}
where 
\begin{eqnarray}
c(\mu,\gamma) &=& 2 \sqrt{\gamma^2 + \mu} ,
\label{cmugamma}\\
u_\mu(\mu,\gamma) &=& {\mu\over \sqrt{\gamma^2 + \mu}},
\label{umugamma}
\\
v_1(\mu,\gamma) &=& {3 - 4 \gamma^2 - \mu \over 12 (\gamma^2 + \mu) }.
\label{v1mugamma}
\end{eqnarray}
As we shall see, $u_\mu(\mu,\gamma)$ and $v_1(\mu,\gamma)$ play the
role of the nonlinear scaling fields associated with $\mu$ and with
the leading irrelevant operator. Note that 
\begin{equation}
v_1(0,\gamma) = 0 \quad {\rm for}\;\;\gamma=\gamma_i = \sqrt{3}/2. 
\label{gammaimp}
\end{equation}
Therefore, provided that $v_1(\mu,\gamma)$ is the correct scaling
field, no corrections of order $L^{-\omega}=L^{-2}$ due to the leading bulk 
irrelevant operator are
expected for the {\em improved} value $\gamma=\gamma_i$ in any
observable (note, however, that corrections of order $L^{-\omega-1}$ 
do not cancel out). 
The identification of $u_\mu$ and $v_1$ as scaling fields
is in full agreement with the CFT results of
Refs.~\onlinecite{Reinicke-87a,Reinicke-87b}, but it goes beyond that,
since it conjectures the expression of the scaling field also outside
the critical point, i.e. for $\mu\not=0$.

Finally, let us discuss duality in the XY model.\cite{FS-78,NO-11} An
exact transformation can be defined for the model with $\gamma = 1$.
In this case, one should consider slightly modified Hamiltonians with
free boundary conditions. One can consider\cite{NO-11}
\begin{equation}
H_{d1}(J,g) = - J \sum_{x=-L/2+1}^{L/2} 
\sigma^{(1)}_x \sigma^{(1)}_{x+1}
- g \sum_{x=-L/2+1}^{L/2-1}\sigma^{(3)}_x,
\label{Ham-dual}
\end{equation}
--- it differs from Hamiltonian (\ref{Isc}) because of the absence
of the magnetic field on site $L/2$---or
\begin{eqnarray}
&&H_{d2}(J,g) = - J \sum_{x=-L/2+1}^{L/2}
\sigma^{(1)}_x \sigma^{(1)}_{x+1}
\nonumber \\
&& \quad
- g \sum_{x=-L/2+1}^{L/2}\sigma^{(3)}_x - 
J \sigma^{(1)}_{L/2},
\label{Ham-dual2}
\end{eqnarray}
in which there is an additional magnetic field along the $x$ direction
at site $x=L/2$.  For these Hamiltonians one can show that there
exists a transformation $U$, such that
\begin{equation}
U H(J,g) U^+ = H(g,J) = g H(1,J/g). 
\label{transfh}
\end{equation}
It follows that there exists an exact correspondence between the
energy level for $g > 1$ and those for $g < 1$ (again we set
$J=1$). The presence of a symmetry can be guessed from the expression
of $E(k)$, see Eq.~(\ref{Ek-XY}). Indeed, for $\gamma = 1$ the energy
levels satisfy
\begin{equation}
E(k,g) = g E(k,1/g), 
\label{appdua}
\end{equation}
where we have written explicitly the $g$ dependence.  It is important
to note that exact duality holds only for Hamiltonians
(\ref{Ham-dual}) and (\ref{Ham-dual2}).  For different types of
boundary conditions, boundary terms break duality, hence there is no
direct correspondence between the states with $g < 1$ and those with
$g > 1$.

It is interesting to understand physically why boundary conditions
break duality. This is due to the different nature of the ground
states for $g > 1$ and $g < 1$. Indeed, if $g$ is large, we expect
Hamiltonian (\ref{Isc}) to have a nondegenerate ground state with the
spins aligned in the $z$ direction.  On the other hand, if $g$ is
small we expect a doubly degenerate ground state, with the spins
aligned either in the $x$ direction or in the $-x$ direction. Since
the degeneracy of the ground state is different for $g > 1$ and $g <
1$, there cannot be an exact duality symmetry. In order to have exact
duality, one must therefore change the model so that (at least) the
degeneracy of the ground state does not depend on $g$.  If we consider
the Hamiltonian (\ref{Ham-dual2}), this condition is realized by
lifting the degeneracy of the ground state for $g < 1$: the magnetic
field along the $x$ direction at site $x=L/2$ makes the ground state
nondegenerate, with all spins pointing in the $+x$ direction. If we
instead consider the Hamiltonian (\ref{Ham-dual}), duality is obtained
at the price of making the ground state doubly degenerate for any
value of $g$. To show this, note that $[\sigma_{L/2}^{(1)},H_{d1}] =
0$. Thus, the Hilbert space can be decomposed into two subspaces
${\cal H}_\pm$, such that $\sigma_{L/2}^{(1)} \psi_\pm = \pm
\psi_\pm$.  If we restrict $H_{d1}$ to ${\cal H}_+$ we obtain
Hamiltonian (\ref{Ham-dual2}) for a chain of length $L-1$.  Hence, the
ground state in ${\cal H}_+$ is nondegenerate for all values of $g$.
If we restrict $H_{d1}$ to ${\cal H}_-$ we obtain $U H_{d2} U^+$,
where $U = \prod_{x=-L/2+1}^{L/2-1} \sigma_x^{(3)}$, hence we obtain
the same spectrum as that of $H_{d1}$ restricted to ${\cal
  H}_+$. Thus, for Hamiltonian (\ref{Ham-dual}) duality is obtained by
making each state doubly degenerate.

Finally, let us note that a remnant of duality is also
present for $\gamma \not = 1$. This guarantees that the 
transition always occurs at $g=1$. Indeed, consider the transformation 
\begin{equation}
\mu = - {\mu'\over 1 + \mu'/\gamma^2}.
\end{equation}
Then, we have 
\begin{equation}
u_\mu(\mu,\gamma) = - u_\mu(\mu',\gamma),
\qquad
c(\mu) = {4\gamma^2\over c(\mu')},
\end{equation}
so that, at points that only differ by the sign of $u_\mu$, the
low-$k$ behavior of $E(k)$ is the same, apart from a change of
normalization.

\subsection{Free energy} \label{III.B}

The free energy of the quantum XY model can be directly related to the
finite-size free energy of the 2D Ising model. If we consider a strip
of width $M$, the Ising free energy density is given by (we use $K$
instead of $\beta$ to avoid confusion with the quantum case and write
$F = - T f_{\rm Is}$)\cite{Onsager-44,FF-69}
\begin{eqnarray}
&&f_{\rm Is}(K,M) = \textstyle{1\over2} \ln (2 \sinh 2K) +
\\
&&\;\; + {1\over 2} \int_0^{2\pi} {dk\over 2\pi} \epsilon(k) 
+ {1\over M} \int_0^{2\pi} {dk\over 2\pi} 
    \ln\left[1 + e^{-M \epsilon(k)}\right],
\nonumber
\end{eqnarray}
where 
\begin{eqnarray}
&& \epsilon(k) = \ln \left[\zeta(k) + \sqrt{\zeta(k)^2 - 1}\right], 
\nonumber \\
&& \zeta(k) = \cosh 2 K \coth 2 K - \cos {k/2}.
\end{eqnarray}
For large values of $M$, the leading behavior of the finite-size
correction term is obtained by expanding $\epsilon(k)$ for $k\to 0$,
since $\epsilon(k)$ is positive and has an absolute minimum at
$k=0$. Close to the critical point $K_c = (1 + \sqrt{2})/2$, if
$\delta = K_c-K > 0$ (paramagnetic phase), we obtain
\begin{equation}
\epsilon(k) = 4 \left(\delta^2 + k^2/64\right)^{1/2}.
\end{equation}
This expression allows us to rewrite 
\begin{eqnarray}
f_{\rm Is}(K,M) = 
F_{\rm reg}(K) - {2 \delta^2\over\pi} \ln \delta^2 + 
\delta^2 g_{\rm Is}(\delta M),
\label{FscalIsing}
\end{eqnarray}
where 
\begin{eqnarray}
F_{\rm reg}(K) &=& {2 G\over \pi} + {\ln 2\over 2} - \delta \sqrt{2} 
\\
&+&
    {2\over \pi} \delta^2 (1 + \ln 2 - \pi/2) + O(\delta^2),
\nonumber \\
g_{\rm Is}(x) &=& {4\over \pi x^2} \int_0^\infty dy
   \ln \{ 1 + \exp[- 4 (x^2 + y^2)^{1/2}]\},
\nonumber 
\end{eqnarray}
and $G$ is Catalan's constant.
For the XY model we obtain a similar result. Defining 
\begin{equation}
F_{\rm XY} = - T\ln {\rm Tr}\, e^{-\beta H}, 
\quad \beta = 1/T,
\label{fxydef}
\end{equation}
we obtain \cite{Katsura-62}
\begin{eqnarray}
F_{\rm XY}(\mu,T,\gamma) &=& 
- {1\over 2} \int_{-\pi}^{\pi} {dk\over 2\pi} E(k) 
\\
&-& {1\over \beta} \int_{-\pi}^{\pi} {dk\over 2\pi} 
    \ln\left[1 + e^{-\beta E(k)}\right],
\nonumber 
\end{eqnarray}
where $E(k)$ is given in Eq.~(\ref{Ek-XY}).  Also in this case the
large $\beta$ behavior is obtained by expanding $E(k)$ around $k=0$,
i.e., by using Eq.~(\ref{dispersion-E}).  At leading order, we
reobtain Eq.~(\ref{FscalIsing}) with some different normalization
constants:
\begin{eqnarray}
F_{\rm XY}(\mu,T,\gamma) &=& F_{\rm XY, reg}(\mu,\gamma) +
\label{FscalXY}\\
&+& {2 a u_\mu^2 \over \pi} \ln u_\mu^2 
- a \,u_\mu^2 \,g_{\rm Is}(b u_\mu/u_t),
\nonumber
\end{eqnarray}
where $u_t = T/c$ with $c$ defined in Eq.~(\ref{cmugamma}), and
\begin{equation}
a = {c\over 16},\qquad  b = {1\over 4}.
\end{equation}
Note the presence of the logarithmic terms, which are due to a
resonance between the identity operator of RG dimension 2 and the
thermal operator of RG dimension 1.\cite{Wegner-76} In principle,
logarithmic terms should appear in all observables and both at leading
and at subleading order. However, extensive analyses of the 2D Ising
model \cite{SS-00,Queiroz-00,Salas-01,ONGP-01,IH-02,CHPV-02,IH-09,%
  CGNP-11,Izmailian-12,Izmailian-13} have identified logarithmic
corrections only in a very few cases.

We wish now to compute the corrections to Eq.~(\ref{FscalXY}). For
this purpose we set $\lambda = u_\mu/u_t$ and consider the expansion
of 
\begin{equation}
B(x,\lambda,T) = {E(xu_t)\over {E}_c(xu_t,u_\mu)} 
\label{ratee}
\end{equation}
where  ${E}_c(x,u_\mu) =
(x^2 + u_\mu^2)^{1/2}$, in powers of $u_t$, keeping $x$ and $\lambda$
fixed. We obtain
\begin{eqnarray}
{1\over c} B(x,\lambda,T) &=& 1 + B_c(x,\lambda,T) 
\nonumber \\
&=& 1 + \sum_{n=2} u_t^n B_{c,n}(x,\lambda).
\label{bcn}
\end{eqnarray}
Since $\beta {E}_c(x u_t,u_\mu) = {E}_c(x,\lambda)/c$
is independent of $u_t$ and $B_c(x,\lambda,T)\sim u_t^2$, we can write
\begin{eqnarray}
&& \beta \int_0^\pi dk \ln (1 + e^{-\beta E(k)}) = 
\int_0^{\beta\pi} dx\Bigl\{ 
    \ln\left[1 + e^{-{E}_c(x,\lambda)}\right]  
\nonumber \\
  && + 
    \ln \Bigl[1 - {e^{-{E}_c(x)} {E}_c(x,\lambda) B_c(x,\lambda,T)
       \over 1 + e^{-{E}_c(x,\lambda)} } + \ldots \Bigr]\Bigr\}
\end{eqnarray}
Each term $B_{c,n}(x,\lambda)$ of Eq.~(\ref{bcn}) increases as a power
of $x$ for $x\to \infty$. Therefore, the integrand
vanishes exponentially as $x\to \infty$ order by order, 
and we can extend the upper
limit of integration to $+\infty$, making an exponentially small
error. The second term in braces can then be expanded in powers of
$u_t$, proving that the free energy admits an expansion in powers of
$u_t$ at $\lambda$ fixed.

Let us now compute the first correction of order $u_t^2$. Proceeding
as discussed above, we obtain
\begin{eqnarray}
&& \beta \int_0^\pi dk \ln\left[1 + e^{-\beta E(k)}\right] =
\int_0^{\infty} dx
    \ln\left[1 + e^{-{E}_c(x,\lambda)}\right] 
\nonumber \\
&&\; - {u^2_t\over2} v_1(\mu,\gamma)
\int_0^{\infty} dx
    {x^4 (x^2 + \lambda^2)^{-1/2} \over 
     1 + \exp \sqrt{x^2 + \lambda^2} }.
\end{eqnarray}
Note that the corrections of order $u_t^2$ are proportional to
$v_1(\mu,\gamma)$. Hence, it is natural to identify this quantity as
the scaling field associated with the leading irrelevant operator. We
will confirm this conjecture in the next sections.

\subsection{Scaling of the energy gap: periodic and antiperiodic boundary
conditions} \label{III.C}

We wish now to compute the finite-size behavior of the difference $\Delta$
between the energy of the lowest excited states and that of the 
ground state, extending the results of Ref.~\onlinecite{Henkel-87}. 
For PBC 
we shall show that $\Delta$ admits an expansion of the form
\begin{equation}
\Delta_P 
= {c(\mu,\gamma)\over 2L} \left[ \Delta_{P0}(\widetilde{w}) + 
    {v_1(\mu,\gamma)\over L^2} \Delta_{P2}(\widetilde{w}) + O(L^{-4})\right]
\label{expansion-Delta-wtilde}
\end{equation}
where 
\begin{equation}
\widetilde{w} = u_\mu L.
\end{equation}
An analogous expansion holds also for antiperiodic boundary conditions (ABC).
Such a result confirms the identification of $u_\mu$ and $v_1$ as 
nonlinear scaling fields. Note that, if
\begin{equation}
w \equiv  {\mu L\over \gamma}
\label{defwsi}
\end{equation}
 is used as scaling variable and $c$ is replaced by its leading
 behavior $2\gamma$, then we have
\begin{equation}
\Delta= {\gamma\over L}\left[ \Delta_{P0}(w) + 
    {1\over L} \Delta_{P1}(w) + O(L^{-2})\right].
\label{expansion-Delta-w}
\end{equation}
The corrections of order $L^{-1}$, which vanish at the critical point 
$w=0$, are due to the expansion of the nonlinear scaling field $u_\mu$ 
and of the sound velocity $c$ (in the 
general case, they would be of order $L^{-1/\nu}$). 

\begin{figure}[tbp]
\includegraphics*[scale=\graphicscale]{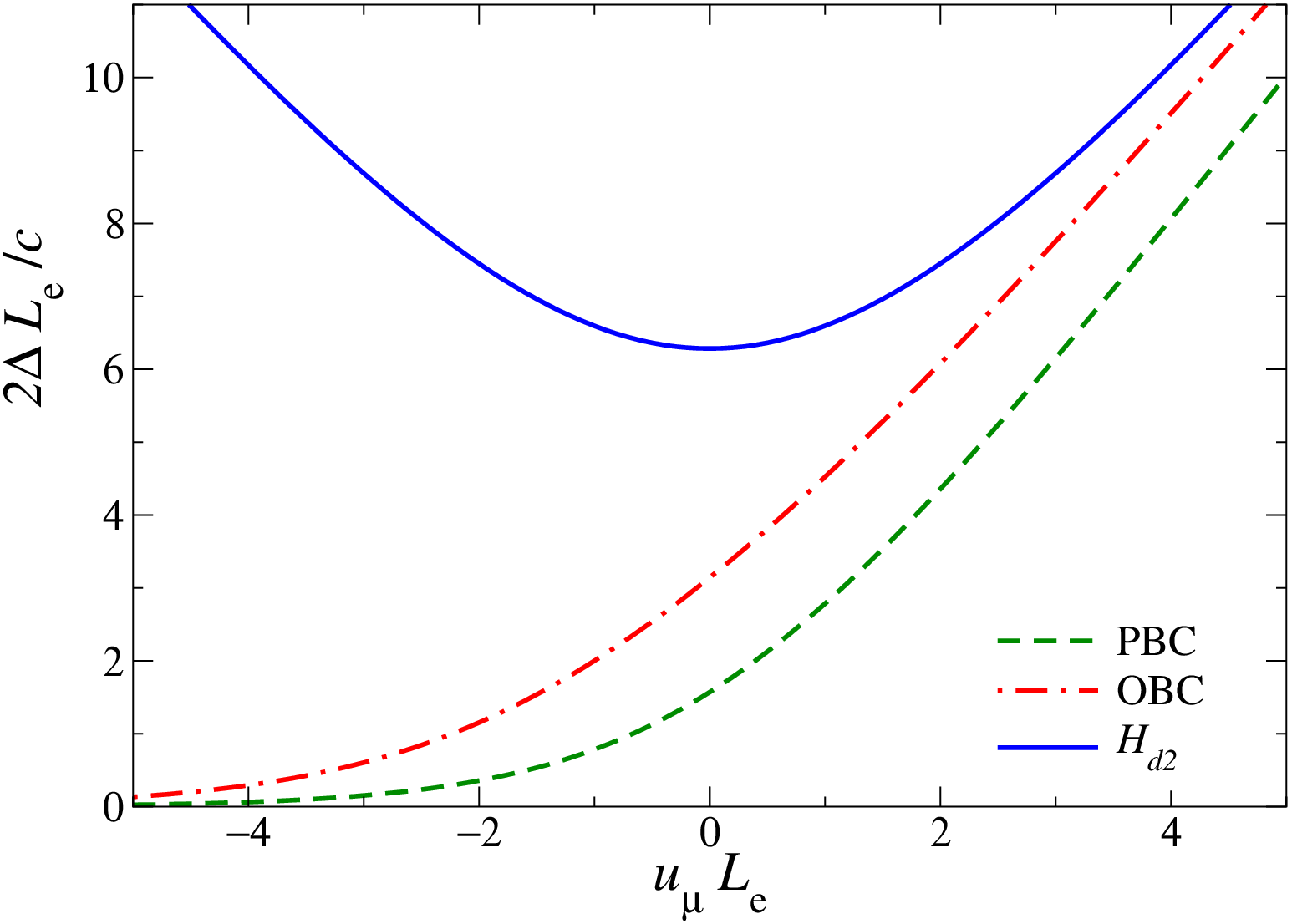}
\caption{(Color online) Plot of $2 \Delta L_{e}/c$ versus
  $\widetilde{w}_{e} = u_\mu L_{e}$ in the scaling limit for PBC
  ($L_{e} = L$ for PBC), OBC, and for the Hamiltonian
  (\ref{Ham-dual2}).  }
\label{Delta1-as}
\end{figure}

To compute the energy levels, we use the results of 
Katsura.\cite{Katsura-62} 
They are obtained by using Eq.~(\ref{H-harmonic}),
with a proper identification of the allowed values of $k$. The energy
levels can be divided in two sectors: the even one in which 
$k = 2 m \pi/L$, $m = 0,\ldots, L-1$, and the odd one in which 
$k = (2 m+1) \pi/L$, $m = 0,\ldots, L-1$.
The corresponding ground-state energies are 
\begin{eqnarray}
{\cal E}_0^{\rm odd} &=& - {1\over 2} \sum_{m=0}^{L-1}
   {E}\left({2 m + 1\over L}\pi\right),
\nonumber \\
{\cal E}_0^{\rm even} &=& - {1\over 2} \sum_{m=0}^{L-1}
   {E}\left({2 m\over L}\pi\right),
\label{groundstate}
\end{eqnarray}
where $E(k)$ is given in Eq.~(\ref{Ek-XY}). Note that, for
$\gamma\not=0$, we have ${\cal E}_0^{\rm odd} < {\cal E}_0^{\rm
  even}$.  Half of the states belong to the odd sector. They can be
written as $a_{k_1}^+ a_{k_2}^+ \ldots a_{k_m}^+|{\rm odd}\rangle$,
where $k_1$, $k_2$, $\ldots$, $k_m$ belong to the odd sector, $m$ is
even for PBC and odd for ABC.  The allowed states in the even sector
can also be written as $a_{k_1}^+ a_{k_2}^+ \ldots a_{k_m}^+|{\rm
  even}\rangle$, but now $m$ depends both on the boundary conditions
and on the value of $g$.  For $g \ge 1$, $m$ is odd (even) for PBC
(ABC). For $g \le 1$ the opposite condition holds: $m$ is even for
PBC, odd for ABC.  For $g=1$ the parity of $m$ can be chosen
arbitrarily, obtaining the same spectrum in all cases as a result of
the fact that $E(0)=0$. Therefore, for $g > 1$ and PBC, the lowest
energy states are
\begin{eqnarray}
E_0^P &=& {\cal E}_0^{\rm odd}, \\
E_1^P &=& {\cal E}_0^{\rm even} + E(0) , 
\nonumber \\
E_2^P &=& {\cal E}_0^{\rm odd} + E(\pi/L) + E(\pi - \pi/L), 
\nonumber
\end{eqnarray}
while for $g \le 1$ we obtain 
\begin{eqnarray}
E_0^P &=& {\cal E}_0^{\rm odd}, \\
E_1^P &=& {\cal E}_0^{\rm even}, 
\nonumber \\
E_2^P &=& {\cal E}_0^{\rm odd} + E(\pi/L) + E(\pi - \pi/L).
\nonumber
\end{eqnarray}
For ABC we have for $g\ge 1$
\begin{eqnarray}
E_0^A &=& {\cal E}_0^{\rm even}, \\
E_1^A &=& {\cal E}_0^{\rm odd} + E(\pi/L)
      = {\cal E}_0^{\rm odd} + E(\pi - \pi/L), 
\nonumber \\
E_2^A &=& {\cal E}_0^{\rm even} + E(0) + E(2 \pi/L)
\nonumber \\
 &=& {\cal E}_0^{\rm even} + E(0) + E(\pi - 2 \pi/L),
\nonumber
\end{eqnarray}
while for $g \le 1$
\begin{eqnarray}
E_0^A &=& {\cal E}_0^{\rm even} + E(0), \\
E_1^A &=& {\cal E}_0^{\rm odd} + E(\pi/L)= {\cal E}_0^{\rm odd} + E(\pi - \pi/L), 
\nonumber \\
E_2^A &=& {\cal E}_0^{\rm even} + E(2 \pi/L)=
 {\cal E}_0^{\rm even} + E(\pi - 2 \pi/L).
\nonumber
\end{eqnarray}
Note that the first two excited states are doubly degenerate.
Then, $\Delta$ and $\Delta^{(2)}$, 
the energy gaps for the first and second excited state, respectively,
are given by
\begin{eqnarray}
\Delta_P &=& {\cal E}_0^{\rm even} - {\cal E}_0^{\rm odd} + 
     \theta(g-1) E(0),
\label{Delta-def}\\
\Delta_A &=& {\cal E}_0^{\rm odd} - {\cal E}_0^{\rm even} + 
     E(\pi/L) - \theta(1-g) E(0) = 
\nonumber \\
   &=& - \Delta_P + E(\pi/L) + 2 (g-1), 
\nonumber \\
\Delta_P^{(2)} &=& 2 {E}(\pi/L) , 
\nonumber \\
\Delta_A^{(2)} &=& E(2\pi/L) + \theta(g-1) E(0), 
\nonumber 
\end{eqnarray}
with $\theta(x) = 1$ for $x\ge 0$, $\theta(x) = 0$ for $x<0$.  The
behavior of these quantities in the FSS limit, in which $\mu = g -
1\to 0$, $L\to\infty$ at $\mu L$ fixed, was considered in
Ref.~\onlinecite{Henkel-87}. We have performed again the calculation,
using the general method discussed in App.~B of
Ref.~\onlinecite{CP-98}. We obtain
\begin{eqnarray}
{1\over \gamma} {\cal E}_0^{\rm odd} &\approx&
 - {L\over \gamma} J + {1\over L}\left[
   {\pi\over3} - w - 4\pi G_1(w/2\pi) \right]
\nonumber \\
&& - {w^2\over 4 \pi L} \left(
     \ln {w^2\over 16\pi^2} + 2 \gamma_E - 1\right),
\\
{1\over \gamma} {\cal E}_0^{\rm even} &\approx&
 - {L\over \gamma} J + {1\over L}\left[
   - {\pi\over6} + 4\pi G_1(w/2\pi) - 2 \pi G_1(w/\pi) \right]
\nonumber \\
&& - {w^2\over 4 \pi L} \left(
     \ln {w^2\over \pi^2} + 2 \gamma_E - 1\right),
\end{eqnarray}
where $w = \mu L/\gamma$, $\gamma_E \approx 0.5772157$ is Euler's constant,
\begin{equation}
J = \int_0^\pi {dk\over 2\pi} E(k),
\end{equation}
and $G_1(x)$ is a remnant function:\cite{FB-72-ARMA}
\begin{equation}
G_1(x) = \sum_{n=1}^\infty 
   \left( \sqrt{n^2 + x^2} - n - {x^2\over 2n}\right).
\end{equation}
For $x\to 0$, $G_1(x) \approx - x^4\zeta(3)/8$, while for $|x|\to \infty$
we have \cite{CP-98}
\begin{eqnarray}
G_1(x) &=& {1\over12} + {x^2\over 4}
\left(- \ln {x^2\over4} + 1 - 2 \gamma_E\right) - 
   {|x|\over2} \nonumber \\
&& - {|x|\over \pi} \sum_{n=1}^\infty 
    {1\over n} K_1(2 \pi n |x|),
\label{G1as-large}
\end{eqnarray}
where $K_1$ is a modified Bessel function.
Using these results we obtain~\cite{footnotehh}
\begin{eqnarray}
L \Delta_P/\gamma &\approx & \Delta_{P0} = 
{\pi\over2} + w + {w^2\over\pi} \ln 2 
\label{Delta1P-scal} \\
&& + 2 \pi G_1(w/\pi) - 8 \pi G_1(w/2\pi),
\nonumber 
\\
L \Delta_A/\gamma &\approx& \Delta_{A0} = - \Delta_{P0} + 
   2 w + 2 \sqrt{\pi^2 + w^2}.
\label{Delta1A-scal} 
\end{eqnarray}
These results are consistent with Eq.~(\ref{expansion-Delta-wtilde})
since $\widetilde{w} \approx w$ and $c\approx 2 \gamma$ for $\mu\to
0$.  The PBC curve is shown in Fig.~\ref{Delta1-as}.  For $w\to 0$,
Eqs.~(\ref{Delta1P-scal}) and (\ref{Delta1A-scal}) give
$\Delta_{P0}(w) = \pi/2$ and $\Delta_{A0}(w) = 3 \pi/2$, in agreement
with Ref.~\onlinecite{BG-85}. For $|w|\to \infty$, using
Eq.~(\ref{G1as-large}) we obtain
\begin{eqnarray}
\Delta_{P0} &=& w + |w| \\
&& - 2 {|w|\over \pi} \sum_{n=1}^\infty
    {1\over n} \left[K_1(2 n |w|) - 2 K_1(n |w|)\right],
\nonumber 
\end{eqnarray}
which shows that $L \Delta_P \approx 2 \mu L$ for $w\to+\infty$ and $L
\Delta_P \approx 0$ for $w\to-\infty$, with exponentially small
corrections. In the same limit, $L \Delta_A$ behaves as $L \Delta_P$,
but corrections are now of order $w^{-2}$.

\begin{figure}[tbp]
\includegraphics*[scale=\graphicscale]{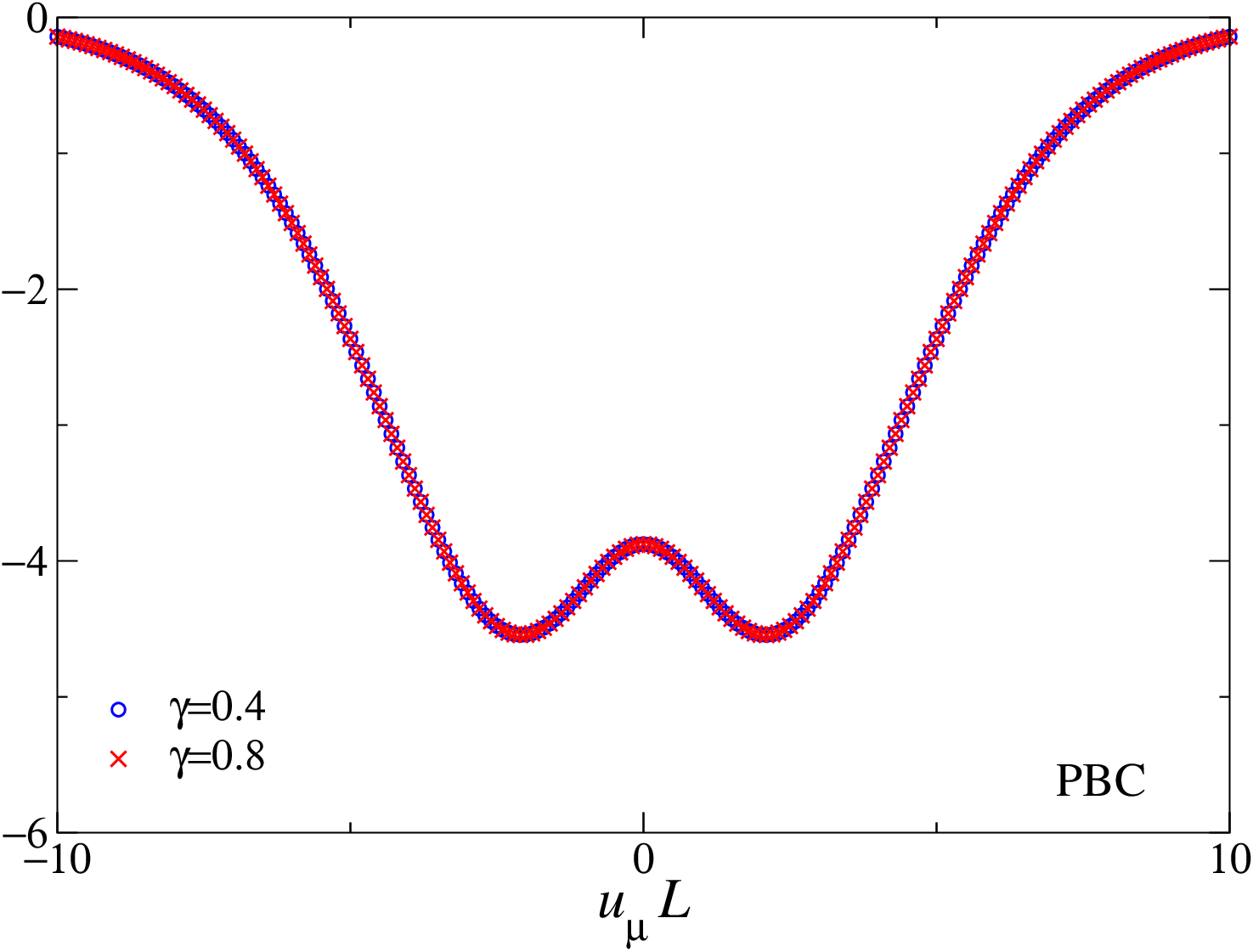}
\caption{(Color online) Plot of  $\widehat{\Delta}_{P2}$ 
versus $\widetilde{w} = u_\mu L$  for $\gamma = 0.4$ and 0.8. 
The two sets of data appear to follow a unique curve.
}
\label{corr-nmeno2-PBC}
\end{figure}

Let us now focus on the corrections. For this purpose, in the PBC case
we consider the combination
\begin{equation}
\widehat{\Delta}_{P2} (\widetilde{w},L,\gamma) = {L^2\over v_1(0,\gamma)} 
   \left[{2 L \Delta_P\over c} - \Delta_{P0} (\widetilde{w}) \right].
\end{equation}
If Eq.~(\ref{expansion-Delta-wtilde}) is correct, such a combination
should converge to ${\Delta}_{P2}(\widetilde{w})$ as $L\to \infty$
at fixed $\widetilde{w}$. Moreover, the limiting curve should be
independent of $\gamma$.  The results for $\gamma = 0.4$ and 0.8 shown
in Fig.~\ref{corr-nmeno2-PBC} are in full agreement, confirming
Eq.~(\ref{expansion-Delta-wtilde}) at order $L^{-3}$.  These curves
are obtained by computing $\Delta$ using high-precision arithmetics for 
values of $L$ in the range $10^3 \lesssim L\lesssim 10^5$ at 
fixed $\widetilde{w}$. On the scale of the figures the results fall 
on top of each other, providing the limiting scaling curve.

\begin{figure}[tbp]
\includegraphics*[scale=\graphicscale]{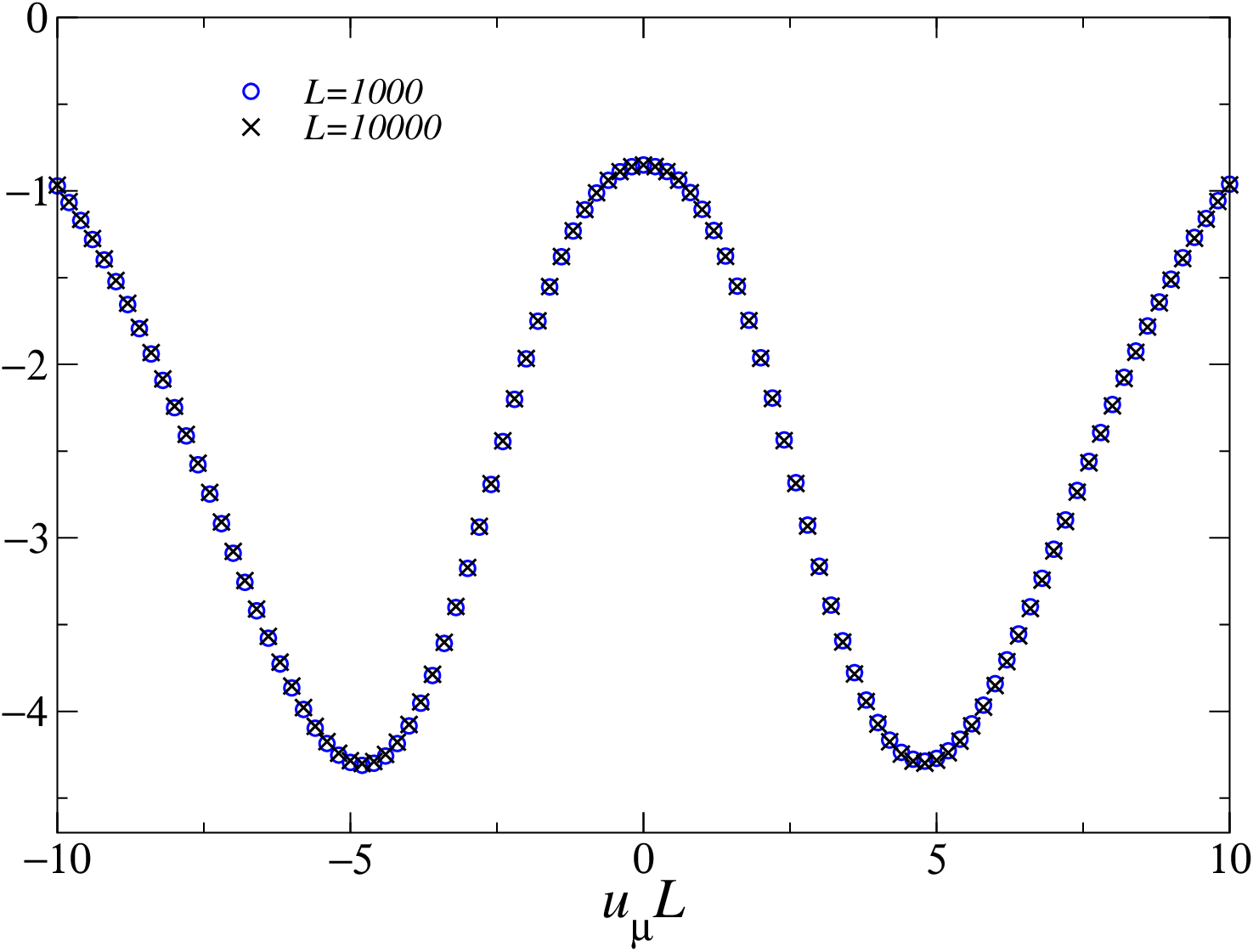}
\caption{(Color online) Plot of $\widehat{\Delta}_{P4}$ defined in
  Eq.~(\ref{delta4def}) versus $\widetilde{w}$ for two values of $L$,
  $L=10^3$ and $L=10^4$.  The two sets of data are hardly
  distinguishable, showing that $\widehat{\Delta}_{P4}$ approaches a
  nontrivial large-$L$ limit.}
\label{delta14}
\end{figure}

To verify that the neglected corrections in
Eq.~(\ref{expansion-Delta-wtilde}) decay as $L^{-4}$, we consider the
case $\gamma = \gamma_i = \sqrt{3}/2$, for which $v_1(\mu,{\gamma}_i)
\approx - \mu/(12\gamma_i^2)$, and compute
\begin{eqnarray}
\widehat{\Delta}_{P4}(\widetilde{w}) = L^4 
    \left[{2 L \Delta_P\over c} - \Delta_{P0} (\widetilde{w}) - 
                     {\widetilde{w} \over 12 \gamma_i L^3} 
                     \Delta_{P2}(\widetilde{w})
    \right].
\label{delta4def}
\end{eqnarray}
If corrections are of order $L^{-4}$, this quantity should have a
finite limit as $L\to \infty$ at fixed $\widetilde{w}$. If corrections
are instead of order $L^{-3}$, this quantity should diverge linearly
in $L$ as $L\to \infty$. The results shown in
Fig.~\ref{delta14} are consistent with a finite limit, hence confirm
that corrections decay as $L^{-4}$.

\subsection{Scaling of the energy gap: open boundary
conditions} \label{III.D}

Let us now consider the OBC case. This case is more complex than the
PBC one, since we must take into account boundary irrelevant
corrections and the fact that $u_l$ is a nontrivial function of
$L$. The latter type of corrections can be taken into account by using
an effective size $L_{e}$. Boundary irrelevant operators give
instead rise to corrections of order $L^{-2}$, hence $\omega_s =
2$. In practice we will show that
\begin{equation}
\Delta= {c(\mu,\gamma)\over 2 L_e} \left[
  \Delta_0(\widetilde{w}_{e}) + 
  {1\over L^2} \Delta_2(\widetilde{w}_{e},\gamma) + o(L^{-2})\right],
\label{expan-Delta-OBC}
\end{equation}
where 
\begin{equation}
   \widetilde{w}_{e} = u_\mu L_{e}.
\end{equation}
The function $\Delta_2(\widetilde{w}_{e},\gamma)$ is not proportional
to $v_1(0,\gamma)$ (for instance, it does not vanish for $\gamma =
\gamma_i = \sqrt{3}/2$), indicating the presence of boundary
contributions with $\omega_s = 2$.  For $\gamma = 1$ the effective
length $L_{e}$ is equal to $L+1/2$.  From the analysis of the
numerical data, we will conjecture a general expression for $L_{e}$,
valid for all values of $\gamma$, see below.

We first consider the case $\gamma = 1$, for which we can use the
analytic expressions reported in Ref.~\onlinecite{Pfeuty-70}.  The gap
$\Delta$ is given by $E(k_0)$, where $k_0$ is the smallest (in
absolute value) $k$ that satisfies the secular equation
\cite{Pfeuty-70}
\begin{equation}
{\sin (L+1) k\over \sin Lk} = {1\over g}.
\end{equation}
In the scaling limit at fixed $w=\mu L/\gamma$, 
the solution of the secular equation depends on $w$. For $w > -1$ , we have 
\begin{equation}
   k_0 = {\delta_1 \over L } + {\delta_2\over L^2} + O(L^{-3}),
\end{equation}
where $\delta_1$ is the solution in $]0,\pi[$ of the equation
\begin{equation}
\delta_1 = - w \tan\delta_1, 
\end{equation}
and $\delta_2$ is given by
\begin{equation}
\delta_2 = - {\delta_1 (\delta_1^2 + 2 w^2)\over 2 (\delta_1^2 + w + w^2)}.
\end{equation}
For $w < - 1$, we have instead
\begin{equation}
   k_0 = {i \delta_1 \over L } + {i \delta_2\over L^2} + O(L^{-3}),
\end{equation}
where $\delta_1$ is the solution in $]0,+\infty[$ of the equation
\begin{equation}
\delta_1 = - w \tanh\delta_1,
\end{equation}
and
\begin{equation}
\delta_2 = - {\delta_1 (\delta_1^2 - 2 w^2)\over 2 (\delta_1^2 - w - w^2)}.
\end{equation}
For $w\to +\infty$, we have $\delta_1 \approx \pi - \pi/w$ and
$\delta_2 \approx - \pi + 2 \pi/w$; for $w \to 0$, we have $\delta_1
\to \pi/2$ and $\delta_2 \to - \pi/4$; for $w\to -1$, $\delta_1$ and
$\delta_2$ both vanish, while for $w\to -\infty$ we obtain
$\delta_1\approx -w [1 + O(e^{-2 |w|})]$ and $\delta_2 \approx w/4$.

Using the expansion of $k_0$ and Eq.~(\ref{dispersion-E}) we obtain in the limit
$L\to \infty$ at $w$ fixed
\begin{equation}
{L^2 \Delta^2\over c^2} = \pm \delta_1^2 + 
 {w}^2 \pm 2 \delta_1\delta_2 {1\over L},
\end{equation}
where the upper signs should be used for $w> -1$ and the lower signs
for $w < -1$.  As in the PBC case, the gap $\delta_1$
vanishes as $w \to -\infty$, while for $w\to+\infty$ we have $L
\Delta/c \approx w$. The resulting curve is reported in
Fig.~\ref{Delta1-as}.

To get rid of the analytic corrections, we express $L^2 \Delta^2/c^2$
as a function of $\widetilde{w}$. Since $w = \widetilde{w} +
\widetilde{w}^2/(2 L)$, we obtain
\begin{equation}
{L^2 \Delta^2\over c^2} = \pm \delta_1^2 + \widetilde{w}^2
   \mp \left( {\delta_1^2 (\delta_1^2 + \widetilde{w}) \over 
    \delta_1^2 \pm \widetilde{w} (1 + \widetilde{w})} \right)  {1\over L},
\label{Delta1-1suL}
\end{equation}
where $\delta_1$ is now a function of $\widetilde{w}$ and we used
\begin{equation}
{d\delta_1\over dw} = {\delta_1 \over w + w^2 \pm \delta_1^2}.
\label{derivatadelta1}
\end{equation}
The $1/L$ correction can be eliminated by rescaling the size $L$. Indeed, if 
we define
\begin{equation}
  L_{e} = L + {1\over2},
\end{equation}
we obtain
\begin{equation}
{L_{e}^2 \Delta^2\over c^2} = \pm \delta_1^2 + 
\widetilde{w}^2_{e} + O(L^{-2}),
\end{equation}
where $\delta_1$ is now a function of $\widetilde{w}_e = u_\mu L_e$.
This result can be derived immediately if we rewrite the secular equation in 
terms of $u_\mu$ and $L_{e}$:
\begin{equation}
{\widetilde{w}_{e}\over L_{e}} = {2 \cot (L_{e} k) \sin k/2 \over 
         \left[1 - (\sin^2k/2) (\sin L_{e} k)^{-2}\right]^{1/2} }.
\end{equation}
This equation is symmetric under $L_{e} \to - L_{e}$ and $k\to -k$,
implying the $k$ has an expansion in odd powers of $1/L_{e}$.  No even
powers appear, confirming the absence of corrections of order
$L_{e}^{-1}$ in the expansion of the gap.

The previous analysis was restricted to the first correction. It is
important to stress that it is not possible to eliminate the
corrections of order $L^{-3}$ in the expansion of $k_0$ by redefining
$L_{e} = L + 1/2 + a/L$, with a suitable $a$.  Indeed, at the critical
point the secular equation gives $k_0 = 2 \pi/(L + 1/2)$
exactly. Therefore, for $\widetilde{w} = 0$, there are no $L^{-3}_{e}$
corrections only if $a=0$.  But, if we take $a=0$, corrections are
present for $\widetilde{w} \not=0$.

\begin{figure}[tbp]
\includegraphics*[scale=\graphicscale]{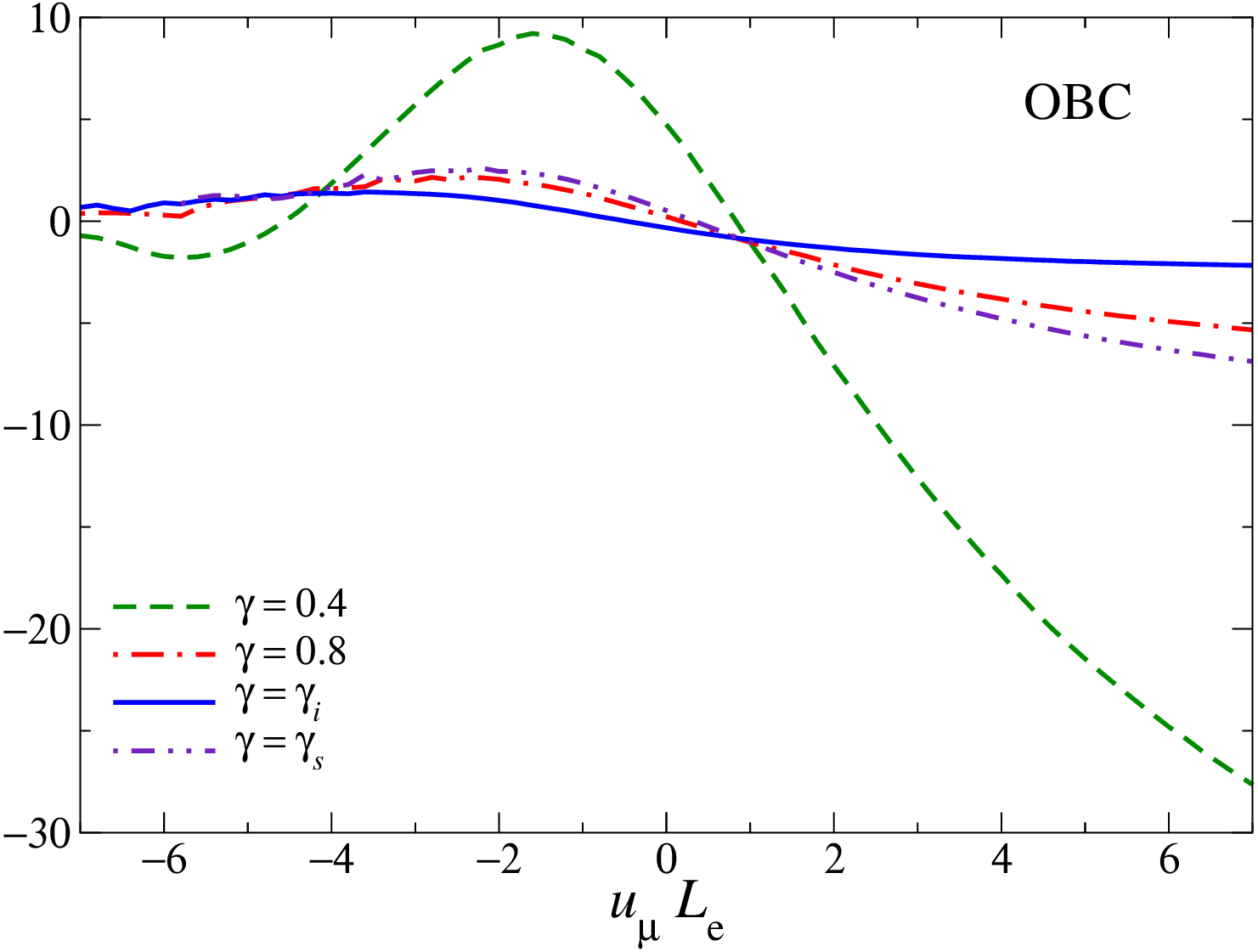}
\caption{(Color online) $\widehat{\Delta}_{2}$
  defined in Eq.~(\ref{delta2obc}) versus $\widetilde{w}_{e}$ for
  OBC. Results for $\gamma = \gamma_i =\sqrt{3}/2$, $\gamma=0.8$,
  0.4, and $\gamma=\gamma_s = \sqrt{3} - 1$ (for $\gamma=\gamma_s$ we
  have $L_{e} = L$).  }
\label{delta12-OBC}
\end{figure}

Let us now consider the energy gap for $\gamma \not= 1$.  In the
absence of analytic results, we compute the difference $\Delta$ of the
two lowest energy levels numerically for $L\le  4096$ 
and for $\gamma = \sqrt{3}/2$, 0.8, and
0.4. Also for these values of $\gamma$ we find that the leading
scaling correction can be eliminated using appropriate
$\gamma$-dependent $L_{e}$. An accurate numerical guess of $L_{e}$ is
\begin{equation}
L_{e} = L + {1\over 2} + {(\gamma + 2) (\gamma - 1)\over 2\gamma}.
\label{legamma}
\end{equation}
With this choice $\Delta$ has an expansion of the form
\begin{equation}
{2 L_{e} \Delta(\mu,L,\gamma) \over c} = 
  \Delta_{0}(\widetilde{w}_{e}) + O(L^{-2}).
\end{equation}
We can estimate the correction term by considering 
\begin{eqnarray}
\widehat{\Delta}_{2} = 
   {4\over 3} \left[
{2 L_{e1} \Delta(\mu_1,L,\gamma) \over c(\mu_1)} 
-   {2 L_{e2} \Delta(\mu_2,2 L,\gamma) \over c(\mu_2)}\right],
\nonumber\\
\label{delta2obc}
\end{eqnarray}
where $L_{{e}1}$ and $L_{{e}2}$ correspond to $L$ and $2L$,
respectively, and $\mu_1$ and $\mu_2$ are obtained by solving
$u_\mu(\mu_1,\gamma) L_{{e}1} = \widetilde{w}_{e}$ and
$u_\mu(\mu_2,\gamma) L_{{e}2} = \widetilde{w}_{e}$. The resulting
quantity has a finite limit for $L\to\infty$ at fixed
$\widetilde{w}_{e}$, reported in Fig.~\ref{delta12-OBC}. Note that
corrections do not vanish for $\gamma_i=\sqrt{3}/2$, where
$v_1(0,\gamma_i)=0$, hence they cannot be only due to the bulk
subleading operator with $\omega = 2$. Moreover, there is no rescaling
that allows us to obtain a collapse of all data onto a single
curve. Therefore the data show the presence of corrections due to a
boundary subleading operator with exponent $\omega_s = 2$.

Finally, it is interesting to consider Hamiltonian
(\ref{Ham-dual2}).  The secular equation turns out to be particularly
simple.  The allowed values of $k$ are simply $k = \pi n/(L+1)$, $n =
1,\ldots L$ for all values of $g$.  Therefore, if we define $L_{e} =
L+1$, we have
\begin{equation}
{L^2_{e} \Delta^2\over c^2} = \pi^2 + \widetilde{w}_{e}^2 + 
   {v_1(\mu,\gamma)\over L^2_{e}} \pi^4 + O(L^{-4}).
\end{equation}
The scaling function for $L\to \infty$ is reported in
Fig.~\ref{Delta1-as}.  Note that $\Delta_1$ does not vanish for
$\widetilde{w}_{e} \to -\infty$, a consequence of the fact that the
degeneracy for $g < 1$ is lifted by the added magnetic field. A second
peculiarity of the result is the absence of boundary corrections, once
length scales are expressed in terms of $L_e$.

\subsection{RG invariant ratios} \label{III.E}

\begin{figure}[tbp]
\includegraphics*[scale=\graphicscale]{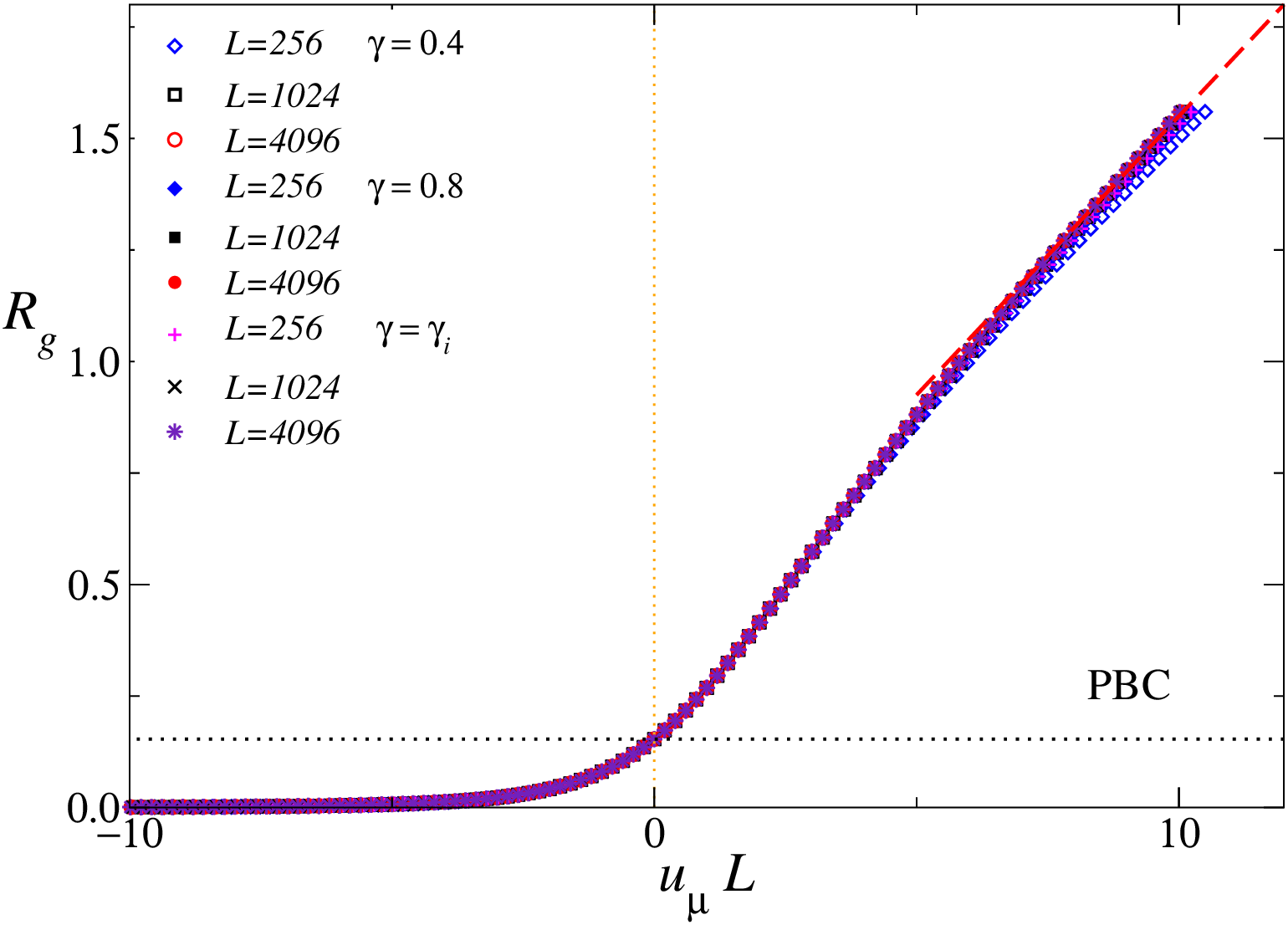}
\includegraphics*[scale=\graphicscale]{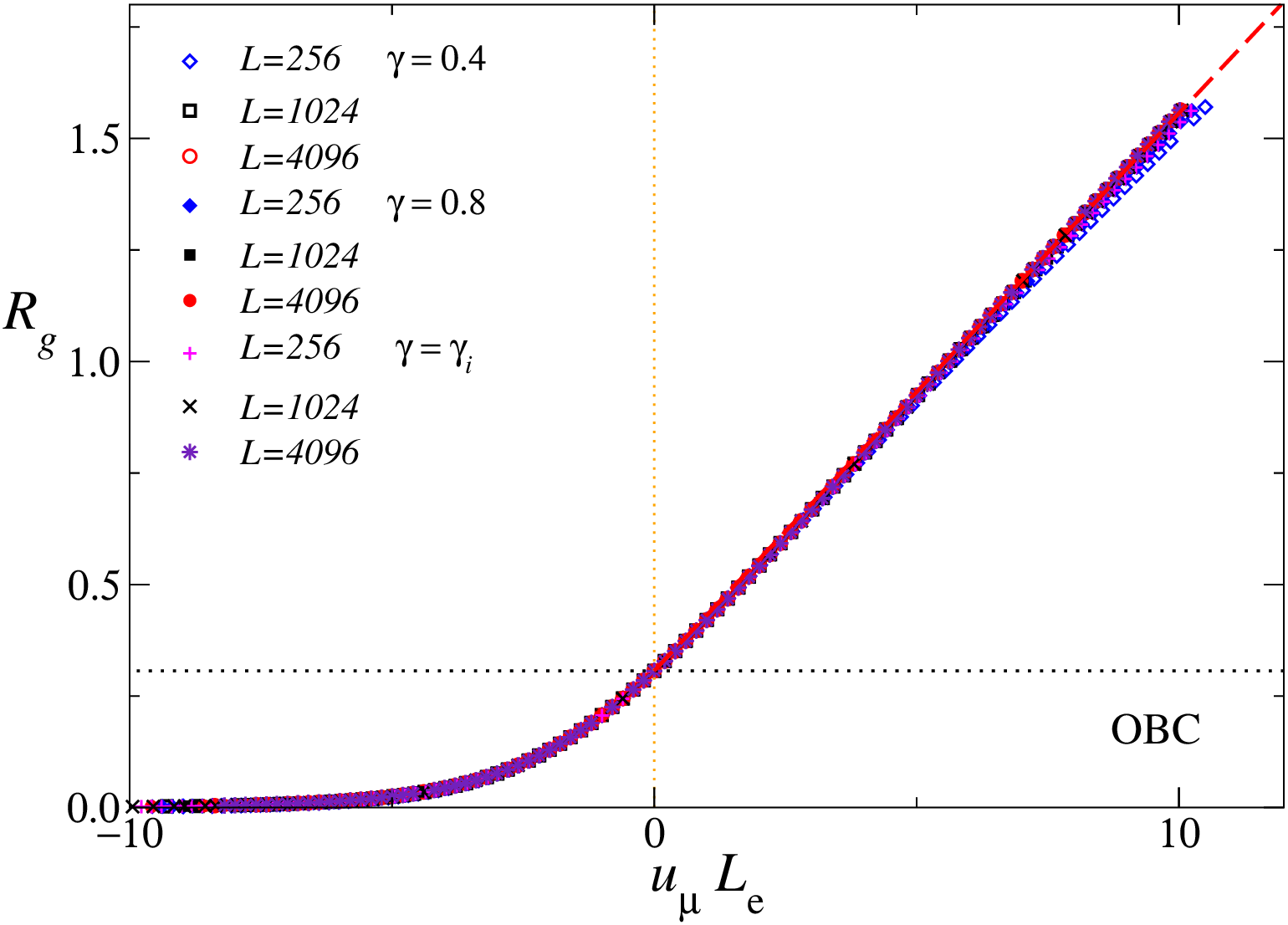}
\caption{(Color online) Plot of $R_g$, defined in Eq.~(\ref{rdef2}),
  versus $\widetilde{w}=u_\mu L$ for PBC (top) and versus
  $\widetilde{w}_{e} = u_\mu L_{e}$ for OBC (bottom).  We report data
  for $\gamma = 0.4$, 0.8, and $\gamma_i=\sqrt{3}/2$. They approach a
  universal curve with increasing $L$.  The horizontal dotted lines
  correspond to the exact value at $\mu=0$.  The dashed lines show the
  asymptotic behavior ${R}_{g}\approx w/8$ for $w\to\infty$.  }
 \label{rgfss}
\end{figure}

In Sec.~\ref{III.B}, \ref{III.C}, and \ref{III.D} we have shown
that the data for the free energy and energy gap are consistent with
the assumption that $u_\mu$ and $v_1$ are nonlinear scaling
fields. Moreover, in the case of OBC the leading boundary correction
can be eliminated by redefining $L\to L_{e}$.

We wish now to verify these conjectures by studying different
observables related to the correlation function of the order parameter
$\sigma_x^{(1)}$. We consider the equal-time correlation function
\begin{equation}
G(x,y) = \langle \sigma^{(1)}_x \sigma^{(1)}_y \rangle,
\label{gxyxy}
\end{equation}
then we define
\begin{eqnarray}
&&\chi \equiv \sum_{x} G(0,x),\label{chidef2}\\
&&\xi^2 \equiv {1\over 2\chi} \sum_{x} x^2 G(0,x),\label{xidef2}
\end{eqnarray}
and the RG invariant quantities 
\begin{equation}
R_\xi\equiv\xi/L,\quad R_g \equiv \ln[G(0,L/8)/G(0,L/4)].
\label{rdef2}
\end{equation}
We compute $R_g$ for PBC and OBC, for several values of $L$ and
$\widetilde{w}$ ($L_{e}$ and $\widetilde{w}_{e}$ in the OBC case) and
extrapolate the results to $L\to\infty$. The results are reported in
Fig.~\ref{rgfss}. Note that in the scaling limit, all scaling
variables are equivalent, i.e., $w\approx \widetilde{w} \approx
\widetilde{w}_{e}$, but this is not true when considering the scaling
corrections.  The value of $R_g$ at the critical point for
$L\to\infty$ can be computed by using the exact expression of the
two-point function at the critical point in the scaling limit. The
numerical values are reported in App.~\ref{cftfo}. We can also predict
the large-$w$ behavior by using the known expression of $G(x)$ in the
infinite-volume limit for $\mu > 0$.  Since $G(x)\sim K_0(x\mu)$ for
$\gamma=1$,\cite{Sachdev-book} where $K_0(x)$ is a modified Bessel
function, we obtain $R_{g} \approx w/8$ for $w \to \infty$.

\begin{figure}[tbp]
\includegraphics*[scale=\graphicscale]{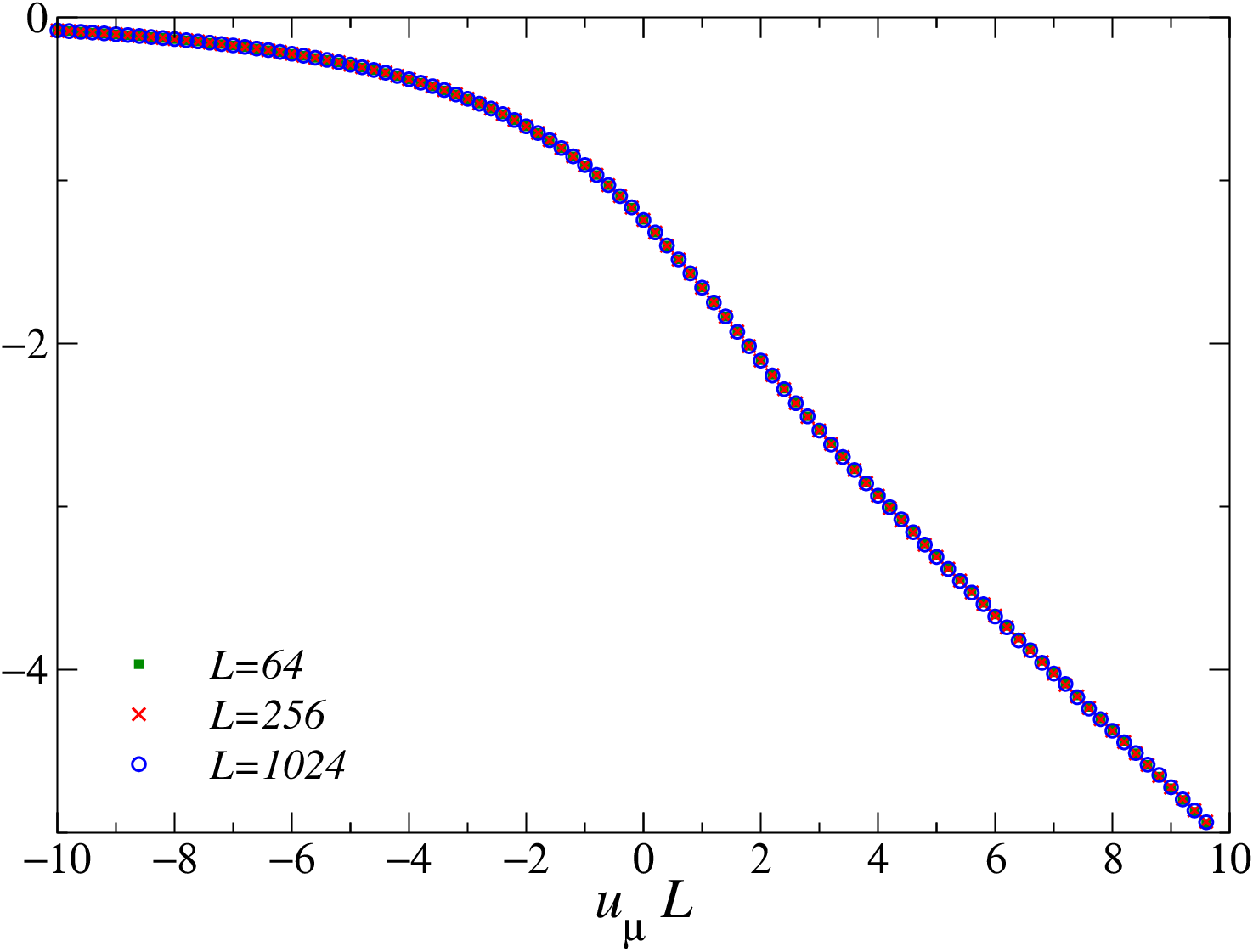}
\caption{(Color online) Plot of $\widehat{R}_{g2}$,
  cf. Eq.~(\ref{whrg2}), versus $\widetilde{w} = u_\mu L$ for PBC.  We
  report data for $\gamma_i = \sqrt{3}/2$, and some values of $L$.  
  Data clearly converge toward an asymptotic large-$L$ curve.
}
\label{rgcPBC}
\end{figure}

Let us now discuss the leading corrections in the PBC case.  According
to the general analysis, in the limit $L\to \infty$, $\mu\to 0$ at
fixed $\widetilde{w}$, we expect corrections of order $L^{-\omega} =
L^{-2}$ due to the leading irrelevant operator and corrections due to
field mixings. We will show that the latter also scale as $L^{-2}$,
obtaining an expansion of the form
\begin{equation}
R_g(\mu,L,\gamma) = R_{g0}(\widetilde{w}) + 
    {1\over L^2} R_{g2}(\widetilde{w},\gamma) + O(L^{-3}),
\label{expan-Rg-PBC}
\end{equation}
with 
\begin{equation}
R_{g2}(\widetilde{w},\gamma) = v_1(\mu,\gamma) R_{g21}(\widetilde{w}) + 
    v_2(\mu,\gamma) R_{g22}(\widetilde{w}).
\end{equation}
where $v_1(\mu,\gamma)$ is the nonlinear scaling field reported in
Eq.~(\ref{v1mugamma}).  To verify this expansion, we consider the
combination
\begin{eqnarray}
\widehat{R}_{g2}(\widetilde{w},L,\gamma) &= &
   {4 L^2\over3} \left[ - R_g(\widetilde{w},L) + 9 R_g(\widetilde{w},2L) \right. 
\nonumber \\
 && \qquad \left. - 
      8 R_g(\widetilde{w},4L) \right].
\label{whrg2}
\end{eqnarray}
If Eq.~(\ref{expan-Rg-PBC}) holds,
$\widehat{R}_{g2}(\widetilde{w},L,\gamma)$ converges to
$R_{g2}(\widetilde{w},\gamma)$ with corrections of order $L^{-2}$. If
instead Eq.~(\ref{expan-Rg-PBC}) does not hold and corrections to the
leading scaling behavior are of order $1/L$,
$\widehat{R}_{g2}(\widetilde{w},L,\gamma)$ diverges as $L\to\infty$.
In Fig.~\ref{rgcPBC} we show the results for $\gamma_i =
\sqrt{3}/2$ for which $v_1(0,\gamma_i)=0$. The combination
$\widehat{R}_{g2}(\widetilde{w},L,\gamma)$ has a finite limit for
$L\to \infty$, confirming that the leading scaling corrections decay
as $1/L^2$.

\begin{figure}[tbp]
\includegraphics*[scale=\graphicscale]{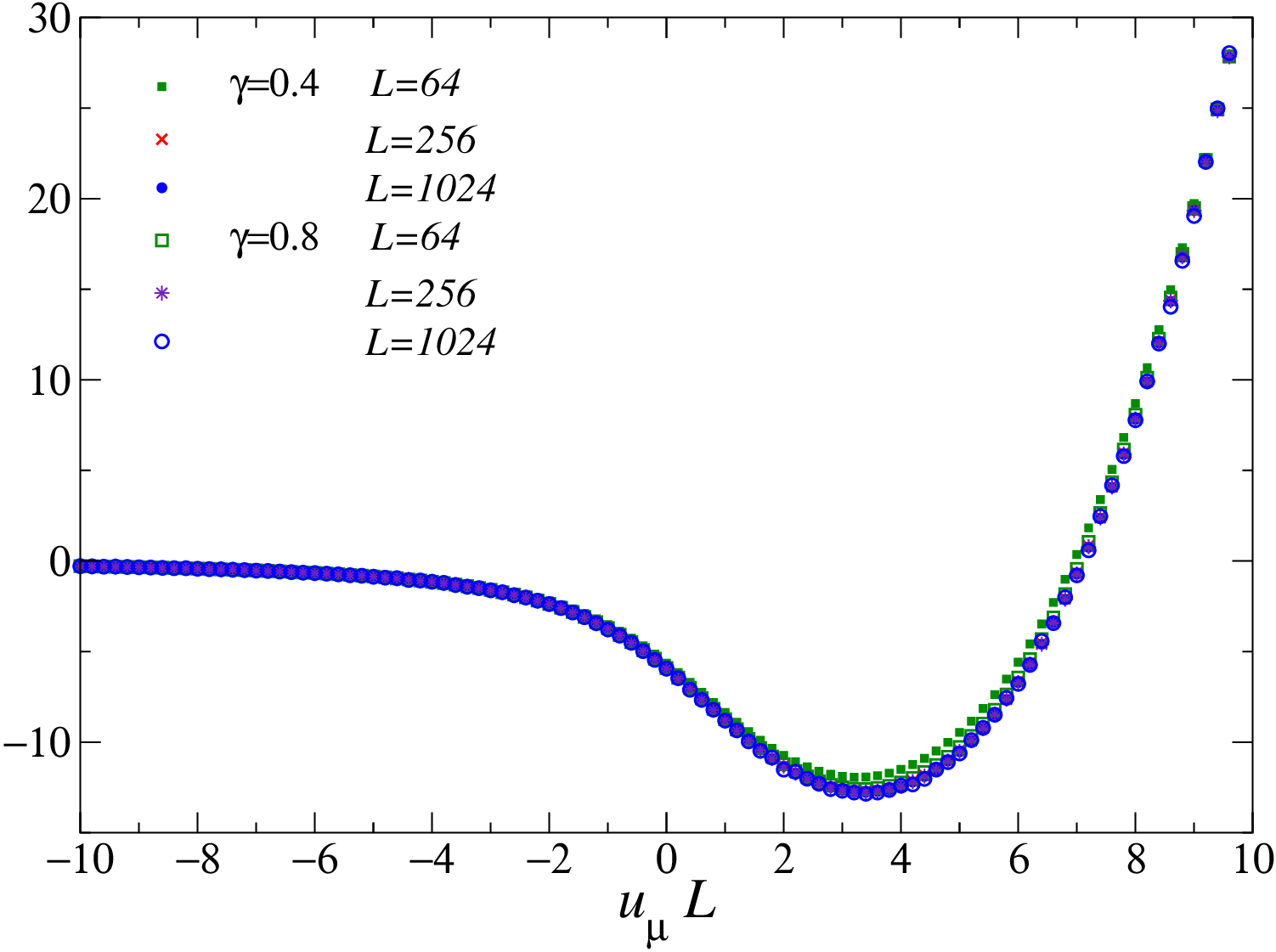}
\caption{(Color online) The ratio
  $\widehat{R}_{2g1}(\mu,L,\gamma)/v_1(0,\gamma)$ versus
  $\widetilde{w} = u_\mu L$ for PBC, cf. Eq.~(\ref{rg21def}).  We use
  Eq.~(\ref{v2cond}).  The data for different values of $\gamma$ appear
  to approach the same curve with increasing $L$.  }
\label{SRg-PBC}
\end{figure}

Since $v_1(0,\gamma_i) = 0$, the corrections we observe cannot be due
to the operator $Q_2^2 + \bar{Q}^2_2$ which controls the leading
scaling correction for the free energy and the spectrum. Corrections
are instead a field-mixing effect.  The lattice operator is a
combination of conformal fields: 
\begin{equation}
\sigma^{(1)}_{\rm LAT} = {\cal O}_\sigma + \sum_{i=1} {\cal O}_{\sigma,i}, 
\label{sig1o}
\end{equation}
where ${\cal O}_\sigma$ is the primary CFT field and ${\cal
  O}_{\sigma,i}$ are the secondary fields that belong to the $\sigma$ family, the
leading one being ${\cal O}_{\sigma1}=L_{-1}|\sigma\rangle$ 
and $y_\sigma-y_{\sigma 1}=1$.
To provide additional
evidence for the validity of Eq.~(\ref{expan-Rg-PBC}), we consider
\begin{eqnarray}
\widehat{R}_{g21}&=&{4 L^2\over 3}
      [ R_g(\mu,L,\gamma)  - R_g(\mu,2 L,\gamma)]
\label{rg21def}\\
&-&
        {4 L^2\over 3} {v_2(0,\gamma)\over v_2(0,\gamma_i)}
       [R_g(\mu,L,\gamma_i) - R_g(\mu,2 L,\gamma_i)].
\nonumber 
\end{eqnarray}
If Eq.~(\ref{expan-Rg-PBC}) holds, then 
\begin{equation}
\widehat{R}_{g21}(\mu,L,\gamma) \approx v_1(0,\gamma) R_{g21}(\widetilde{w}).
\end{equation} 
Since $v_1(\mu,\gamma)$ is known, this relation gives us a recipe to
identify $v_2(0,\gamma)$.  We determine $v_2(0,\gamma)$ by requiring
$\widehat{R}_{g21}(\mu,L,\gamma)/v_1(0,\gamma)$ to be independent of $\gamma$. 
By using numerical results for $\gamma = 0.4$ and $\gamma=0.8$, we find
that this condition is satisfied by simply taking
\begin{equation}
v_2(0,\gamma)/v_2(0,\gamma_i) = 1.
\label{v2cond}
\end{equation}
This is shown by the data in Fig.~\ref{SRg-PBC}.  The scaling field
$v_2(\mu,\gamma)$ is independent of $\gamma$ for $\mu\to 0$.

\begin{figure}[tbp]
\includegraphics*[scale=\graphicscale]{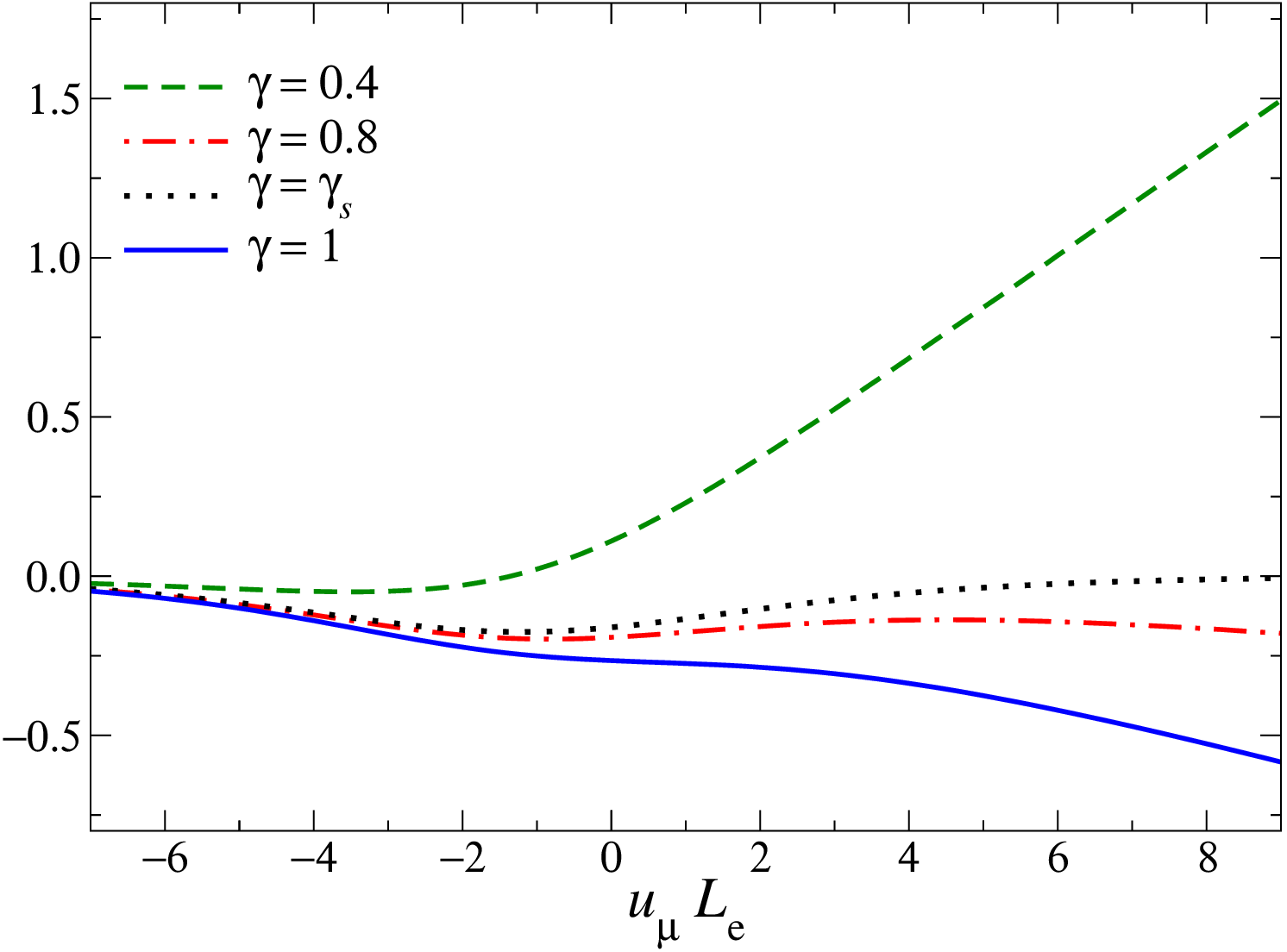}
\caption{(Color online) Plot of
  $\widehat{R}_{g1}(L,\widetilde{w}_{e},\gamma)$, defined in
  Eq.~(\ref{def-T}), for $L = 512$ and some values of $\gamma$. 
}
\label{plot-T}
\end{figure}

Let us now consider the scaling corrections in the OBC case. In this
case we find that scaling corrections of order $1/L$ are present even
if the limit $L\to \infty$ is taken at fixed $\widetilde{w}_{e}$.
Indeed, the numerical data are consistent with
\begin{equation}
R_g(\mu,L,\gamma) = R_{g0}(\widetilde{w}_{e}) + 
   {1\over L} R_{g1}(\widetilde{w}_{e},\gamma).
\end{equation}
To estimate the $1/L$ correction, we consider
\begin{eqnarray}
&&\widehat{R}_{g1}(L,\widetilde{w}_{e},\gamma) =
  {2L\over 3} \Bigl[R_g(L,\widetilde{w}_{e}) 
  -   13 R_g(2 L,\widetilde{w}_{e}) +
\nonumber\\
&&\qquad +
    44 R_g(4 L,\widetilde{w}_{e}) - 
    32 R_g(8 L,\widetilde{w}_{e}) \Bigr].
\label{def-T}
\end{eqnarray}
For $L\to \infty$, we have 
\begin{equation}
\widehat{R}_{g1}(L,\widetilde{w}_{e},\gamma) \to
R_{g1}(\widetilde{w}_{e},\gamma)
\label{rg1sca}
\end{equation}
 with corrections of order $L^{-3}$.~\cite{footnote-T} We have computed 
$\widehat{R}_{g1}(L,\widetilde{w}_{e},\gamma)$ for
$64\le L \le 512$, obtaining, for all values of $\gamma$ a nonzero
result. The function $\widehat{R}_{g1}(L,\widetilde{w}_{e},\gamma)$ for $L=512$ (it
is essentially asymptotic) is reported in Fig.~\ref{plot-T}. Note that
it has a nontrivial dependence on $\gamma$: no rescaling exists that
makes the curves corresponding to different values of $\gamma$ fall
one on top of the other. This implies that such correction cannot be
ascribed to a single subleading operator. We can also exclude that the
$1/L$ correction can be eliminated by using $L_{e}$ in the definition
of $R_g$, i.e., by defining
\begin{equation}
R_g' \equiv \ln[G(0,L_{e}/8)/G(0,L_{e}/4)].
\end{equation}
Indeed, for $\gamma = \gamma_s = \sqrt{3} - 1$, we have $L_{e} = L$,
hence $R_g = R_g'$. But also in this case $1/L$ corrections are
present.  They may be explained by the presence of field mixings with
the boundary operators.

\begin{figure}[tbp]
\includegraphics*[scale=\graphicscale]{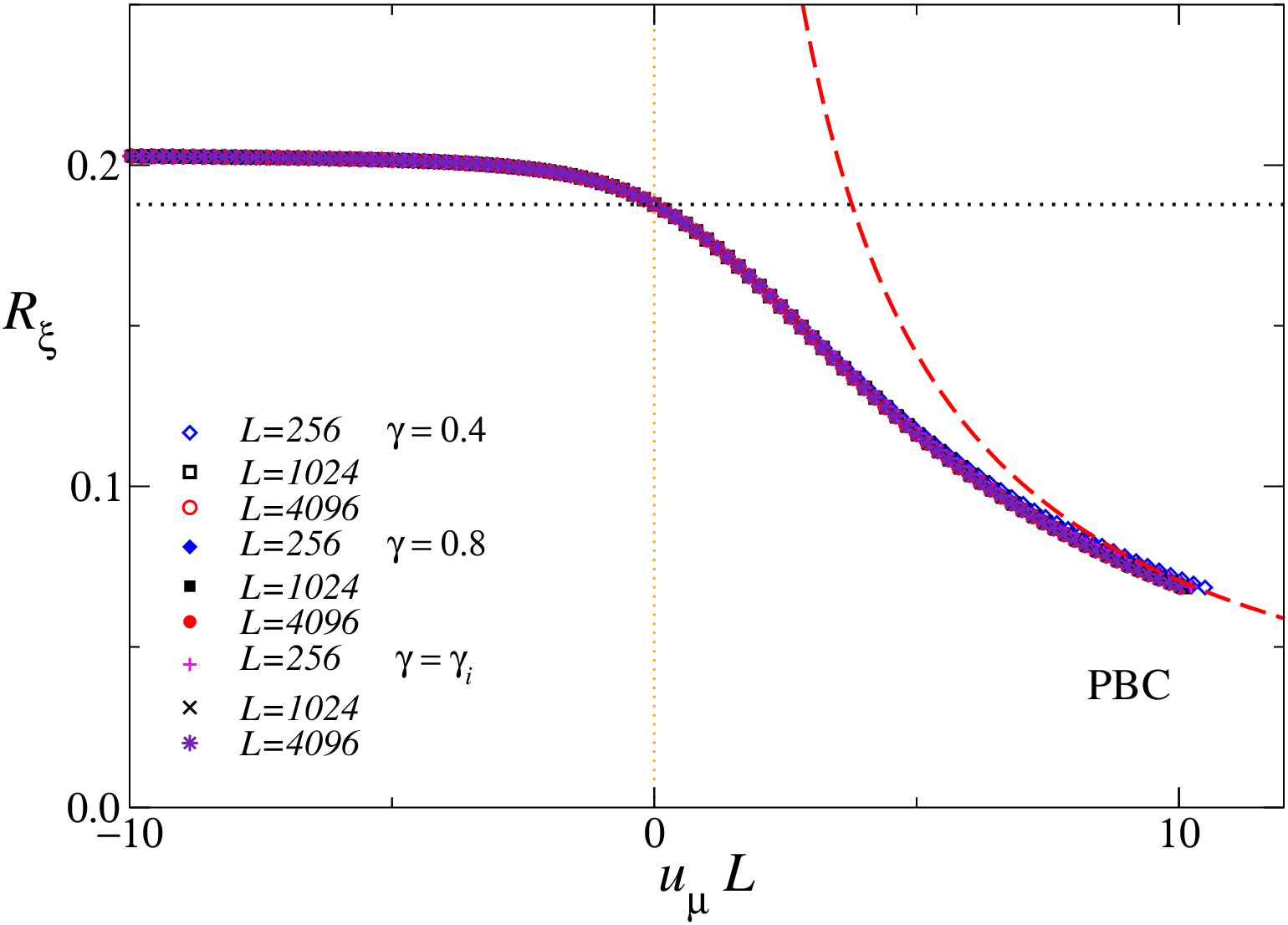}
\includegraphics*[scale=\graphicscale]{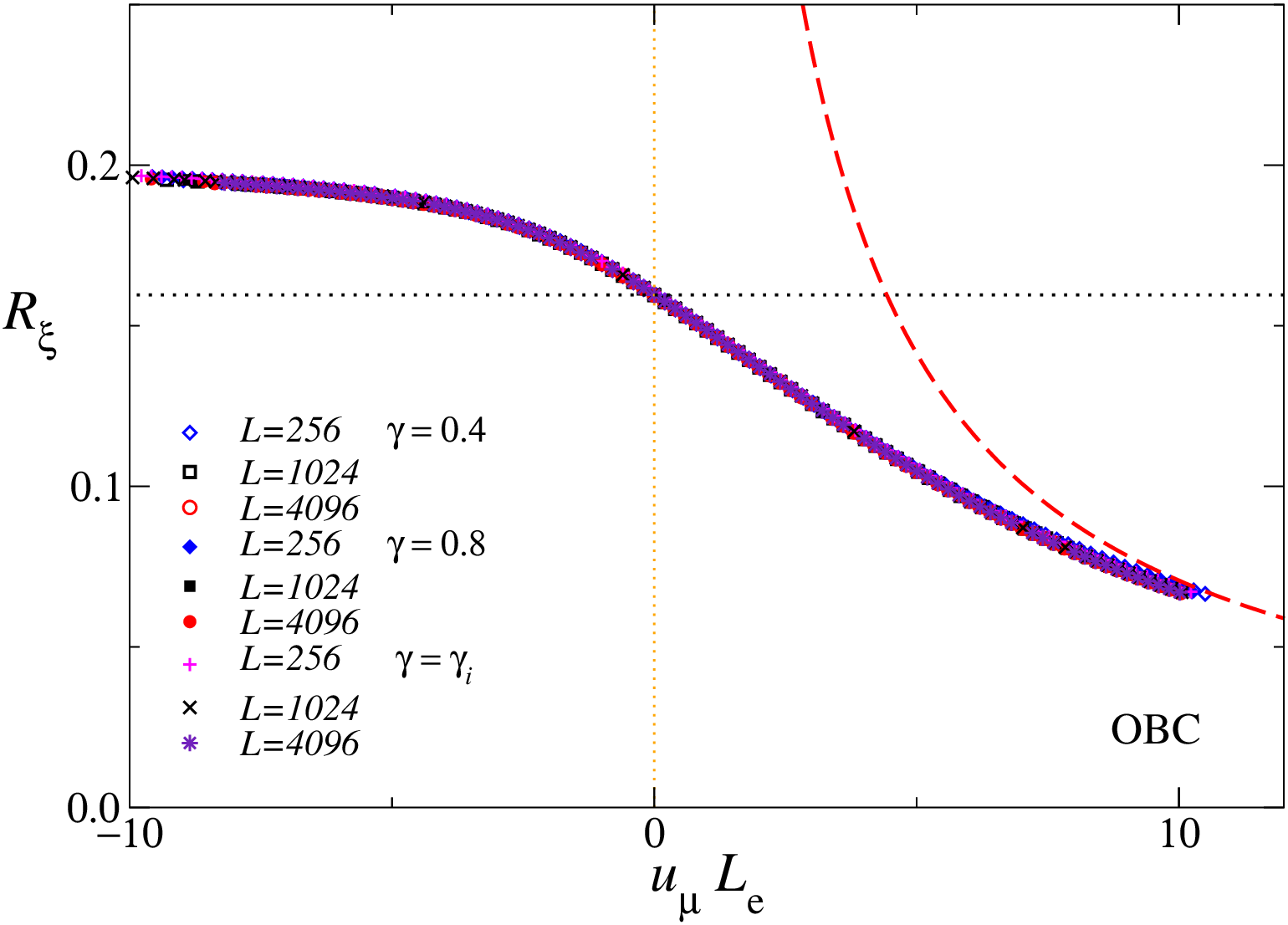}
\caption{(Color online) Plot of $R_\xi\equiv \xi/L$ versus
  $\widetilde{w} = u_\mu L$ for PBC (top) and versus
  $\widetilde{w}_{e} = u_\mu L_{e}$ for OBC (bottom). We report data
  for $\gamma = 0.4$ and 0.8. For both OBC and PBC the data for
  different values of $\gamma$ approach a universal curve with
  increasing $L$.  The horizontal dotted lines correspond to the exact
  value at $\mu=0$.  The dashed lines show the asymptotic behavior
  $R_\xi\approx 1/(\sqrt{2} \widetilde{w})$ for $\widetilde{w} \to
  \infty$ (for OBC we should replace $\widetilde{w}$ with
  $\widetilde{w}_{e}$).  }
\label{rxifss}
\end{figure}

\begin{figure}[tbp]
\includegraphics*[scale=\graphicscale]{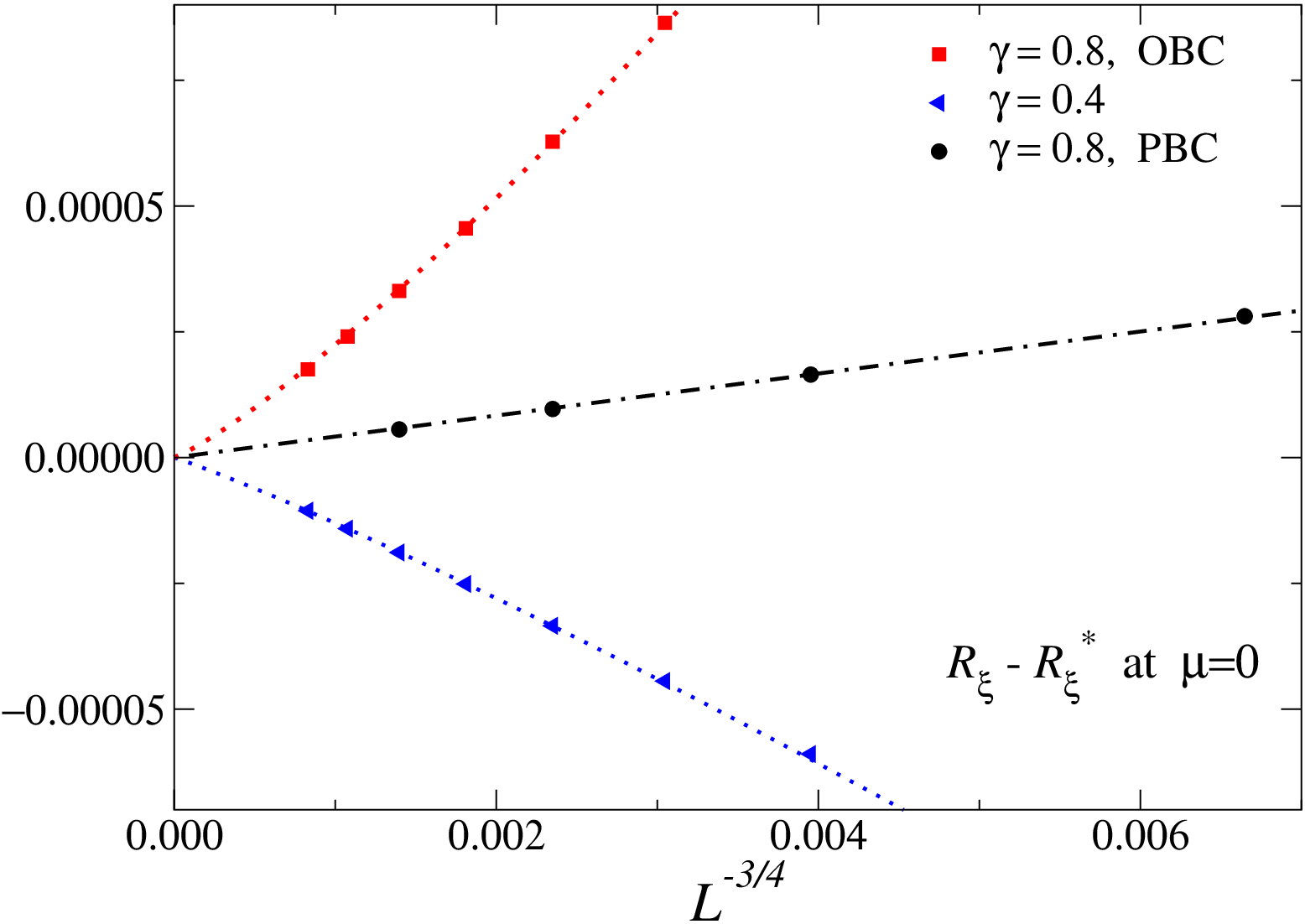}
\caption{(Color online) Plot of $R_\xi-R_\xi^*$ at $\mu = 0$ vs
  $L^{-3/4}$.  The dotted lines are fits to $aL^{-3/4} + b L^{-1}$,
  while the dot-dashed one is a fit to $aL^{-3/4}$ only.  Results for
  OBC and PBC and for $\gamma = 0.4$ and 0.8.  }
\label{scalcorxi}
\end{figure}

Let us finally consider $R_\xi$. Its behavior in the scaling limit is
shown in Fig.~\ref{rxifss}.  The finite-size behavior of $R_\xi$ is
more complex, since one must also take into account the background
term which gives corrections of order $L^{-2+z+\eta} = L^{-3/4}$,
independent of the type of boundary conditions. For OBC
next-to-leading corrections are of order $L^{-1}$, while for PBC, if
the scaling limit is taken at fixed $\widetilde{w}$, they are of order
$L^{-7/4}$ and are due to the $L$ dependence of the background term.
These predictions are well confirmed by the data shown in
Fig.~\ref{scalcorxi}.

\section{Finite-size scaling of bipartite entanglement entropies}
\label{entang}

In a quantum system the reduced density matrices of spatial
subsystems, and, in particular, the corresponding entanglement
entropies and spectra, provide effective probes of the nature of the
quantum critical behavior, see, e.g.,
Refs.~\onlinecite{AFOV-08,CCD-09,ECP-10,AP-06,LW-06}.  Their
dependence on the finite size of the system may be exploited to
determine the critical parameters of a quantum
transition.\cite{IL-08,MA-10,XA-11,XA-12,DLLS-12,LDS-13}

In this section we discuss the finite-size behavior of the
entanglement entropy of spatial bipartitions of the system.  We
restrict the discussion to zero temperature and to one-dimensional
systems with an isolated quantum critical point with $z=1$ and central
charge $c$.  The general FSS behavior is then compared with exact and numerical
results for the XY chain.

\subsection{FSS in 1D systems at a quantum critical point}
\label{eprobes}

We divide the chain into two connected parts of length $\ell_A$ and
$L-\ell_A$, and consider the R\'enyi entropy ($\alpha > 0$)
\begin{equation}
S_\alpha(\ell_A,L) = S_\alpha(L-\ell_A,L) = 
{1\over 1-\alpha} \ln {\rm Tr} \rho_A^\alpha,
\label{renyientropies}
\end{equation}
where $\rho_A$ is the reduced density matrix of one of the two
subsystems. For $\alpha\to 1$, the R\'enyi entropy coincides with the
von Neumann (vN) entropy
\begin{equation}
S_1(\ell_A,L)=S_1(L-\ell_A,L) = -{\rm Tr}\,\rho_A \ln \rho_A.
\label{vNen}
\end{equation} 
The asymptotic behavior of bipartite entanglement entropies is known
at the critical point $\mu=0$.\cite{HLW-94,CC-04,JK-04} We have
\begin{equation}
S_\alpha(\ell_A,L)\approx c \,q\, {1+\alpha^{-1}\over 12} \left[ \ln L 
+ \ln \sin( \pi \ell_A/L) + e_\alpha \right],
\label{ccfo}
\end{equation}
where $c$ is the central charge, $q$ counts the number of boundaries
between the two parts, thus $q = 2$ in the case of PBC and $q=1$ in
the case of OBC.  The constant $e_\alpha$ is nonuniversal and depends on
the boundary conditions.\cite{CC-04,JK-04,IJ-08}  

The corrections to Eq.~(\ref{ccfo}) may have various origins. Beside
the corrections discussed in Sec.~\ref{generalFSS}, there are
additional corrections. Within CFT they are related to the operators
associated with the conical singularities at the boundaries between
the two parts, which appear in the $\alpha$-sheeted Riemann surface
introduced to compute ${\rm Tr}\,\rho_A^\alpha$.~\cite{CC-10,CCP-10}
In the limit $L,\ell_A\to \infty$ at fixed $\ell_A/L$, these new
operators give rise to terms of order $L^{-\varepsilon/\alpha}$ in the
case of OBC\cite{CC-10,CCP-10} and of order $L^{-2
  \varepsilon/\alpha}$ in the case of PBC.\cite{CCEN-10,FC-11} Here
$\varepsilon>0$ is the RG dimension~\cite{CC-10,CCP-10} of the leading
conical operator.  The results for a number of 1D models suggest that
the energy operator plays a major role in this
respect,\cite{CCEN-10,FC-11,XA-12,DLF-12} hence $\varepsilon = 1/\nu$.
Moreover, the analysis of exactly solvable models shows the presence
of other corrections suppressed by integer powers of
$L$.~\cite{CCEN-10} The general predictions are confirmed by the exact
results for the XY chain at the critical point, for both OBC and
PBC. They are summarized in App.~\ref{App.B}.

The asymptotic behavior of the bipartite entanglement entropies is
also known in the thermodynamic limit close to the transition
point,\cite{CC-04,CCP-10} i.e., for $L,\ell_A\ll \xi$, where $\xi$ is
the length scale of the critical modes. One obtains
\cite{CC-04,FIK-08,CCP-10}
\begin{equation}
S_\alpha(\ell_A,L;\mu)\approx cq {1+\alpha^{-1}\over 12}\ln \xi + a_\alpha,
\quad \xi\ll \ell_A,L,
\label{salthl}
\end{equation}
where again $q = 2$ in the case of PBC and $q=1$ in the case of OBC,
and $a_\alpha$ is a nonuniversal constant.  The corrections to the
asymptotic behavior (\ref{salthl}) are expected to be\cite{CCP-10} of
order $\xi^{-\varepsilon/\alpha}$, where $\varepsilon$ is the same
exponent controlling the finite-size corrections at the critical
point. Additional corrections of order $\xi^{-2}$ should also be
present, see e.g. Ref.~\onlinecite{EEFR-12}

In the general FSS regime, the bipartite entanglement entropy has been
conjectured to satisfy the asymptotic scaling equation~\cite{CC-04}
\begin{equation}
 S_\alpha(\ell_A,L;\mu) - S_\alpha(\ell_A,L;0) \approx  
\Sigma_\alpha(\ell_A/L,\mu L^{1/\nu}).
\end{equation}
Consistency with Eqs.~(\ref{ccfo}) and (\ref{salthl}) implies 
that for $w\to\infty$
\begin{equation}
\Sigma_\alpha(\ell_A/L,w) \approx - \nu\, c\, 
q \,{1 + \alpha^{-1}\over 12} \ln w\,.
\label{asywe}
\end{equation}

If we include the scaling corrections, we expect
\begin{eqnarray}
&& S_\alpha(\ell_A,L;\mu) - S_\alpha(\ell_A,L;0) \approx
\Sigma_\alpha(\ell_A/L,u_\mu/u_l^{1/\nu}) 
 + \nonumber \\
&& \;\; + \; b_\alpha u_l^{\varepsilon/\alpha} 
\Sigma_{\alpha,c}(\ell_A/L,u_\mu/u_l^{1/\nu}) + \ldots
\label{fsssal}
\end{eqnarray}
where the dots correspond to other corrections of order
$u_l^{\omega}$, $u_l^{\omega_s}$, $\ldots$, which may be more
relevant than the conical ones in some cases.

Starting from the entanglement entropies, one can define RG invariant
quantities, which can be used to determine the critical behavior in a
finite volume. For this purpose, we consider
\begin{equation}
Q_{\alpha}(X,Y) = {12 \over q(1+\alpha^{-1})} \left[{S_\alpha(XL,L,\mu)
    - S_\alpha(YL,L,\mu)\over \ln\sin(\pi X)-\ln\sin(\pi Y)}\right]
\label{qalpha}
\end{equation}
with $0<Y<X<1$.  According to Eq.~(\ref{ccfo}), at the critical point
$\mu=0$
\begin{equation}
{\rm lim}_{L\to\infty} \; Q_\alpha(X,Y)= c.
\label{critqa}
\end{equation}
On the other hand, for $\mu\not=0$ and $\xi\ll L$, since
$S_\alpha(\ell_A,L,\mu)$ is independent of $\ell_A$ in this limit, we have
$Q_{\alpha}(X,Y) = 0$.

The quantity $Q_\alpha$ may be used to determine the transition point
and critical exponents, as the RG invariant quantities $R$ considered
in Sec.~\ref{drgi}.  For any boundary condition, Eq.~(\ref{fsssal})
implies
\begin{equation}
Q_\alpha(\mu,L) = {\cal Q}_\alpha(u_\mu u_l^{-1/\nu}) + b_\alpha
u_l^{\varepsilon/\alpha} {\cal Q}_{\alpha,c}(u_\mu u_l^{-1/\nu}) + \ldots
\label{qalfss}
\end{equation}
with ${\cal Q}_\alpha(0)=c$, where the dependence on the interval
coordinates $X,Y$ is understood.  The scaling functions ${\cal
  Q}_\alpha$ and ${\cal Q}_{\alpha,c}$ depend only on $X$,$Y$, and the
boundary conditions, apart from a trivial normalization of their
argument, while $b_\alpha$ is a nonuniversal constant. In the PBC case,
we have 
\begin{equation}
{\cal Q}_{\alpha,c}(0) = 0,
\label{qac0}
\end{equation}
since corrections decay as $L^{-2\varepsilon/\nu}$ at the critical point.
Beside the
corrections of order $L^{-\varepsilon/\alpha}$, one should
also consider the standard corrections related to the usual bulk and
boundary irrelevant operators, and analytic corrections.

\subsection{FSS in the XY chain}
\label{xychainent}

\begin{figure}[tbp]
\includegraphics*[scale=\graphicscale]{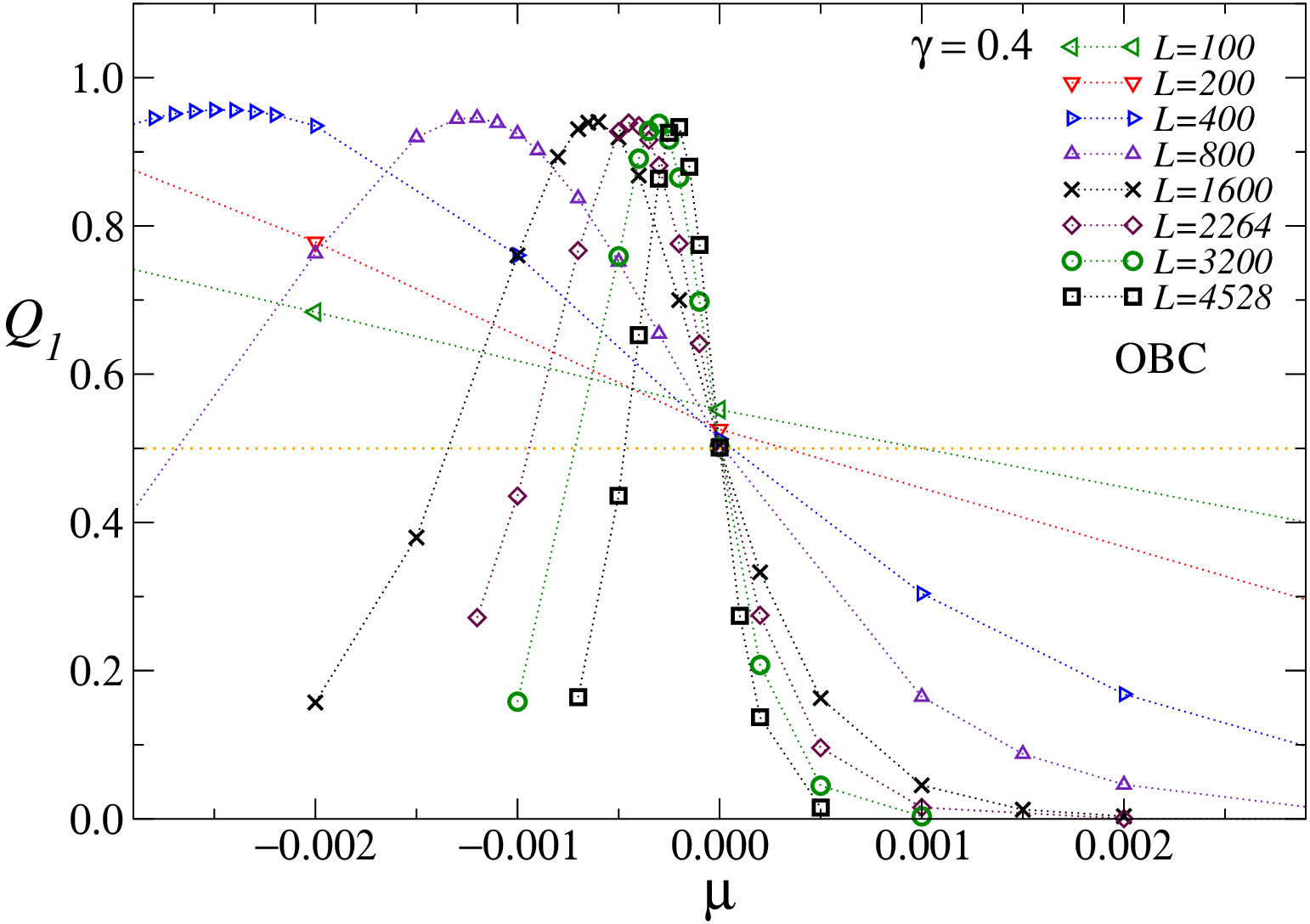}
\caption{(Color online) The quantity $Q_1$, derived from the vN
  entanglement entropy using Eqs.~(\ref{qalpha}) and (\ref{qadef}),
  for $\gamma=0.4$ and OBC for several values of $L$.  The dotted
  lines connecting the data corresponding to the same value of $L$ are
  only meant to guide the eye.}
\label{q1vsmu}
\end{figure}

\begin{figure}[tbp]
\includegraphics*[scale=\graphicscale]{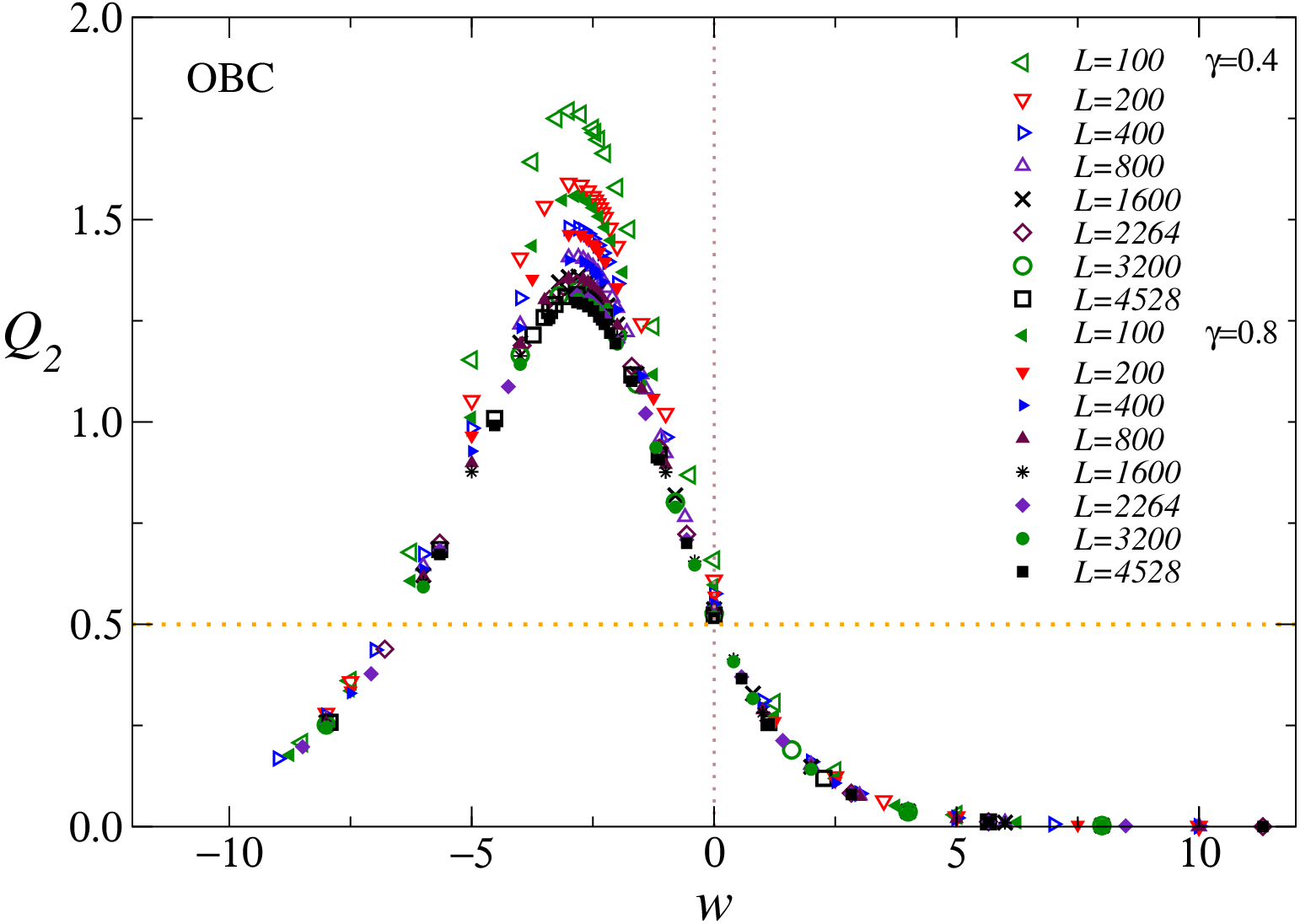}
\includegraphics*[scale=\graphicscale]{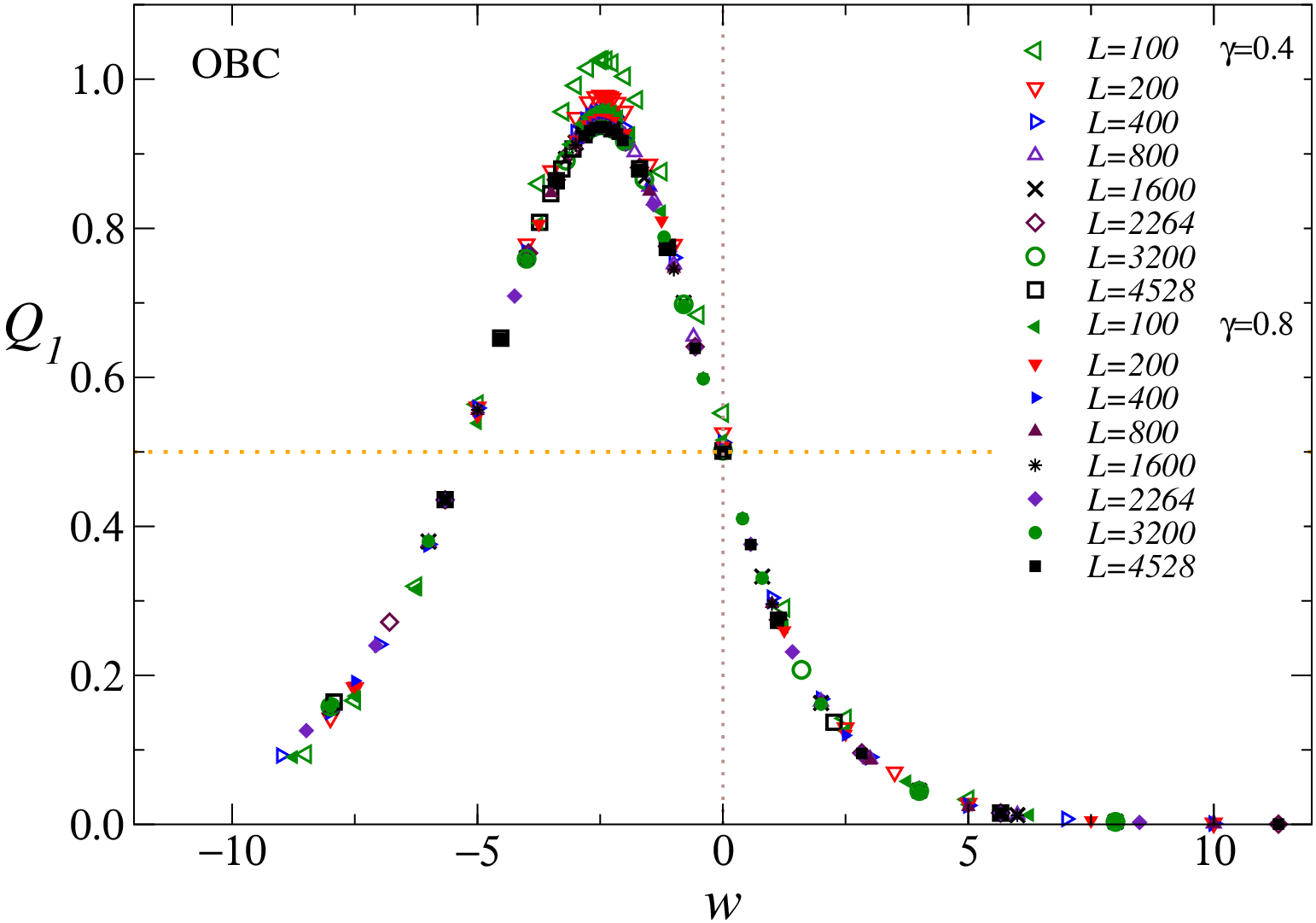}
\caption{(Color online) Plot of $Q_1$ (bottom) and $Q_2$ (top) for OBC
  vs $w \equiv \mu L/\gamma$, for $\gamma=0.4$ and $\gamma=0.8$.  
In both cases  the data for different values of $\gamma$ approach 
a universal large-$L$ curve.}
\label{qfssobc}
\end{figure}

\begin{figure}[tbp]
\includegraphics*[scale=\graphicscale]{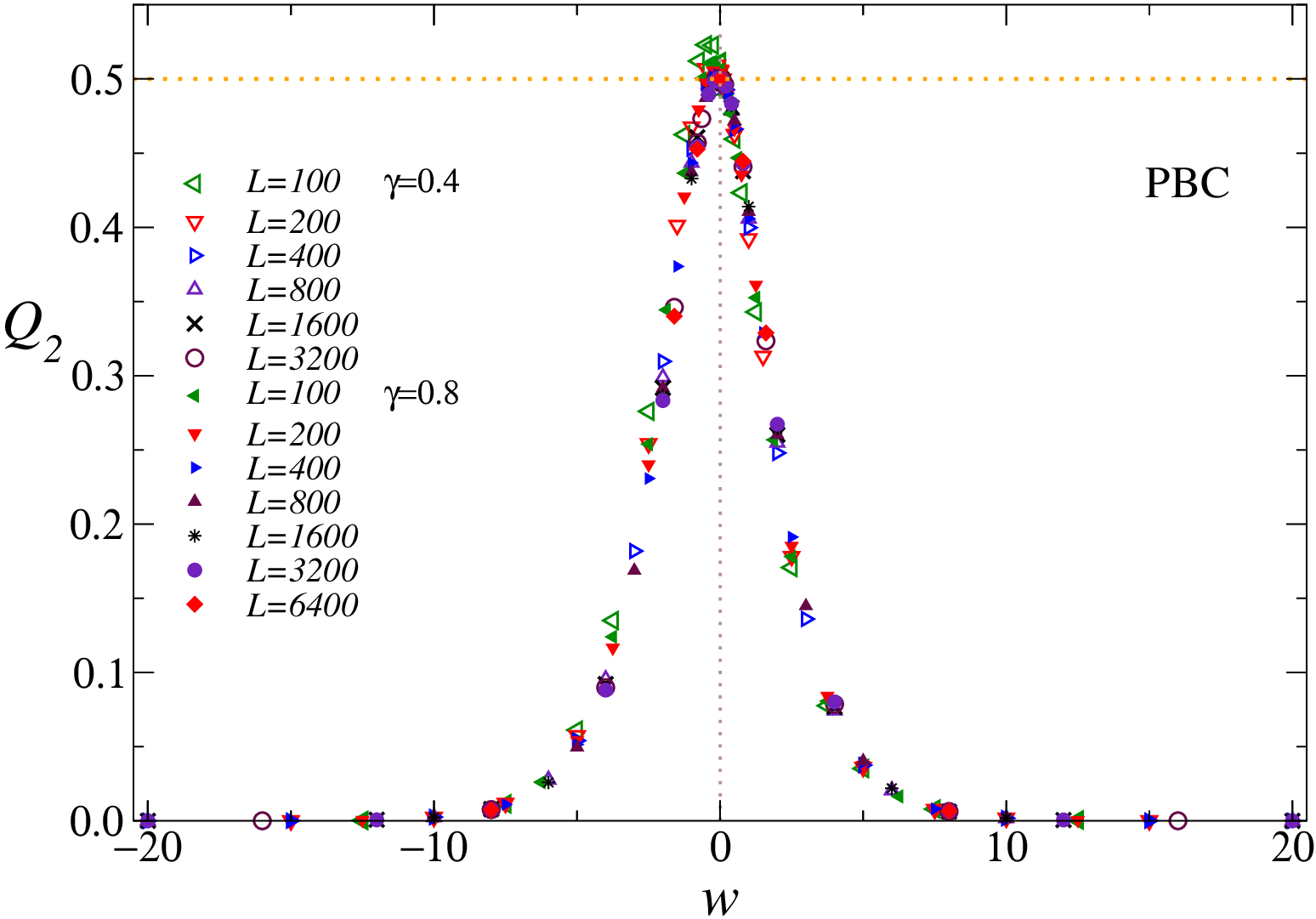}
\includegraphics*[scale=\graphicscale]{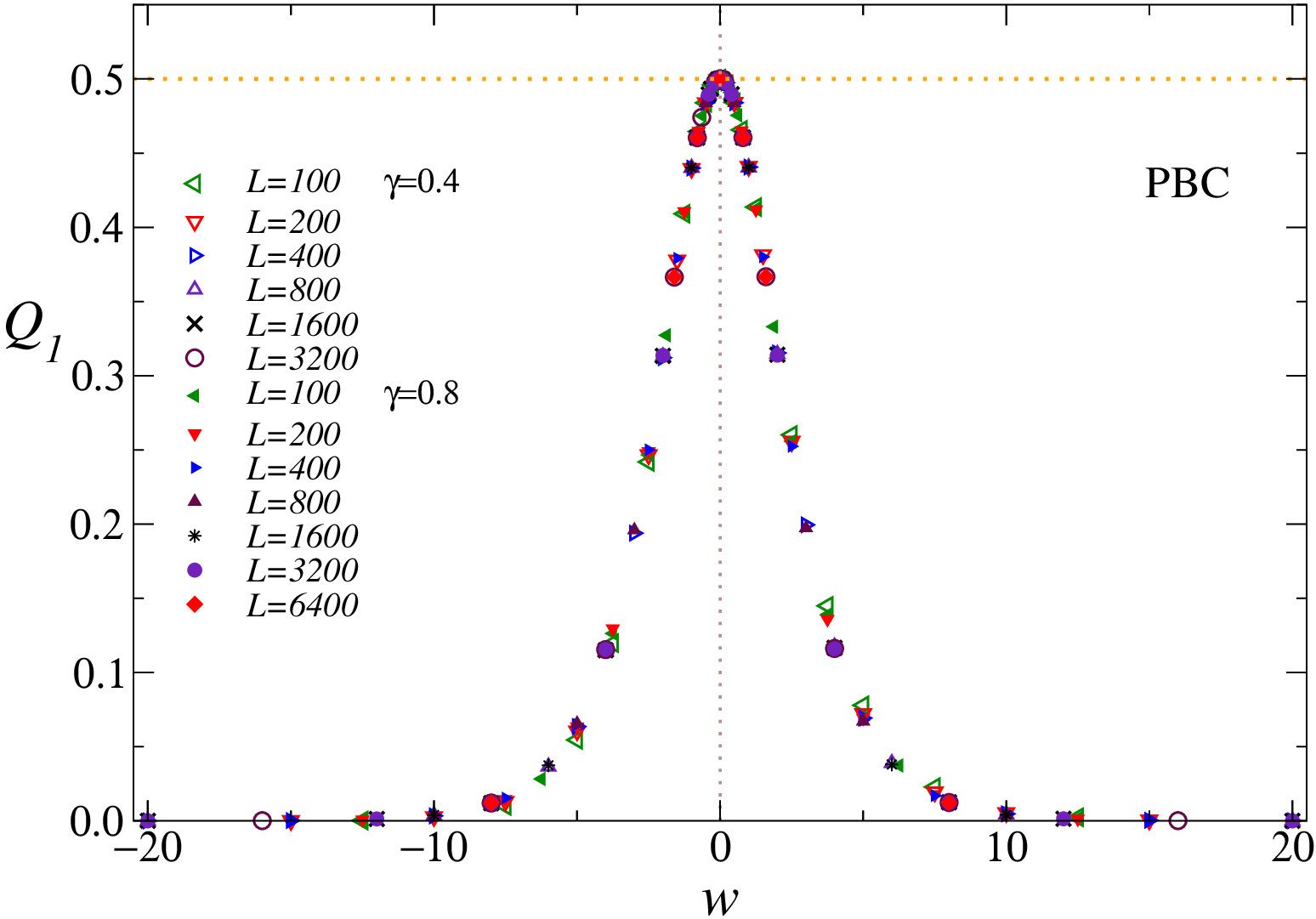}
\caption{(Color online) $Q_1$ (bottom) and $Q_2$ (top) for OBC vs $w =
  \mu L/\gamma$, for $\gamma=0.4$ and $\gamma=0.8$.  In both cases the
  data clearly converge toward an asymptotic large-$L$ curve which is
  independent of $\gamma$.}
\label{qpbcfss}
\end{figure}

To verify the general FSS behaviors presented in Sec.~\ref{eprobes},
we consider again the XY chain. In this case $\nu = 1$, so that
$\varepsilon=1$.  Therefore, for $\alpha > 1$ the corrections
associated with the R\'enyi entanglement entropies of order
$L^{-1/\alpha}$ are stronger than the standard ones discussed in the
previous sections, which scale as $1/L$ at least. We consider the
quantity
\begin{equation}
Q_\alpha \equiv Q_\alpha(X=1/2,Y=1/4),
\label{qadef}
\end{equation}  
i.e., we take $X=1/2$ and $Y=1/4$ in Eq.~(\ref{qalpha}).  In the
following we present results derived from the Renyi and vN
entanglement entropies of XY chains, for several values of $\gamma$,
OBC and PBC, and lattice sizes up to $L=O(10^4)$.

Fig.~\ref{q1vsmu} shows $Q_1$ for $\gamma=0.4$ and several values of
$L$ with OBC.  The curves show a maximum for $\mu<0$ and cross each
other approximately at $\mu=0$.  Using Eq.~(\ref{qalfss}), one can
easily establish that the crossing point $\mu_{\rm cross}(L)$, defined
by
\begin{equation}
Q_\alpha[\mu_{\rm cross}(L),L] = Q_\alpha[\mu_{\rm cross}(L),2L],
\label{crossp}
\end{equation}
approaches the critical point as
\begin{equation}
\mu_{\rm cross} = O(L^{-1/\nu - 1/\alpha}).
\label{mucrossq}
\end{equation}
This is confirmed by the results for $Q_\alpha$, see e.g.
Fig.~\ref{q1vsmu}.

Figs.~\ref{qfssobc} and \ref{qpbcfss} show plots of $Q_1$ and $Q_2$,
for OBC and PBC respectively, versus the scaling variable $w=\mu
L/\gamma$ for $\gamma=0.4$ and $\gamma=0.8$.  The data appear to
approach universal curves with increasing $L$, clearly supporting the
universality of the asymptotic function ${\cal Q}_\alpha(w)$,
cf. Eq.~(\ref{qalfss}), for both OBC and PBC.  Note that the maximum
of the PBC scaling curve is at $w\equiv\mu L/\gamma=0$ and equals
$c=1/2$, while the OBC maximum is larger than $c=1/2$---we obtain
$Q_{1,{\rm max}}\approx 0.9358$ and $Q_{2,{\rm max}}\approx
1.248$---and it is located in the region $w<0$.  The scaling curves
vanish exponentially for $|w|\to\infty$.  As expected, scaling
corrections appear larger for $Q_2$ than $Q_1$.

\begin{figure}[tbp]
\includegraphics*[scale=\graphicscale]{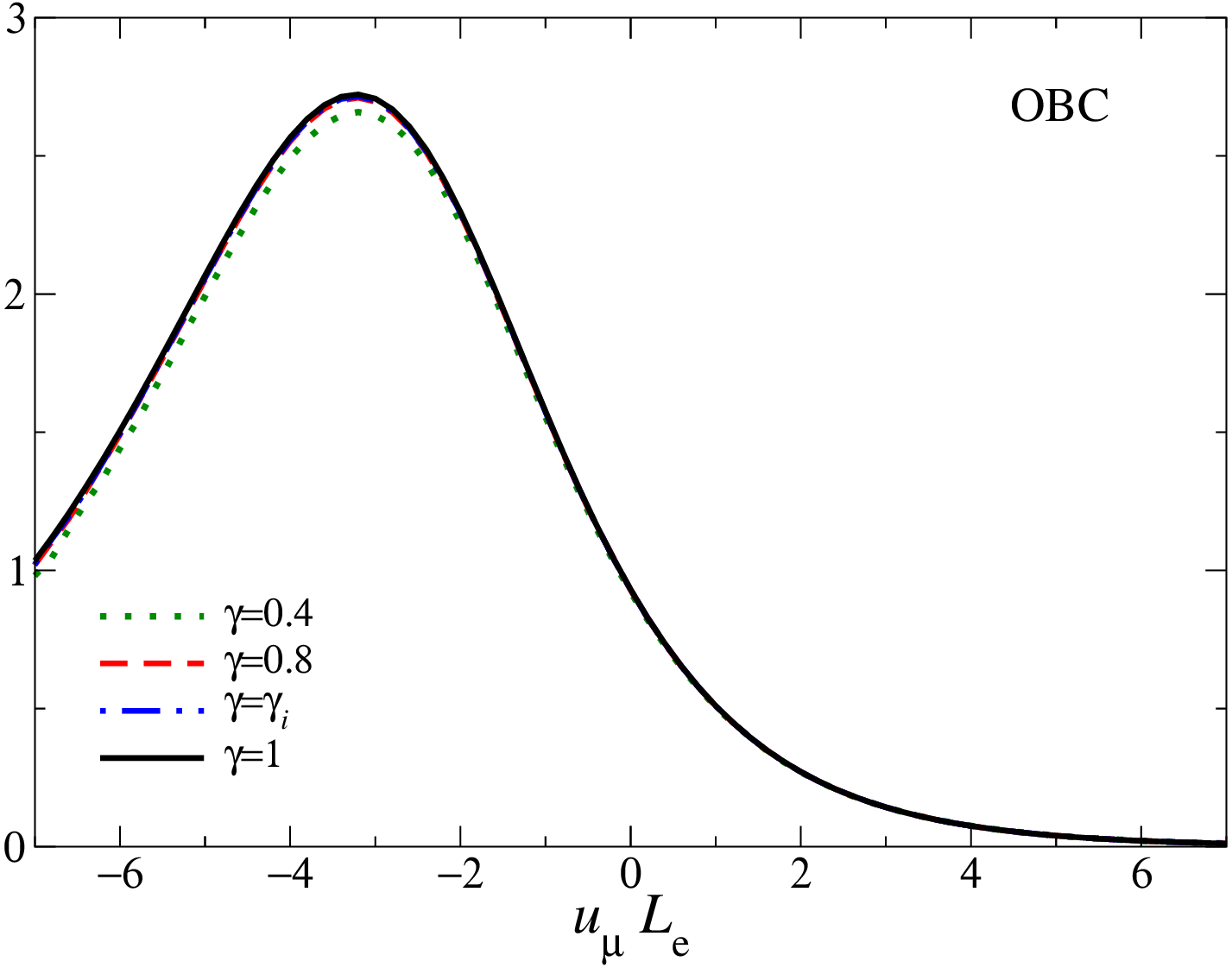}
\caption{(Color online) The large-$L$ limit of $\widehat{Q}_{2}$ for OBC, defined
  in Eq.~(\ref{Q1corr-PBC-def}), vs $\widetilde{w}_e = u_\mu L_e$,
  for several values of $\gamma$, in particular $\gamma_i=\sqrt{3}/2$.
  The different curves are hardly distinguishable: the small
  differences are within the accuracy of the large-$L$ extrapolation
  of the data up to $L=1024$.  }
\label{q2corrOBC}
\end{figure}

\begin{figure}[tbp]
\includegraphics*[scale=\graphicscale]{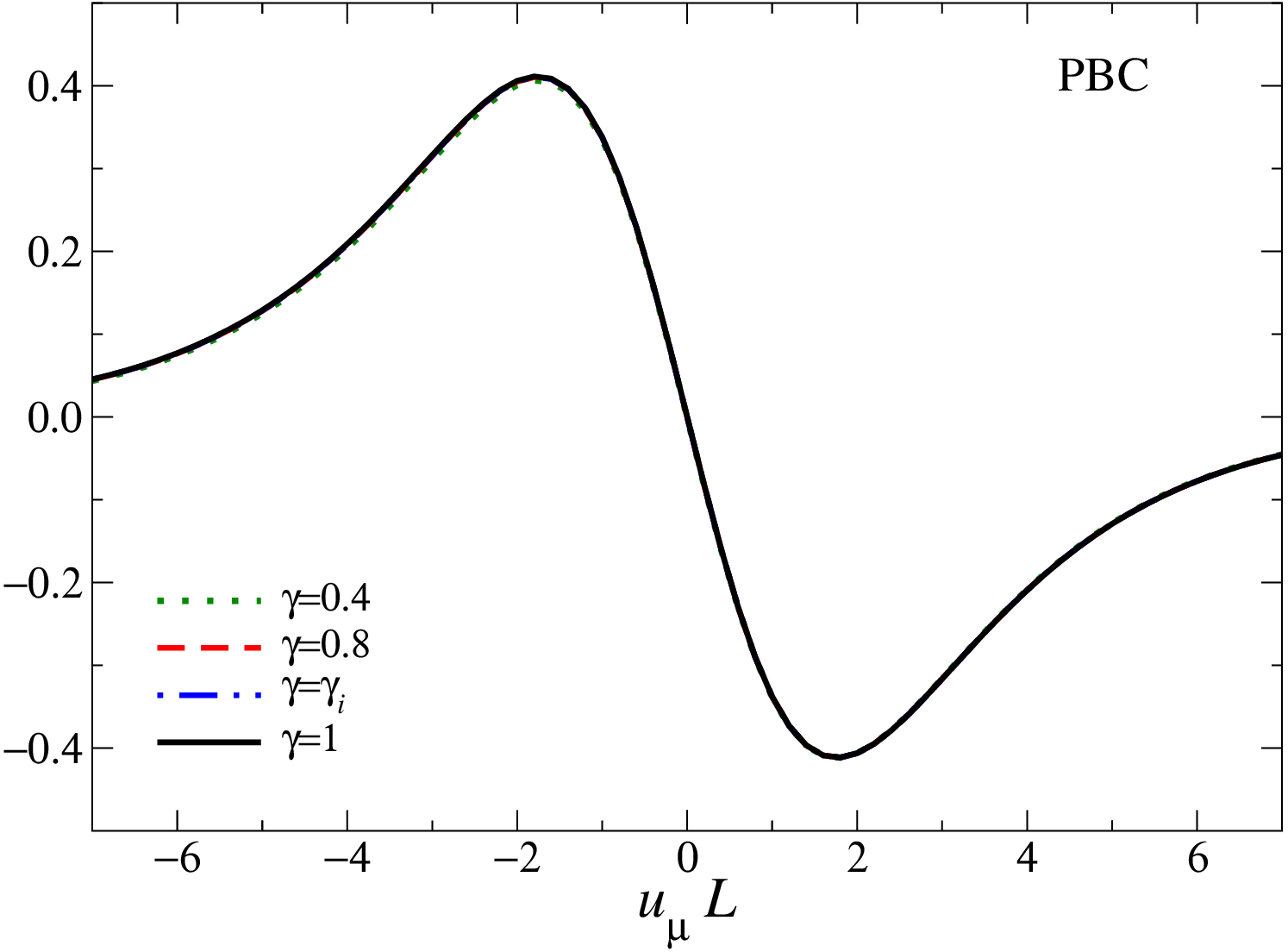}
\caption{(Color online) The large-$L$ limit of $\widehat{Q}_{2}$ for
  PBC, defined in Eq.~(\ref{Q1corr-PBC-def}), vs $\widetilde{w} =
  u_\mu L$, for several values of $\gamma$. Here
  $\gamma_i=\sqrt{3}/2$.  On the scale of the figure, the different
  curves are hardly distinguishable. The small differences are within
  the extrapolation errors.  For $w=0$ the numerical data are
  consistent with zero, i.e., with the absence of $L^{-1/2}$
  corrections, in agreement with the exact results at the critical
  point.}
\label{q2corrPBC}
\end{figure}

Let us now investigate the corrections to the leading term.  To begin
with, we consider the Renyi entanglement entropies for $\alpha>1$,
whose leading corrections are expected to be due to the conical
singularities, i.e. the $O(L^{-1/\alpha})$ term explicitly reported in
Eq.~(\ref{qalfss}).  We use the asymptotic formulas reported in
App.~\ref{App.B} to derive the finite-size behavior of $Q_\alpha$ at
$\mu=0$. We obtain
\begin{eqnarray}
Q_\alpha = 1/2 + b_\alpha\, L^{-1/\alpha}+O(L^{-2/\alpha}) + O(L^{-1}),
\label{qacritobc}
\end{eqnarray}
where
\begin{eqnarray}
&&b_\alpha = \bar{b}_\alpha \gamma^{-1/\alpha},
\label{gdepba}\\
&&\bar{b}_{\alpha} = 
{12 (\pi/8)^{1/\alpha} \Gamma[1/2+1/(2\alpha)] (2^{1/(2\alpha)} - 1) \over  
(1+\alpha) \Gamma[3/2-1/(2\alpha)] \ln 2}.
\nonumber
\end{eqnarray}
In particular $\bar{b}_2=0.925049...$ for $\alpha=2$.
Instead, for PBC at $\mu=0$, we find
\begin{eqnarray}
&&Q_\alpha = 1/2 + p_\alpha L^{-2/\alpha}+O(L^{-4/\alpha}) + O(L^{-2}), 
\label{qacritpbc}\\
&&p_{\alpha} =
{ 3 (\alpha-1)(\pi/4)^{2/\alpha} (2^{1/\alpha} -1 ) 
\Gamma[1/2+1/(2\alpha)]^2
\over \alpha (\alpha+1)
\Gamma[3/2-1/(2\alpha)]^2 {\rm ln}2}
\nonumber 
\end{eqnarray}
for $\gamma=1$. 
In particular $p_2(\gamma=1)= 0.428928...$.

The FSS limit is taken at fixed $\widetilde{w}_e = u_\mu L_e$ for OBC
and $\widetilde{w} = u_\mu L$ for PBC (see Eqs.~(\ref{umugamma}) and
(\ref{legamma}) for the definitions of $u_\mu$ and $L_e$), to avoid
analytic corrections due to the expansion of the scaling fields.  The
numerical data of the $\alpha=2$ Renyi entropy are in full agreement
with Eq.~(\ref{qalfss}): for both PBC and OBC scaling corrections
decay as $L^{-1/2}$.  Moreover, for both OBC and PBC, the corrections
are proportional to $\gamma^{-1/2}$ as found at the critical point,
cf. Eq.~(\ref{gdepba}).  This is clearly demonstrated by the analysis
of the large-$L$ behavior of the quantity
\begin{eqnarray}
\widehat{Q}_{2,c} \equiv
   2 (\gamma L)^{1/2} 
   [Q_2(\widetilde{w},L,\gamma) - Q_2(\widetilde{w},4 L,\gamma)].
\label{Q2corr-PBC-def}
\end{eqnarray}
If corrections are of order $(\gamma L)^{-1/2}$, in the limit
$L\to\infty$ at fixed $\widetilde{w}$ or $\widetilde{w}_e$,
$\widehat{Q}_{2,c}$ converges to a nontrivial $\gamma$-independent
scaling function, i.e. to the function ${\cal Q}_{2,c}(\widetilde{w})$
appearing in Eq.~(\ref{qalfss}).  Figs.~\ref{q2corrOBC} and
\ref{q2corrPBC} show the extrapolation of $\widehat{Q}_{2,c}$ for OBC
and PBC, respectively. They are obtained by using results for chains
of length $L\le 4096$. The resulting tiny differences that are hardly
visible in Figs.~\ref{q2corrOBC} and \ref{q2corrPBC} are plausibly due
to tiny numerical errors affecting the raw data and to the
extrapolation uncertainty.  The curves for different values of
$\gamma$ appear to approach a unique curve, thus supporting our
general scenario.  Analogous results are expected for any $\alpha>1$.

\begin{figure}[tbp]
\includegraphics*[scale=\graphicscale]{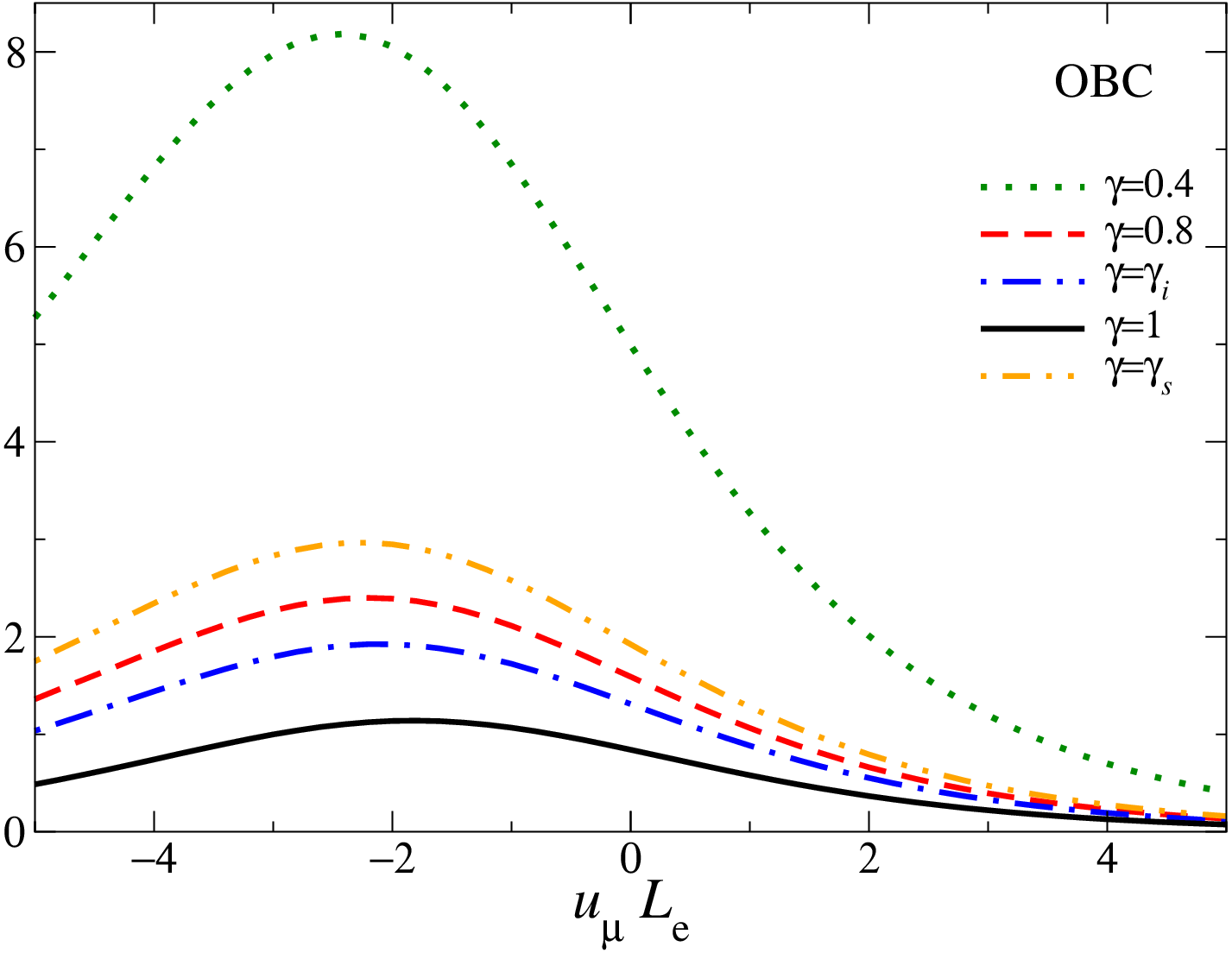}
\caption{(Color online) The large-$L$ limit of $\widehat{Q}_{1,c}$
  defined in Eq.~(\ref{Q1corr-OBC-def}) vs
  $\widetilde{w}_e = u_\mu L_e$, for several values of $\gamma$.
  Here $\gamma_i=\sqrt{3}/2$ and $\gamma_s = \sqrt{3} - 1$.}
\label{q1corrOBC}
\end{figure}

The analysis of the leading corrections for the vN entanglement
entropy is more complicated, essentially because, in the 
limit $\alpha\to 1$, the leading corrections may have different origins.  This is
already shown by the results at the critical point.  The asymptotic
expansion of the vN entanglement entropy at the critical point for OBC
and $\gamma=1$ is reported in App.~\ref{App.B}. This allows us to
derive
\begin{eqnarray} 
&&Q_1(0,L) = 1/2 + b_{\rm vN} L^{-1} + O(L^{-2}),
\label{q1obccr} \\
&&b_{\rm vN} = {\pi(6\sqrt{2}-7)\over 8{\rm ln}2}  
\quad {\rm for}\;\;\gamma=1.
\nonumber
\end{eqnarray}
Note that $b_{\rm vN}$ does not coincide with the $\alpha\to 1$ limit
of the coefficient $b_\alpha$ appearing in Eq.~(\ref{gdepba}).  Thus
other corrections contribute at order $1/L$. To understand better the
subleading FSS behavior, we computed the corrections of order $L^{-1}$
at fixed $\widetilde{w}_e = u_\mu L_e$. They can be estimated by
considering
\begin{eqnarray}
 \widehat{Q}_{1,c} = 2 L 
   [Q_1(\widetilde{w}_e,L,\gamma) - Q_1(\widetilde{w}_e,2L,\gamma)].
\label{Q1corr-OBC-def}
\end{eqnarray}
This quantity is constructed so that it approaches a nontrivial
function if the leading corrections are of order $L^{-1}$.
Fig.~\ref{q1corrOBC} shows the large-$L$
extrapolations~\cite{footnoteextrapo} of $\widehat{Q}_{1}$ for several
values of $\gamma$. We verify that the $\gamma$ dependence cannot be
eliminated by rescaling $\widehat{Q}_{1,c}(\widetilde{w}_e,L,\gamma)$
by a function of $\gamma$. Hence, beside the conical contribution,
there must be other corrections due to the boundaries. They may be
interpreted as analytic corrections related to the length $\ell$ of
the domain. Analogously to the nonlinear scaling field $u_l$
associated with $1/L$ which has an expansion in powers of $1/L$,
cf. Eq.~(\ref{ull}), it is natural to introduce a scaling field
$u_\ell$ associated with $\ell$, with $u_\ell \approx 1/\ell +
a/\ell^2$. The expansion of $u_\ell$ would contribute additional
boundary corrections of order $1/L$, when the limit is taken at fixed
$\ell/L$. This is confirmed by the asymptotic behavior of the vN
entanglement entropy at the critical point, see
App.~\ref{App.B}. Indeed, for $\gamma=1$ it can be written as
\begin{eqnarray}
S_1(\ell,L) &=& 
{1\over 12} \Bigl[ \ln L_e + \ln\sin{\pi\ell_e\over L_e} + e_1 \Bigr]
\nonumber\\
&&- {\pi \over 16\sin(\pi \ell_e/L_e)} {1\over L_e}
+O(L_e^{-2}),
\label{salpha1t} 
\end{eqnarray}
where $L_e = L+1/2$ and $\ell_e = \ell + 1/4$, $e_1$ is a constant,
and the term of order $L^{-1}$ is the $\alpha\to 1$ limit of the
corrections of order $L^{-1/\alpha}$ occurring for generic
$\alpha>1$. Eq.~(\ref{salpha1t}) allows us to identify the origin of
the correction terms: there are conical corrections that give rise to
the $L_e^{-1}$ term appearing in Eq.~(\ref{salpha1t}), and boundary
terms that can be allowed for by introducing $L_e$ and $\ell_e$.

\begin{figure}[tbp]
\includegraphics*[scale=\graphicscale]{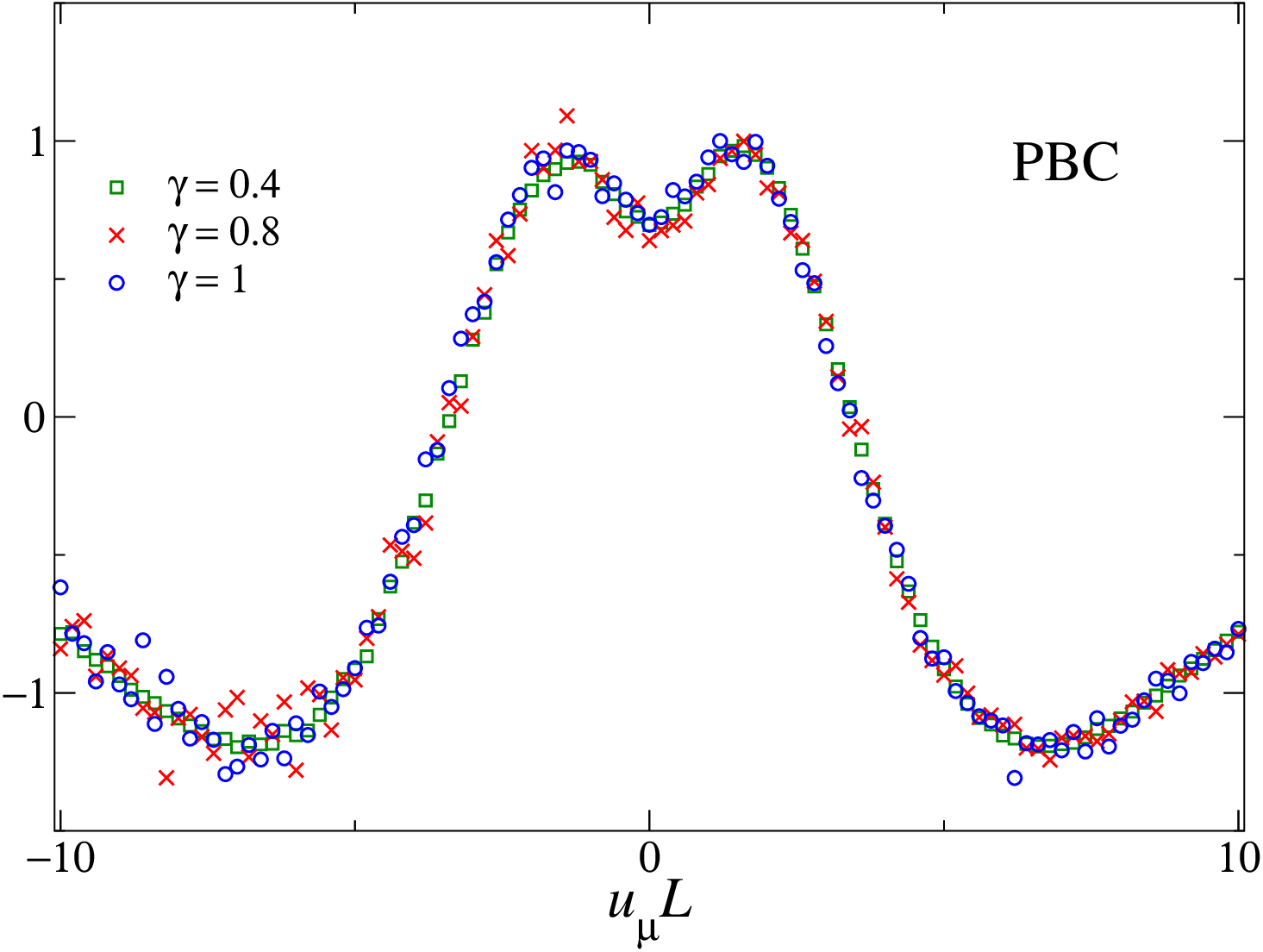}
\caption{(Color online) The large-$L$ limit of $\widehat{Q}_{1,c2}$
  defined in Eq.~(\ref{Q1corr-PBC-def}) for various values of $\gamma$
  vs $\widetilde{w} = u_\mu L$.  The data collapse along a
  unique curve; the apparent oscillations, that are particularly
  visible for $\widetilde{w}<0$ and $\gamma = 0.8$, are essentially
  due to numerical errors in the computation of the raw data for $L\le
  512$. }
\label{q1corrPBC}
\end{figure}

In the case of PBC, the results of App.~\ref{App.B} 
at the critical point lead to
\begin{eqnarray}
&&Q_1(0,L,\gamma) = 1/2 +  p_{\rm vN} L^{-2} + O(L^{-4}),
\label{q1pbccr}\\
&&p_{\rm vN} = {\pi^2\over 80 {\rm ln} 2} \quad {\rm for}\;\;\gamma=1.
\nonumber
\end{eqnarray}
The constant $p_{\rm vN}$ is
unrelated to the constant $p_\alpha$ defined in
Eq.~(\ref{qacritpbc}). Indeed, $p_\alpha$ vanishes for
$\alpha=1$. These results apparently indicate that conical
singularities are not related to the $L^{-2}$ corrections at the
critical point. Let us now extend the analysis outside the critical
point, computing $Q_1$ in the FSS limit at fixed $\widetilde{w}$.  A
detailed numerical analysis shows that there are no scaling
corrections of order $1/L$. The function ${\cal Q}_{1,c}$ appearing in
Eq.~(\ref{qalfss}) vanishes identically: corrections for
$\widetilde{w}\not=0$ decay as $L^{-2}$ as it occurs at the critical
point. A detailed analysis of the numerical data for several values of
$\gamma$ shows that we can write
\begin{eqnarray}
&& Q_1(\mu,L,\gamma) = {\cal Q}_1(\widetilde{w}) + 
\label{scal-Q1-PBC} \\
&& \, L^{-2} 
\left[
{\cal Q}_{1,c1}(\widetilde{w}) + v_1(0,\gamma) {\cal Q}_{1,c2}(\widetilde{w})
\right] + O(L^{-3}).
\nonumber 
\end{eqnarray}
where $v_1(\mu,\gamma)$, defined in Eq.~(\ref{v1mugamma}), is the
scaling field associated with the leading bulk subleading corrections.
Notice that if we replace $\widetilde{w}\equiv u_\mu L$ with its
linear approximation $w \equiv \mu L/\gamma$ in
Eq.~(\ref{scal-Q1-PBC}), the leading term does not change but now the
corrections are of order $L^{-1}$. They are due to the next-to-leading
term appearing in the expansion of $\widetilde{w}$ in powers of $w$.
To verify Eq.~(\ref{scal-Q1-PBC}), we consider
\begin{equation}
\widehat{Q}_{1,c2}(\widetilde{w},L,\gamma) = {L^2\over v_1(0,\gamma)} 
   \left[Q_1(\widetilde{w},L,\gamma) - 
         Q_1(\widetilde{w},L,\gamma_i)\right],
\label{Q1corr-PBC-def}
\end{equation}
where $\gamma_i = \sqrt{3}/2$ (we remind the
reader that $v_1(0,\gamma_i) = 0$).
If Eq.~(\ref{scal-Q1-PBC}) holds,
$\widehat{Q}_{1,c2}(\widetilde{w},L,\gamma)$ should converge to ${\cal
  Q}_{1,c2}(\widetilde{w})$, hence it should be independent of
$\gamma$ for large values of $L$. As shown in Fig.~\ref{q1corrPBC}, a
straightforward extrapolation~\cite{footnoteextrapo} of data up to
$L=4096$ supports it.  The second correction term in
Eq.~(\ref{scal-Q1-PBC}) is associated with the bulk irrelevant
operator. The origin of the first term, which is independent of
$\gamma$, is instead less clear. As we have already discussed, 
at the critical point the conical corrections of order $L^{-2/\alpha} =
L^{-2}$ vanish, hence ${\cal Q}_{1,c1}(0)$ can only be an analytic
correction. It is natural to conjecture that the same is true outside
the critical point. Indeed, if conical and analytic corrections were
both present, one would expect them to have different $\gamma$
dependencies. Hence, one would expect two different scaling functions
with different $\gamma$-dependent coefficients.

\section{Summary and conclusions}
\label{conclusions}

We study FSS at quantum zero-temperature transitions, focusing on the
corrections to the leading asymptotic behavior.  This issue is
relevant for numerical and experimental studies of quantum transition,
where the data are generally available for a limited range of system
sizes, which are often relatively small.  In these cases, the FSS
predictions are subject to sizable scaling corrections, which must be
taken into account to obtain reliably accurate estimates of the
critical parameters and, if needed, to identify the universality class
of the transition.

We present a RG analysis of FSS at quantum zero-temperature
transitions of $d$-dimensional systems characterized by two relevant
parameters $\mu$ and $h$, which are respectively even and odd with
respect to an assumed parity-like symmetry. Well known examples of
such quantum transitions are those occurring in quantum XY (Ising)
systems and general O$(N)$-symmetric spin models, superfluid or
metallic transitions in particle systems, etc.; see, e.g.,
Ref.~\onlinecite{Sachdev-book}.

To characterize the scaling corrections, we generalize Wegner's
scaling Ansatz~\cite{Wegner-76} to quantum transitions.  This allows
us to predict the type of subleading corrections that are expected in
finite systems and/or at finite temperature.  First, there are
corrections associated with the bulk and boundary irrelevant RG
perturbations, that decay as $L^{-\kappa}$, where $\kappa$ is
generally a noninteger exponent (for example, in the case of the
quantum transitions of two-dimensional quantum Ising and Heisenberg
models, the leading bulk $O(L^{-\omega})$ corrections have
$\omega\approx 0.8$, see e.g. Ref.~\onlinecite{PV-02}).  Then, one
should consider analytic corrections due to the regular
backgrounds. Finally, since the RG predictions are expressed in terms
of the nonlinear scaling fields, one should also consider the
correction terms arising from their expansion in powers of the
Hamiltonian parameters, the spatial size $L$, and the temperature $T$.
These corrections are also named analytic, though they decay as
$L^{-\rho}$, with noninteger $\rho$.

To check the general predictions, we consider the quantum XY chain in
a transverse field with Hamiltonian (\ref{Isc}), which is a standard
theoretical laboratory to understand issues concerning quantum
transitions. In particular, it is an ideal testing ground, since its
Hamiltonian can be exactly diagonalized,\cite{Katsura-62} allowing us
to compute several interesting quantities either exactly or very
accurately by using numerical methods.

The analytic computation of the finite-size behavior of the energy
spectrum and of the free energy allows us to infer the exact form of
the nonlinear scaling fields related to the relevant Hamiltonian
parameter $\mu\equiv g-1$ and to the leading irrelevant operator with
RG dimension $-2$. Moreover, we can also determine the {\em speed of
  sound} $c$ which enters the relation between the temperature $T$ and
the corresponding scaling field.  We provide a complete analysis of
the asymptotic FSS behavior of the energy gap $\Delta$ (i.e., the
difference between the energies of the two lowest levels) up to
$O(L^{-4})$ for PBC and to $O(L^{-2})$ for OBC,
cf. Eq.~(\ref{expansion-Delta-wtilde}) and
Eq.~(\ref{expan-Delta-OBC}), respectively. In the PBC case, we show
that all terms up to $L^{-4}$ are due to the expansion of the
nonlinear scaling field $u_\mu$ associated with $\mu$ and to the
leading irrelevant RG perturbation.  In the OBC case the corrections
of order $L^{-1}$ in the expansion of $\Delta$ are due to the $L$
dependence of the nonlinear scaling field $u_l$ associated with the
spatial size $L$; they can be eliminated by introducing an effective
spatial size $L_e=L+l(\gamma)$, cf. Eq.~(\ref{legamma}).  Instead, the
corrections of order $L^{-2}$ show contributions associated with
boundary irrelevant RG perturbations of RG dimension
$\widetilde{y}_1=-2$.  Then, we perform an analogous analysis for some
RG invariant quantity derived from the two-point function of the order
parameter, i.e.  $G(x,y) = \langle \sigma^{(1)}_x \sigma^{(1)}_y
\rangle$, pointing out the presence of further corrections, arising
from mixings of the operator $\sigma^{(1)}_x$ with other odd
subleading operators.  These results for the XY chains are in full
agreement with the general RG framework put forward in
Sec.~\ref{generalFSS}, which generalizes Wegner's scaling theory to
quantum transitions.

Finally, we discuss the FSS behavior of bipartite entanglement R\'enyi
and vN entropies of one-dimensional systems with an isolated quantum
critical point with $z=1$ and central charge $c$.  They presents
further peculiar corrections to the asymptotic FSS behavior predicted
by CFT, arising from operators associated with conical singularities
in the corresponding conformal mapping.\cite{CC-10} The FSS
predictions are compared with results for the XY chain. We show that
the leading FSS corrections for the R\'enyi entropies with $\alpha>1$
are always of order $L^{-1/\alpha}$, for any boundary conditions, see
Eq.~(\ref{fsssal}).  In particular, in the PBC case corrections are of
order $L^{-2/\alpha}$ only at the critical point $\mu=0$. The behavior
of the vN entanglement entropy is more complex. In the OBC case, the
leading correction of order $L^{-1}$ is the sum of terms of different
origin: we find contributions from the conical operators and boundary
corrections as well.  In the PBC case, the expected corrections of
order $L^{-1}$ vanish: the leading FSS corrections are of order
$L^{-2}$ also for $\mu\not=0$. Apparently, they are the sum of an
analytic contribution and of a term due to the bulk irrelevant RG
operator.

In our FSS study of the entanglement properties we introduce the RG
invariant quantity $Q_\alpha$. It is defined in terms of the R\'enyi
entanglement entropy $S_\alpha$, see Eq.~(\ref{qalpha}), in such a way
to have a universal FSS behavior (in particular, it approaches the
central charge $c$ at the critical point).  The quantity $Q_\alpha$
may be useful to investigate 1D quantum transitions exploiting
entanglement properties.

\medskip

{\em Acknowledgements.} We thank Pasquale Calabrese for useful
discussions.

\appendix

\section{Useful CFT formulas}
\label{cftfo}

The 2D Ising universality class
can be associated with a CFT with central charge $c=1/2$. 
CFT provides the asymptotic FSS behavior of the two-point function at
the critical point.\cite{It-Dr-book}  We report
some useful CFT formulas for the critical two-point function which are
used in the paper.  We consider strips $L\times\infty$ with PBC and
OBC, i.e. with coordinates $-L/2\le x \le L/2$ and $y \in {\mathbb
  R}$.

\subsection{Open boundary conditions}
\label{apobc}

Setting $z_i \equiv x_i+L/2$, and 
\begin{equation}
Z_\pm\equiv (z_2\pm z_1)/L,  \quad Y \equiv (y_2-y_1)/L, 
\label{xydef}
\end{equation}
the critical two-point function on a strip with 
OBC reads~\cite{It-Dr-book,CFT-book}
\begin{eqnarray}
&&G_{\rm cft}(\vec{r}_1,\vec{r}_2) = 
{(\pi/L)^{1/4} \over \left[ \sin(\pi z_2/L)\sin(\pi z_1/L)\right]^{1/8}} \times
\label{gcftobcis}\\
&& \left[ {|\sin\pi(Z_++iY)/2|^{1/2}\over |\sin\pi(Z_-+iY)/2|^{1/2}}
- {|\sin\pi(Z_-+iY)/2|^{1/2}\over |\sin\pi(Z_++iY)/2|^{1/2}} \right]^{1/2}.
\nonumber
\end{eqnarray}
The two-point function $G(x_1,x_2)$ at the critical point is obtained
by setting $Y=0$.  This result allows us to exactly compute the
universal large-$L$ limit of the RG invariant quantities $R_\xi\equiv
\xi/L$ and $R_g$, defined in Eqs.~(\ref{xidef2}) and (\ref{rdef2}),
respectively, at the critical point. We obtain the critical value
\begin{equation}
R_g^*=0.306462\ldots, \qquad R_\xi^* = 0.159622\ldots
\label{rstarobc}
\end{equation}
For the XY chain we  may also consider the connected equal-time
two-point function of the operator $\sigma^{(3)}_x$, i.e.
\begin{equation}
G_n(x,y) = \langle \sigma^{(3)}_x \sigma^{(3)}_y \rangle
- \langle \sigma^{(3)}_x \rangle \langle \sigma^{(3)}_y \rangle.
\label{g3xy}
\end{equation}
For $T=h=0$, we obtain
\begin{equation}
G_n(0,x) \sim { {\rm cos}(\pi x/L) \over L^2 \,{\rm sin}^2(\pi x/L) }.
\label{gn0x}
\end{equation}
Note that $G_n(0,x) \sim x^{-2}$
for $|x|\ll L$. Therefore, the integral of $G_n(0,x)$ with respect to $x$ 
is infinite.

\subsection{Periodic boundary conditions}
\label{appbc}

In the case of PBC we have
\begin{eqnarray}
G_{\rm cft}(\vec{r}_1,\vec{r}_2) =  
{(\pi/L)^{1/4} \over |\sin\pi(Z_-+iY)|^{1/4}}.
\label{gcftpbcis}
\end{eqnarray}
Again, setting $Y=0$, we obtain the two-point function $G(x_1,x_2)$ at
the critical point, from which we can compute
\begin{equation}
R_g^*=0.153493..., \quad R_\xi^* = 0.187790... 
\label{rstarpbc}
\end{equation}
for $\xi$ defined as in Eq.~(\ref{xidef2}).
If instead the correlation length is defined as 
\begin{equation}
\xi^2 \equiv  
{\widetilde{G}(0)  - \widetilde{G}(k_{\rm min})\over 
k_{\rm min}^2 \widetilde{G}(k)},
\label{xidefpbc}
\end{equation}
where $\widetilde{G}$ is the Fourier transform of $G$, and $k_{\rm
  min}=2 \pi/L$, we obtain $R_\xi^*=0.389848\ldots$

\section{Some exact results for the entanglement entropies}
\label{App.B}

In this appendix we report some exact results for the entanglement
entropies of the XY chain at the critical point.  For this purpose, we
also exploit known results for the XX model,
\begin{eqnarray}
H_{\rm XX} = - {1\over 2} \sum_{i=1}^L [ \sigma^{(1)}_i \sigma^{(1)}_{i+1}
+ \sigma^{(2)}_i \sigma^{(2)}_{i+1} ] ,
\label{XXmodel}
\end{eqnarray}
and the exact relation~\cite{IJ-08}
\begin{equation}
S_\alpha^{\rm XY}(\ell,L)={1\over 2}S_\alpha^{\rm XX}(2\ell,2L) 
\label{exisxx}
\end{equation}
between the entanglement entropies of the XY model (\ref{Isc}) with
$g=1$ and $\gamma=1$, and those of the XX model (\ref{XXmodel}).

Some results for the corrections to the leading behavior within OBC
were already reported in Ref.~\onlinecite{CV-10-jstat}. Using also the
results of Ref.~\onlinecite{FC-11} for the XX model, we can write the
large-$L$ behavior of the R\'enyi entropy with $\alpha>1$ at fixed
$\ell/L$ as
\begin{eqnarray}
&&S_\alpha(\ell,L) = 
C_\alpha\left[ \ln L + \ln\sin(\pi X) + e_\alpha(\gamma)
\right]  \nonumber \\
&&\;\; - {\Gamma[1/2+1/(2\alpha)] \over 2 \alpha \Gamma[3/2-1/(2\alpha)]} 
\left[ {\pi \over 8 \gamma L \sin(\pi X)}\right]^{1/\alpha}
\nonumber \\
&&\;\;+\,O(L^{-2/\alpha}) \,+ \, O(L^{-1}),
\label{salpha} 
\end{eqnarray}
where
\begin{eqnarray}
&&C_\alpha\equiv  c {1+\alpha^{-1}\over 12},\quad c=1/2, 
\qquad X\equiv\ell/L,
\label{Calpha}\\
&&e_\alpha(\gamma) =  \ln\gamma + \ln(8/\pi) + \int_0^\infty \frac{{\rm d} t}t 
 \Bigl[{6\over 1-\alpha^{-2}} \times \nonumber \\
&&\quad\times \left(\frac1{\alpha\sinh t/\alpha} -
      \frac1{\sinh t}\right) \frac1{\sinh t}- {e}^{-2t}\Bigr].
\label{egamma}
\end{eqnarray}
Note that scaling corrections are proportional to
$\gamma^{-1/\alpha}$, a property which also holds outside the critical
point, see Sec.~\ref{xychainent}.
Eq.~(\ref{salpha}) does not allow us to compute the
corrections for the vN entropy. Indeed, for $\alpha = 1$, there are
two sources of $L^{-1}$ terms: the conical corrections and the
analytic boundary corrections.  For $\gamma=1$ we have~\cite{CV-10-jstat}
\begin{eqnarray}
&&S_1(\ell,L) = 
{1\over 12} \Bigl[ \ln L_e + \ln\sin{\pi X_e} + e_1(1)\Bigr]
\nonumber \\
&&\; - {\pi \over 16 L \sin(\pi X)} + O(L^{-2}),
\label{salpha1} 
\end{eqnarray}
where $L_e = L+1/2$, and $X_e\equiv \ell_e/L_e$ with $\ell_e = \ell +
1/4$. Thus, after appropriately shifting $\ell$ and $L$ to $\ell_e$
and $L_e$ respectively, the remaining $L^{-1}$ correction term turns
out to be equal to the limit $\alpha\to 1$ of the correction of order
$L^{-1/\alpha}$ appearing in Eq.~(\ref{salpha}).  Therefore, at the
critical point the leading $O(L^{-1})$ correction in the vN entropy
shows both conical and boundary contributions.  However, the latter
can be reabsorbed be redefining both length scales $L$ and $\ell$.
Actually, for $\gamma=1$, replacing $L\to L_e=L+1/2$ and $\ell \to
\ell_e=\ell+1/4$ in Eq.~(\ref{salpha}), we also obtain the $O(L^{-1})$
corrections for general $S_\alpha$.

In the case of PBC, using the results for the XX model reported in
Refs.~\onlinecite{CCEN-10,CMV-11-jstat}, we obtain for $\alpha>1$ and
$\gamma=1$
\begin{eqnarray}
&&S_\alpha(\ell,L)|_{\gamma=1} = 
2 C_\alpha \left[ \ln L + \ln\sin(\pi X)
+ \widetilde{e}_\alpha(1) \right]  \nonumber \\
&&\; - (\alpha-1){\Gamma[1/2+1/(2\alpha)]^2\over
4 \alpha^2 \Gamma[3/2-1/(2\alpha)]^2} 
\left[ {\pi \over 4 L \sin(\pi X)}\right]^{2/\alpha}
\nonumber \\
&&\; + \; O(L^{-4/\alpha}) \;+ \; O(L^{-2}),
\label{salphap} 
\end{eqnarray}
where $\widetilde{e}_\alpha(\gamma) = e_\alpha(\gamma) - \ln 2$. 
For the vN entropy we instead obtain 
\begin{eqnarray}
&&S_1(\ell,L)|_{\gamma=1} = 
{1\over 6} \left[ \ln L + \ln\sin(\pi X) + 
\widetilde{e}_1(1) \right]  \nonumber\\
&&\quad - {\pi^2\over 480 L^2 \sin^2(\pi X)} + {\pi^2\over 144 L^2} + 
O(L^{-4}).
\label{salpha1p} 
\end{eqnarray}
Note that the limit $\alpha\to 1$ of the corrections of order
$L^{-2/\alpha}$ in Eq.~(\ref{salphap}) vanishes, hence in the PBC case
the leading conical singularities do not contribute at the
critical point.


\begin{thebibliography}{99}

\bibitem{FB-72}
M. E. Fisher and M. N. Barber, Phys. Rev. Lett. {\bf 28}, 1516 (1972).

\bibitem{Barber-83}
M. N. Barber, in {\em Phase Transitions and Critical Phenomena},
edited by C. Domb and J. L. Lebowitz (Academic Press, New York, 1983),
Vol. 8.

\bibitem{Privman-90} V. Privman ed., 
{\em Finite Size Scaling and Numerical Simulation of Statistical Systems}
\/ (World Scientific, Singapore, 1990). 

\bibitem{PHA-91} 
V.~Privman, P.~C.~Hohenberg, and A.~Aharony, 
in {\em Phase Transitions and Critical Phenomena}, Vol.\ 14,
edited by C.~Domb and J.~L.~Lebowitz (Academic Press, New York, 1991).

\bibitem{PV-02} A. Pelissetto and E. Vicari, Phys. Rep. {\bf 368}, 549 (2002).

\bibitem{SGCS-97}
S. L. Sondhi, S. M. Girvin, J. P. Carini, and D. Shahar,
Rev. Mod. Phys. {\bf 69}, 316 (1997).

\bibitem{GKMD-08} 
F. M. Gasparini, M. O. Kimball, K. P. Mooney, and M. Diaz-Avilla,
Rev. Mod. Phys. {\bf 80}, 1009 (2008).

\bibitem{CW-02}
E. A. Cornell and C. E. Wieman, Rev. Mod. Phys. {\bf 74}, 875 (2002).

\bibitem{Ketterle-02}
N. Ketterle, Rev. Mod. Phys. {\bf 74}, 1131 (2002).

\bibitem{DRBOKS-07}
T. Donner, S. Ritter, T. Bourdel, A. \"Ottl, M. K\"ohl, and
T. Esslinger, Science {\bf 315}, 1556 (2007).

\bibitem{BDZ-08} 
I. Bloch, J. Dalibard, and W. Zwerger, 
Rev.\ Mod.\ Phys.\ {\bf 80}, 885 (2008).

\bibitem{CV-09} M. Campostrini and E. Vicari, 
Phys. Rev. Lett. {\bf 102}, 240601 (2009); (E) {\bf 103}, 269901 (2009).

\bibitem{Sachdev-book} S. Sachdev, {\em Quantum Phase Transitions}, (Cambridge
  Univ. Press, Cambridge, MA, 1999).

\bibitem{Wegner-76} 
F. J. Wegner, in
{\em Phase Transitions and Critical Phenomena},
edited by C.~Domb and M.~S.~Green 
(Academic Press, New York, 1976), Vol.\ 6.

\bibitem{LSM-61}
E. Lieb, T. Schultz, and D. Mattis, Ann. Phys. (NY) {\bf 16}, 407 (1961).

\bibitem{Katsura-62}
S. Katsura, Phys. Rev. {\bf 127}, 1508 (1962).

\bibitem{CC-10} P. Calabrese and J. Cardy,  J. Stat. Mech. P04023 (2010).

\bibitem{CCP-10}
P. Calabrese, J. Cardy, and I. Peschel, J. Stat. Mech. P09003 (2010).

\bibitem{Affleck-86}
I. Affleck, Phys. Rev. Lett. {\bf 56}, 746 (1986).

\bibitem{CSY-94}
A. V. Chubukov, S. Sachdev, and J. Ye,
Phys. Rev. B {\bf 49}, 11919 (1994).

\bibitem{Zia-review}
B. Schmittmann and R. K. P. Zia, in
{\em Phase Transitions and Critical
Phenomena}, edited by C. Domb and J.  L. Lebowitz (Academic Press, N.Y., 1995)
Vol. 17

\bibitem{Diehl-86} H. W. Diehl, 
{\em Field Theoretical Approach at Surfaces in Phase Transitions and 
Critical Phenomena}, 
edited by C. Domb and J. L. Lebowitz, Vol. 10 (Academic Press, London, 1986)
p. 76.

\bibitem{SS-00}
J. Salas and A. D. Sokal,
J. Stat. Phys.  {\bf 98}, 551  (2000); for an extensive discussion of FSS, 
see the longer arXiv version of this paper, cond-mat/9904038v1 (1999).

\bibitem{Hasenbusch-09}
M. Hasenbusch, J. Stat. Mech. P02005 (2009); 
J. Stat. Mech. P07031 (2009).

\bibitem{Hasenbusch-12}
M. Hasenbusch, Phys. Rev. B {\bf 85}, 174421 (2012).

\bibitem{DGHHRS-12}
H. W. Diehl, D. Gr\"uneberg, M. Hasenbusch, A. Hucht, S. B. Rutkevich, and
F. M. Schmidt, Europhys. Lett. {\bf 100}, 10004 (2012).

\bibitem{DC-09}
H. W. Diehl and H. Chamati, 
Phys. Rev. B {\bf 79}, 104301 (2009).

\bibitem{Kastening-12}
B. Kastening, Phys. Rev. E {\bf 86}, 041105 (2012).

\bibitem{Fisher-74-Temple}
M. E. Fisher, in 
{\em Renormalization Group in Critical Phenomena  and Quantum Field Theory}, 
edited by J. D. Gunton and M. S. Green
(Temple Univ. Press, Philadelphia, 1974) p.~65.

\bibitem{BKT-73} 
J. M. Kosterlitz and  D. J. Thouless,
J.\ Phys. C: Solid State {\bf 6},  1181 (1973);
V. L. Berezinskii, 
Zh. Eksp. Theor. Fiz. {\bf 59}, 907 (1970) 
[Sov. Phys. JETP {\bf 32}, 493 (1971)];
J. M. Kosterlitz, J. Phys. C {\bf 7}, 1046 (1974);
J. V. Jos\'e, L. P. Kadanoff, S. Kirkpatrick, and D. R. Nelson,
Phys. Rev. B {\bf 16}, 1217 (1977).

\bibitem{AGG-80}
D. J. Amit, Y. Y. Goldschmidt, and G. Grinstein,
J. Phys. A {\bf 13}, 585 (1980).

\bibitem{Hasenbusch-05}
M. Hasenbusch,  J. Phys. A {\bf 38}, 5869 (2005). 

\bibitem{PV-13} 
A. Pelissetto and E. Vicari, Phys. Rev. E {\bf 87}, 032105 (2013);
G. Ceccarelli, J. Nespolo, A. Pelissetto, and E. Vicari,
Phys. Rev.  B  {\bf 88}, 024517 (2013).

\bibitem{CHPV-02}
M. Caselle, M. Hasenbusch, A. Pelissetto, 
and E. Vicari, J. Phys. A {\bf 35}, 4861 (2002).

\bibitem{PV-99}
A. Pelissetto and E. Vicari,
Nucl. Phys. B {\bf 519}, 626 (1998);
Nucl. Phys. B {\bf 540}, 639 (1999).

\bibitem{CV-10-XX} M. Campostrini and E. Vicari, 
Phys. Rev. A {\bf 81},  063614 (2010).

\bibitem{CMV-11-jstat}
P. Calabrese, M. Mintchev, and E. Vicari,
J.  Stat. Mech. P09028 (2011).

\bibitem{CTV-12}
G. Ceccarelli, C. Torrero, and E. Vicari,
Phys. Rev. A {\bf 85}, 023616 (2012).

\bibitem{Henkel-87}
M. Henkel, J. Phys. A {\bf 20}, 995 (1987).

\bibitem{Reinicke-87a}
P. Reinicke, J. Phys. A {\bf 20}, 4501 (1987).

\bibitem{Reinicke-87b}
P. Reinicke, J. Phys. A {\bf 20}, 5325 (1987). 

\bibitem{footnote-Reinicke}
Ref.~\onlinecite{Reinicke-87a} reports $u_\mu \approx |\mu|/\gamma$, 
a result which cannot be correct since $u_\mu$ must be an analytic 
function of the model parameters. The absolute value is a consequence 
of an erroneous term $|z|$ appearing in Eq.~(5.3) of
Ref.~\onlinecite{Henkel-87}. The correct result does not have the absolute 
value, hence $u_\mu \approx \mu/\gamma$.

\bibitem{Pfeuty-70}
P. Pfeuty, Ann. Phys. {\bf 57}, 79 (1970).

\bibitem{BG-85} 
T. W. Burkhardt and I. Guim, J. Phys. A {\bf 18}, L33 (1985).

\bibitem{FS-78}
E. Fradkin and L. Susskind, Phys. Rev. D {\bf 17}, 2637 (1978).

\bibitem{NO-11} 
H. Nishimori and G. Ortiz, 
{\em Elements of Phase Transitions and Critical Phenomena},
Chap. 10 (Oxford Univ. Press, Oxford, 2011).

\bibitem{Onsager-44}
L. Onsager, Phys. Rev. {\bf 65}, 117 (1944).

\bibitem{FF-69}
A. E. Ferdinand and M. E. Fisher,
Phys. Rev. {\bf 185}, 832 (1969).

\bibitem{Queiroz-00}
S. L. A. de Queiroz, J. Phys. A {\bf 33}, 721 (2000).

\bibitem{Salas-01}
J. Salas, J. Phys. A {\bf 34}, 1311 (2001).

\bibitem{ONGP-01}
W. Orrick, B. Nickel, A. J. Guttman, and J. H. H. Perk,
J. Stat. Phys. {\bf 102}, 795 (2001).

\bibitem{IH-02}
N. Sh. Izmailian and C.-H. Hu,
Phys. Rev. E {\bf 65}, 036103 (2002).

\bibitem{IH-09}
N. Sh. Izmailian and C.-H. Hu,
Nucl. Phys. B {\bf 808}, 613 (2009).

\bibitem{CGNP-11}
Y. Chan, A. J. Guttman,  B. Nickel, and J. H. H. Perk,
J. Stat. Phys. {\bf 145}, 549 (2011).

\bibitem{Izmailian-12}
N. Sh. Izmailian,
Nucl. Phys. B {\bf 854}, 184 (2012).

\bibitem{Izmailian-13}
X. Wu, N. Izmailian, and W. Guo,
Phys. Rev. E {\bf 86}, 041149 (2012);
Phys. Rev. E {\bf 87}, 022124 (2013);
arXiv:1308.2040.

\bibitem{CP-98}
S. Caracciolo and A. Pelissetto, 
Phys. Rev. D {\bf 58}, 105007 (1998).

\bibitem{FB-72-ARMA}
M. E. Fisher and M. N. Barber,
Arch. Rat. Mech. Anal. {\bf 47}, 205 (1972).

\bibitem{footnotehh}
Eq.~(\ref{Delta1P-scal}) is essentially in agreement with the results
of Ref.~\onlinecite{Henkel-87}, see his Eq.~(5.3).  The only difference
concerns the linear term, which is reported erroneously as $|w|$. This
is probably due to the fact that the relevant states were incorrectly
identified for $g < 1$.

\bibitem{footnote-T}
More precisely, if $R_g = \sum a_n L^{-n}$, we have 
$T = a_1 + 15 a_4/(64 L^3) + \ldots$.

\bibitem{AFOV-08} L. Amico, R. Fazio, A. Osterloh, and V. Vedral, 
Rev. Mod. Phys.   {\bf 80}, 517 (2008).

\bibitem{CCD-09}
{\em Entanglement entropy in extended systems},
edited by P.~Calabrese, J.~Cardy, and B.~Doyon,
J. Phys. A {\bf 42}, 500301 (2009).

\bibitem{ECP-10}
J. Eisert, M. Cramer, and M. B. Plenio, 
Rev. Mod. Phys.  {\bf 82}, 277 (2010).

\bibitem{AP-06}
A. Kitaev and J. Preskill,
Phys. Rev. Lett. {\bf 96}, 110404 (2006).

\bibitem{LW-06}
M. Levin and X.-G. Wen,
Phys. Rev. Lett. {\bf 96}, 110405 (2006).

\bibitem{IL-08}
F. Igloi and Y.-C. Lin, J. Stat. Mech. P06004 (2008)

\bibitem{MA-10}
A. Montakhab and A. Asadian, Phys. Rev. A {\bf 82}, 062313 (2010)

\bibitem{XA-11} 
J.C. Xavier and F. C. Alcaraz, 
Phys. Rev. B {\bf 84}, 094410  (2011).

\bibitem{XA-12} 
J.C. Xavier and F. C. Alcaraz, Phys. Rev. B {\bf 85}, 024418
  (2012).

\bibitem{DLLS-12}
G. De Chiara, L. Lepori, M. Lewenstein, and A. Sampera,
Phys. Rev. Lett. {\bf 109}, 237208 (2012).

\bibitem{LDS-13}
G. De Chiara, L. Lepori, and A. Sampera, 
Phys. Rev. B {\bf 87}, 235107 (2013).

\bibitem{HLW-94} C. Holzhey, F. Larsen, and F. Wilczek, Nucl. Phys. B {\bf
  424}, 443 (1994).

\bibitem{CC-04}
P. Calabrese and J. Cardy, J. Stat. Mech. P06002 (2004).

\bibitem{JK-04}
B.-Q. Jin and V. E. Korepin,
J. Stat.\ Phys. {\bf 116}, 79 (2004).

\bibitem{IJ-08}
F. Igl\'oi  and R. Juh\'asz,
Europhys.\ Lett. {\bf 81}, 57003 (2008).

\bibitem{CCEN-10} 
P. Calabrese, M. Campostrini, F. H. L.  Essler and B. Nienhuis,
Phys. Rev. Lett. {\bf 104} 095701 (2010);
P. Calabrese and F. H. L. Essler, J. Stat. Mech. P08029 (2010).

\bibitem{FC-11}
M. Fagotti and P. Calabrese, J. Stat. Mech. P01017 (2011).

\bibitem{DLF-12}
A. De Luca and F. Franchini, 
Phys. Rev. B {\bf 87}, 045118 (2013).

\bibitem{FIK-08}
F. Franchini, A. R. Its, and V. E. Korepin,
J. Phys. A {\bf 41}, 025302 (2008).

\bibitem{EEFR-12}
E. Ercolessi, S. Evangelisti, F. Franchini, and F. Ravanini,
Phys. Rev. B {\bf 85}, 115428 (2012).

\bibitem{footnoteextrapo} 
We use data in the range $128\le L \le 4096$ to obtain the large-$L$ limit.
The accuracy of the results, which is generally good, is 
limited by the numerical precision of the raw data.

\bibitem{It-Dr-book} C. Itzykson and J. M. Drouffe,
{\em Statistical Field Theory} (Cambridge Univ. Press, Cambridge 1989).


\bibitem{CFT-book}
P. Di Francesco, P. Mathieu, and D. Senechal,
{\em Conformal Field Theory} (Springer Verlag, New York, 1997).

\bibitem{CV-10-jstat} M. Campostrini and E. Vicari, 
 J. Stat. Mech. P08020 (2010); E04001 (2010).

\end{thebibliography}
\end{document}